\documentclass{PoS}
\usepackage{wrapfig}
\usepackage{tikz}
\usepackage{graphicx}
\usepackage{sidecap}

\title{New results on spectra and fluctuations from NA61/SHINE}

\ShortTitle{New results on spectra and fluctuations from NA61/SHINE}

\author{\speaker{Katarzyna Grebieszkow}
\hspace{0.1cm} for the  NA61/SHINE Collaboration \\
        Warsaw University of Technology \\
        E-mail: \email{kperl@if.pw.edu.pl}}

\abstract{
The NA61/SHINE experiment aims to discover the critical point of strongly interacting matter and study the properties of the onset of deconfinement. For this purpose we perform a two-dimensional scan of the ($T-\mu_{B}$) phase diagram by varying the energy ($5.1 < \sqrt{s_{NN}} < 16.8/17.3$ GeV) and the system size (p+p, p+Pb, Be+Be, Ar+Sc, Xe+La, Pb+Pb) of the collisions.

In this article the NA61/SHINE results on particle spectra as well as fluctuations and correlations in p+p, Be+Be, Ar+Sc, and Pb+Pb collisions are presented. In particular, the latest results on charged kaons spectra, charged pions ratios (electromagnetic effects), proton intermittency, and anisotropic flow
are discussed. 
Finally, the motivation, NA61/SHINE plans, and the first measurements of open charm production in heavy ion collisions at the Super Proton Synchrotron energies are shown.
}

\FullConference{Corfu Summer Institute 2018 "School and Workshops on Elementary Particle Physics and Gravity"\\
		(CORFU2018)\\
		31 August - 28 September, 2018\\
		Corfu, Greece}

\begin{document}

\section{Introduction and NA61/SHINE program}

NA61/SHINE at the CERN Super Proton Synchrotron (SPS) is a fixed-target experiment pursuing a rich physics program including measurements for strong interactions, neutrino, and cosmic ray physics.

Originally, for neutrino and cosmic ray programs, we planned to obtain precision data on hadron production (spectra) as reference measurements of p+C interactions for the T2K experiment for computing neutrino fluxes from the T2K beam targets, and reference measurements of p+C, $\pi$+p, p+p, and $\pi$+C interactions for cosmic-ray physics (Pierre-Auger and KASCADE experiments) for improving air shower simulations. Currently, those programs are extended by h+$A$ measurements for the Fermilab neutrino program and analysis for experiments located in space, namely the measurement of Nuclear Fragmentation Cross Sections (NFCS) of intermediate mass nuclei needed to understand the propagation of cosmic rays in our Galaxy (background for dark matter searches with space-based experiments as AMS).

\begin{wrapfigure}{r}{6.5cm}
\centering
\includegraphics[width=0.4\textwidth]{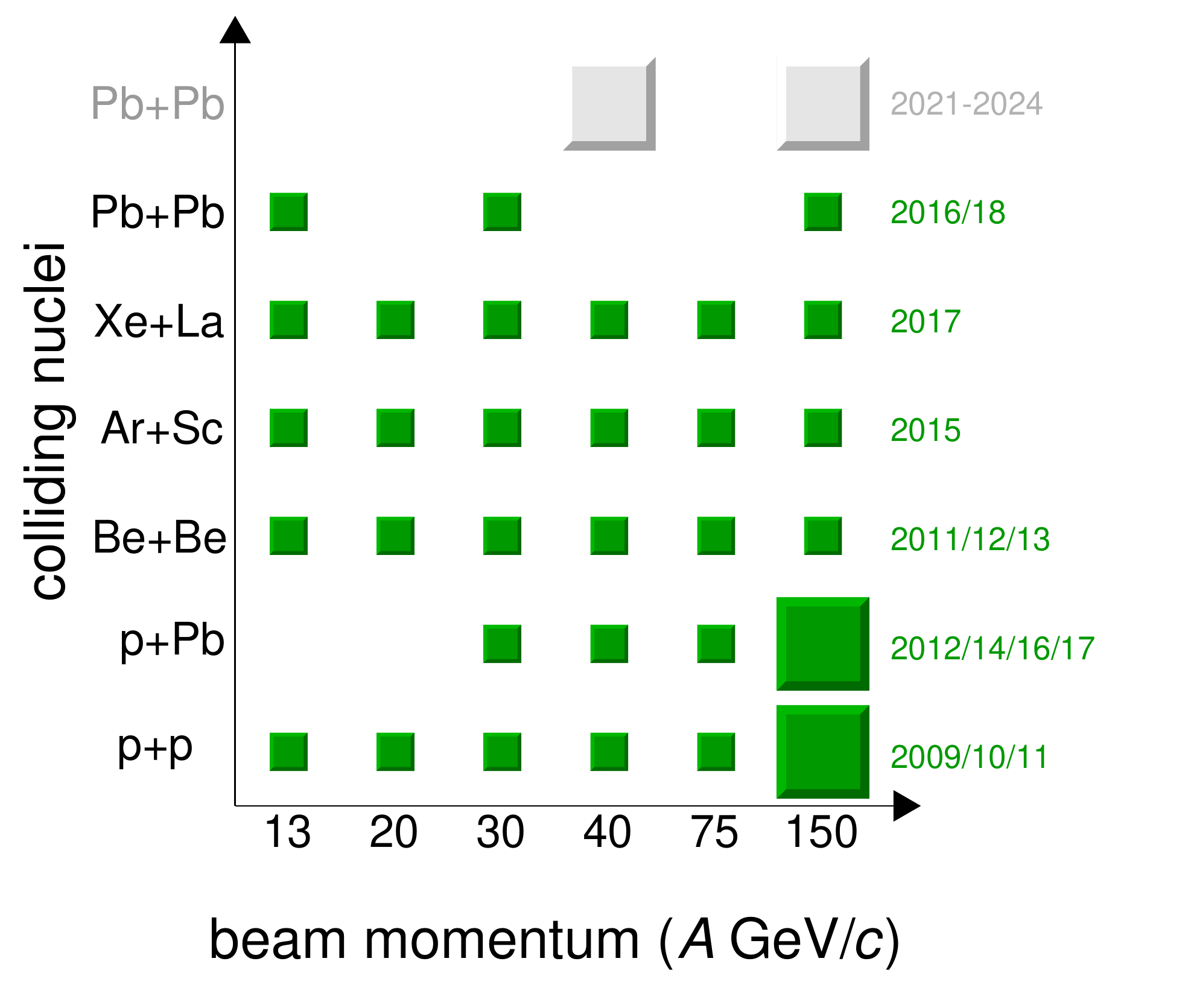}
\vspace{-0.2cm}
\caption[]{\footnotesize Data sets recorded within the strong interactions program of NA61/SHINE, and those planned to be recorded after year 2020.}
\label{box_plot}
\end{wrapfigure}

The main goal of the strong interactions program is to study the characteristics of the onset of deconfinement and search for the signatures of the critical point (CP). Presently, there are several more projects at energies corresponding to the SPS energy range: the RHIC Beam Energy Scan (BNL, Brookhaven), NICA (JINR, Dubna), J-PARC-HI (J-PARC, Tokai), and SIS-100 (FAIR GSI, Darmstadt). NA61/SHINE is the first experiment to perform a two-dimensional scan, in beam momentum (13$A$--150$A$/158$A$ GeV/c) and mass number (p+p, p+Pb, Be+Be, Ar+Sc, Xe+La, Pb+Pb) of colliding nuclei. For the study of the properties of the onset of deconfinement we search for the onset of the \textit{horn}, \textit{kink} and \textit{step}~\cite{Gazdzicki:1998vd} in collisions of light nuclei. In order to search for the critical point we look for a non-monotonic behavior of CP signatures such as fluctuations of transverse momentum and multiplicity, intermittency, etc., when the system freezes out close to the CP. The planned strong interactions program includes also the study of high $p_T$ particles (energy dependence of nuclear modification factor) in p+p and p+Pb interactions to better understand cold nuclear matter effects at SPS energies. 

The NA61/SHINE strong interactions program is currently being extended by Pb+Pb interactions. It will allow to obtain the first measurements of open charm production in heavy ion collisions at SPS energies. Moreover, the anisotropic flow effects, spectator-induced electromagnetic phenomena, as well as fluctuations are also planned to be measured in Pb+Pb events. The recorded systems and those planned to be recorded in the nearest future are presented in Fig.~\ref{box_plot}, where the size of the box is proportional to the number of recorded events.

\section{Can we see the onset of deconfinement in light and intermediate mass systems?}

\begin{figure}
\includegraphics[width=0.35\textwidth]{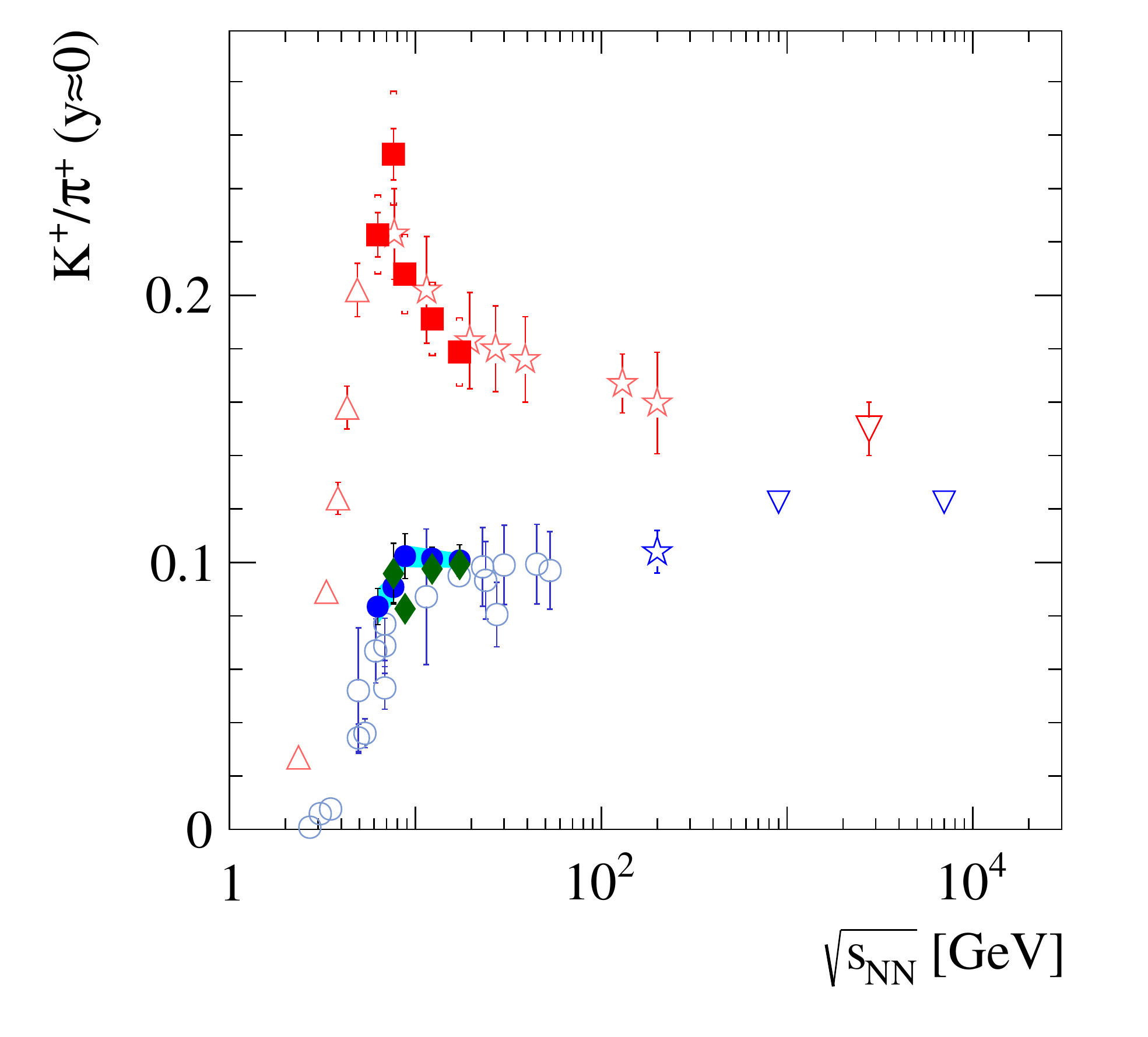}
\includegraphics[width=0.35\textwidth]{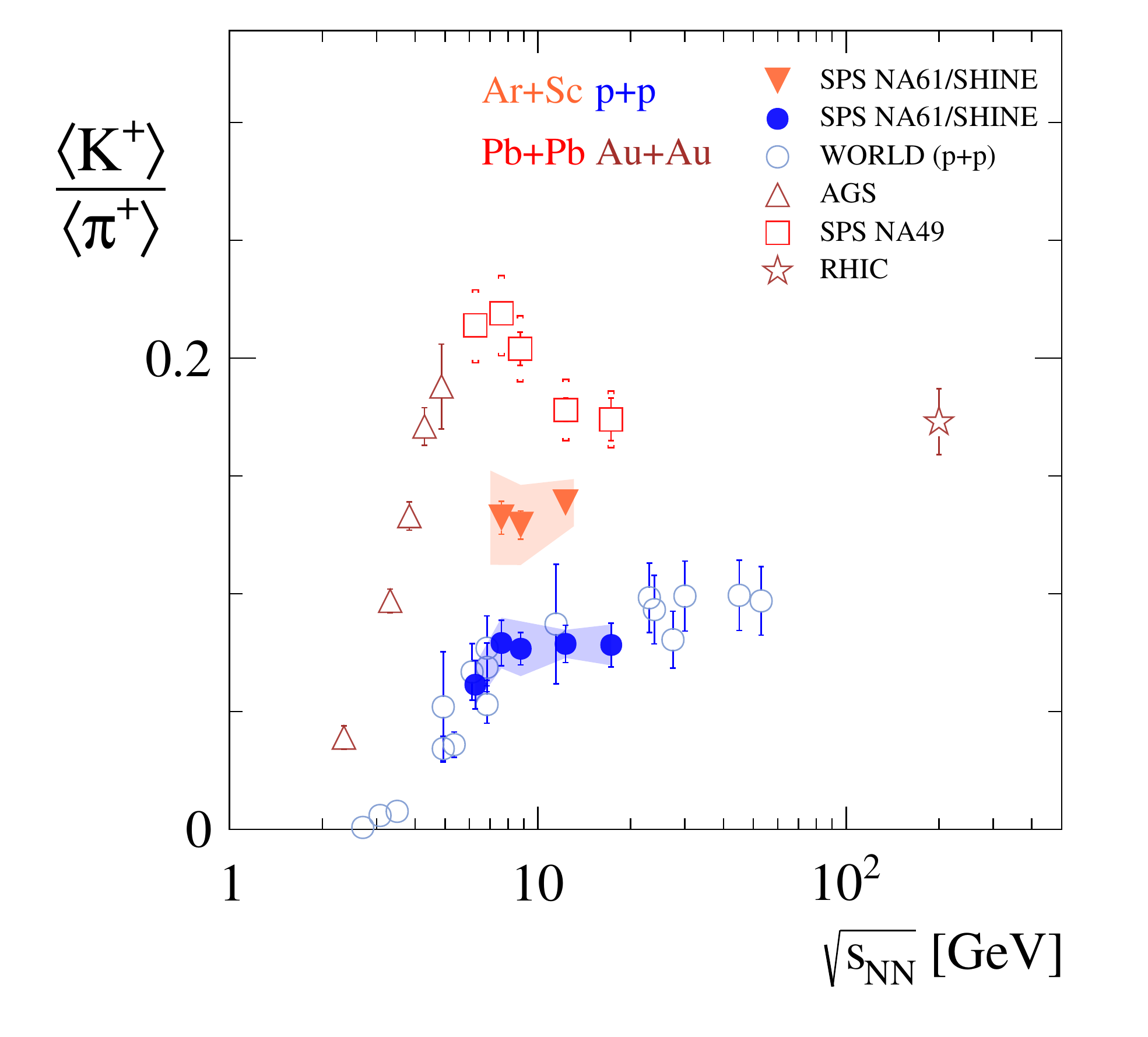}
\newline
\includegraphics[width=0.35\textwidth]{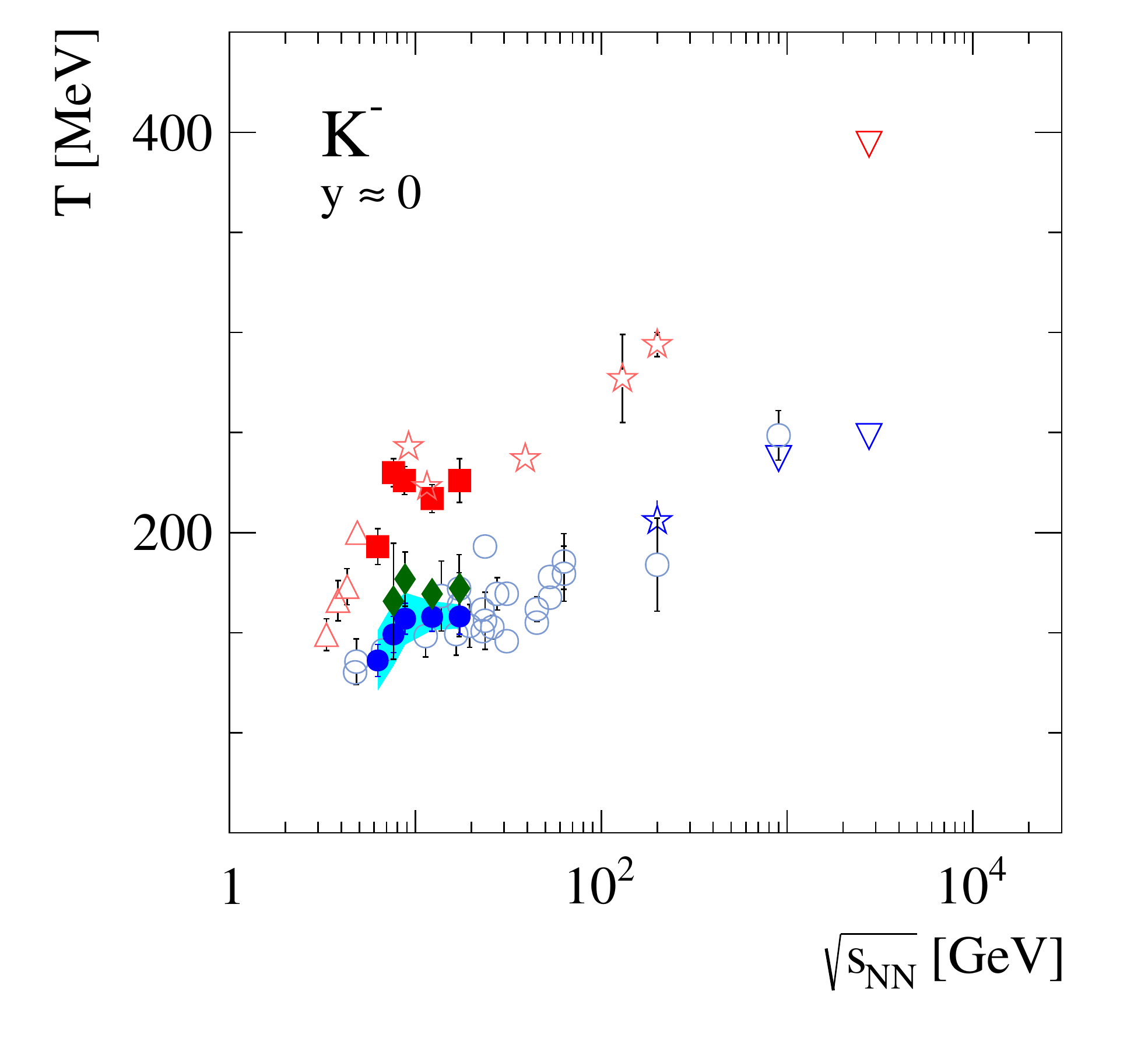}
\includegraphics[width=0.35\textwidth]{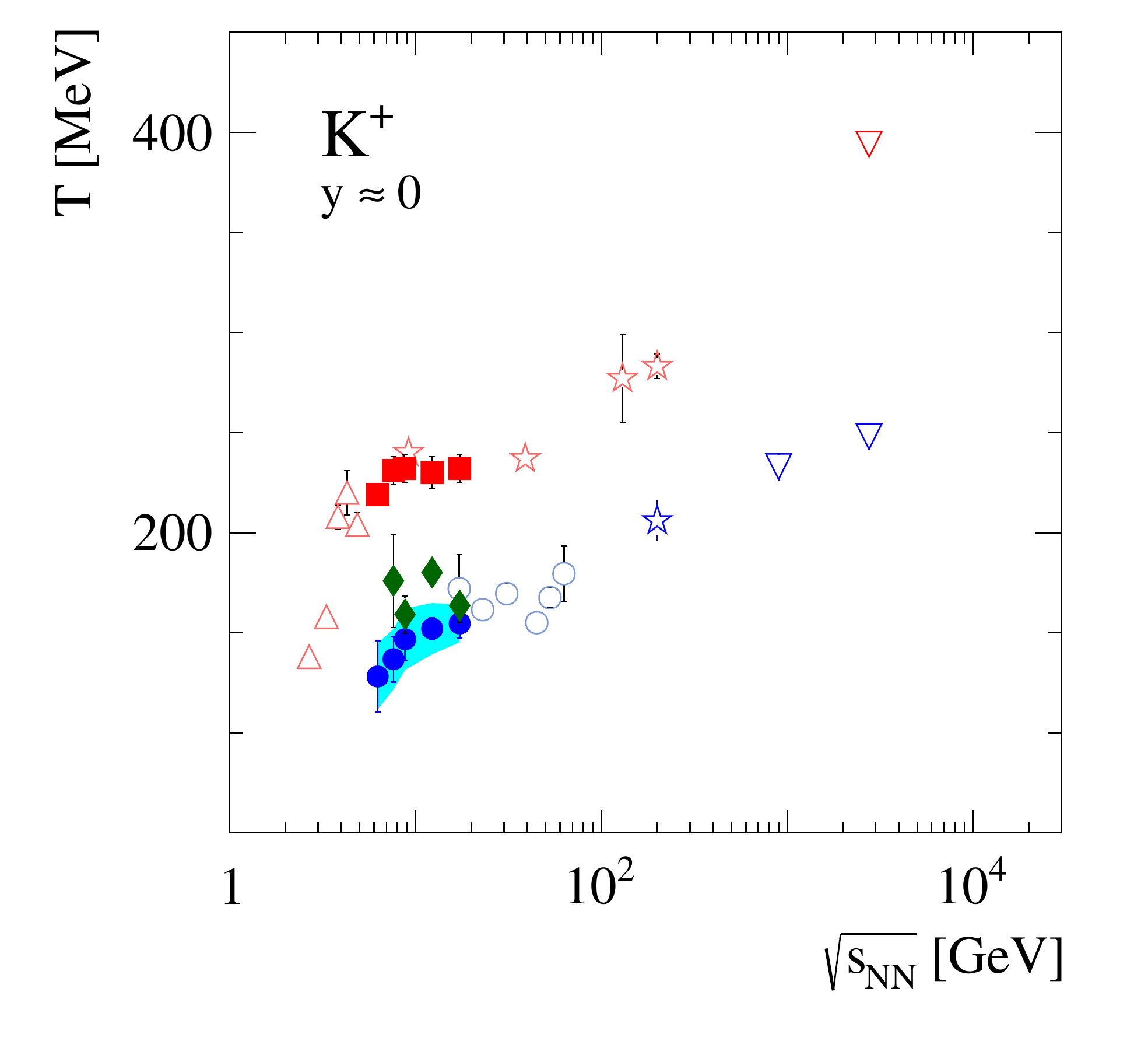}
\includegraphics[width=0.25\textwidth]{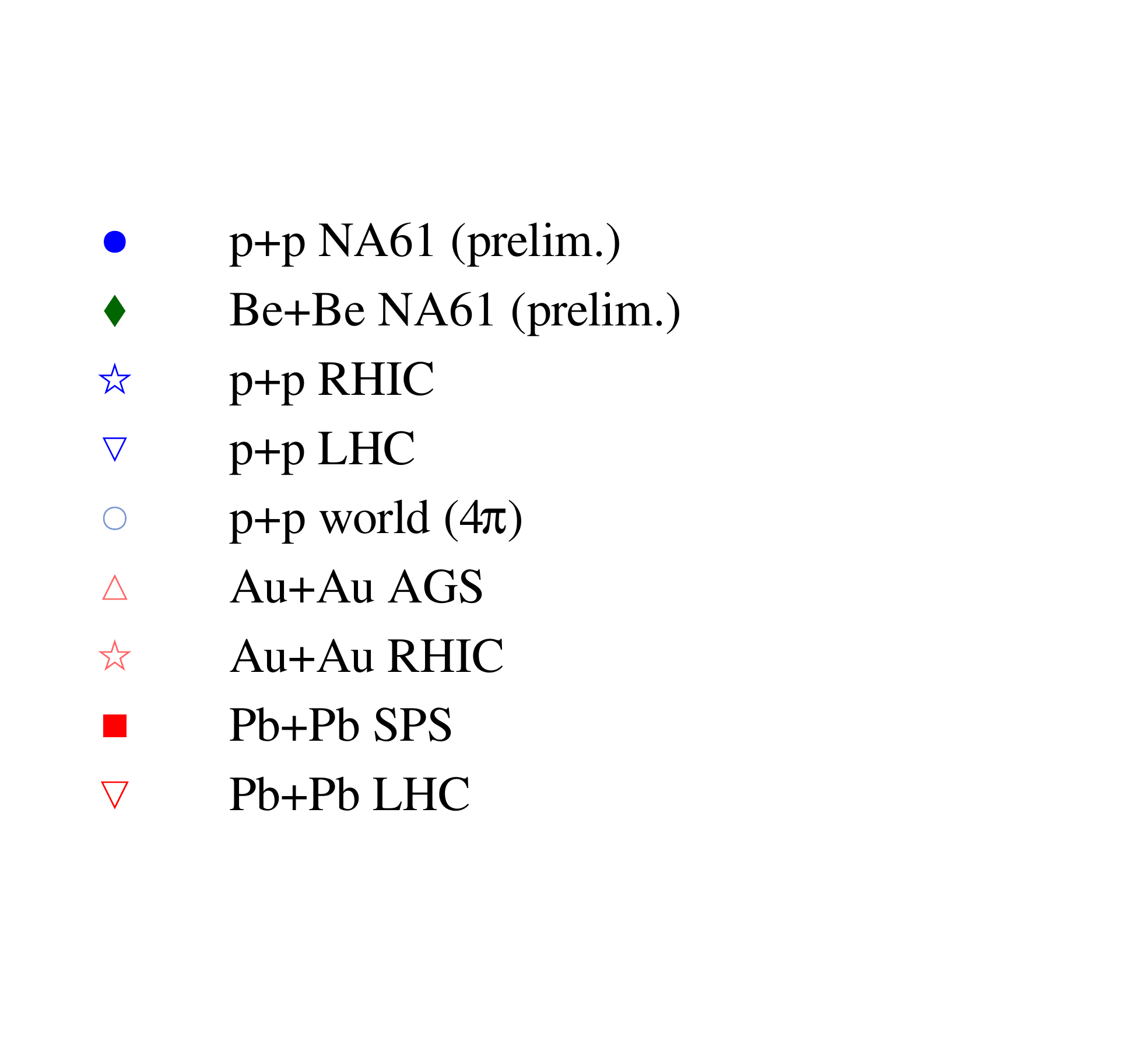}
\vspace{-0.2cm}
\caption[]{\footnotesize Upper panel: \textit{horn} plots -- $K^{+}/\pi^{+}$ ratio versus energy at mid-rapidity (left) and in $4\pi$ (right). Lower panel: \textit{step} plots -- inverse slope parameter ($T$) of $m_T$ spectra of $K^{-}$ (left) and $K^{+}$ (right) versus energy (status of plots as available before the CPOD 2018 conference). Kaon and pion multiplicities in p+p data are taken from Ref.~\cite{Aduszkiewicz:2017sei}, the rest of NA61/SHINE results are preliminary. See Ref.~\cite{Pulawski:2015tka} for references to world data.}
\label{horn_step_before_CPOD18}
\end{figure}

The spectra and yields analyzed in NA61/SHINE allow to study the properties of the onset of deconfinement by looking whether the \textit{kink}, \textit{horn}, and \textit{step}~\cite{Gazdzicki:1998vd} structures appear also in collisions of small and intermediate mass nuclei. For Pb+Pb interactions the NA49 experiment reported a sharp peak (\textit{horn}) in the $K^{+}/\pi^{+}$ ratio, which was interpreted as due to onset of deconfinement~\cite{Gazdzicki:2014sva}. Additionally, a plateau (\textit{step}) in the inverse slope parameter ($T$) of $m_T$ spectra was also observed as expected for constant temperature and pressure in a mixed phase. The status of \textit{horn} and \textit{step} plots, as available before the CPOD 2018 conference, is presented in Fig.~\ref{horn_step_before_CPOD18}. 

The recent NA61/SHINE results show that even in p+p collisions the energy dependences of the $K^{+}/\pi^{+}$ ratio and $T$ exhibit rapid changes (however without a maximum for $K^{+}/\pi^{+}$) in the SPS energy range. Moreover, the results for Be+Be are very close to the ones for p+p, independently of collision energy. For the currently available three energies the $\langle K^{+} \rangle / \langle \pi^{+}\rangle$ ratio in Ar+Sc collisions lies between the values for Pb+Pb and p+p. 


\begin{figure}
\centering
\includegraphics[width=0.35\textwidth]{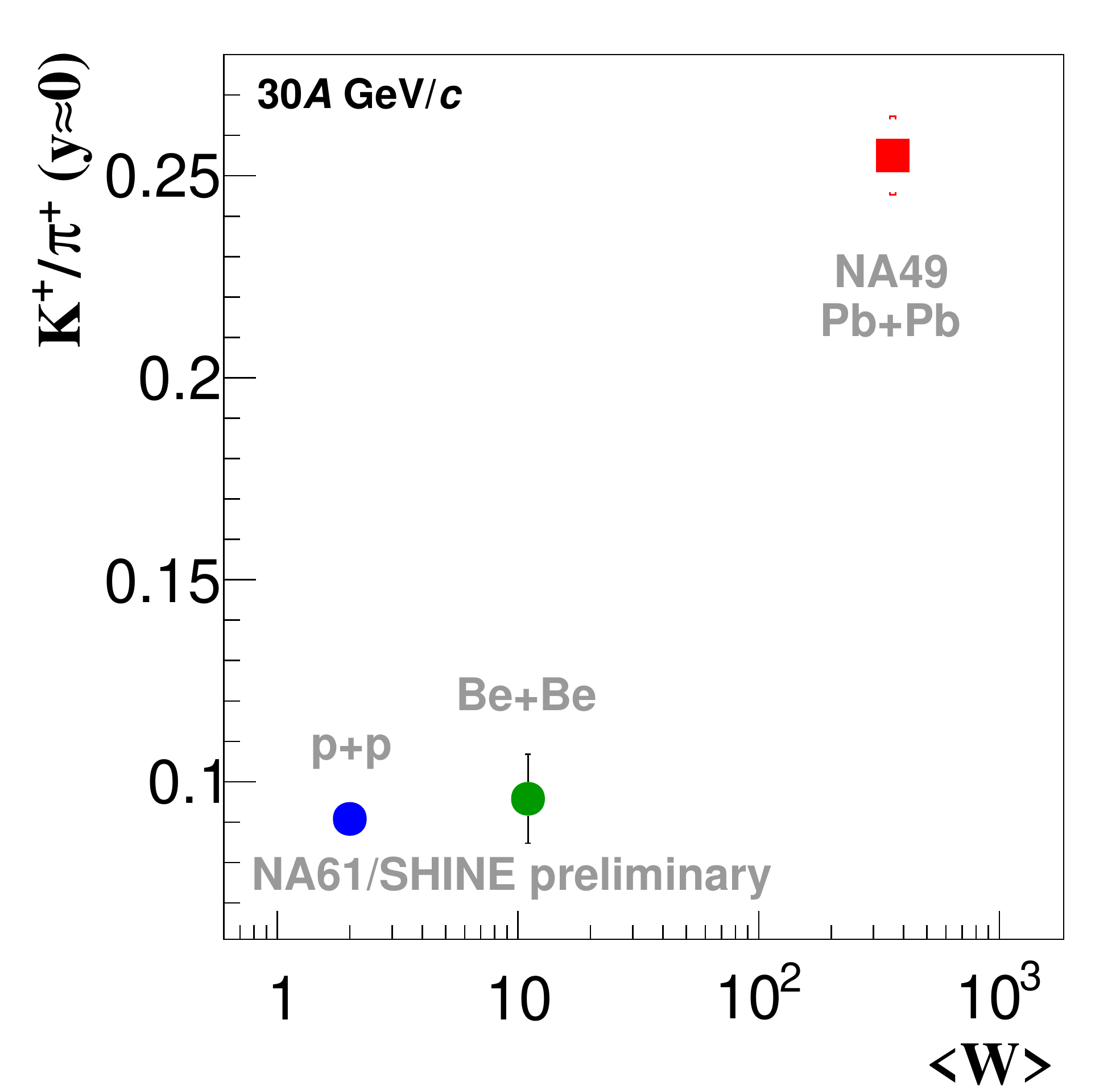}
\includegraphics[width=0.35\textwidth]{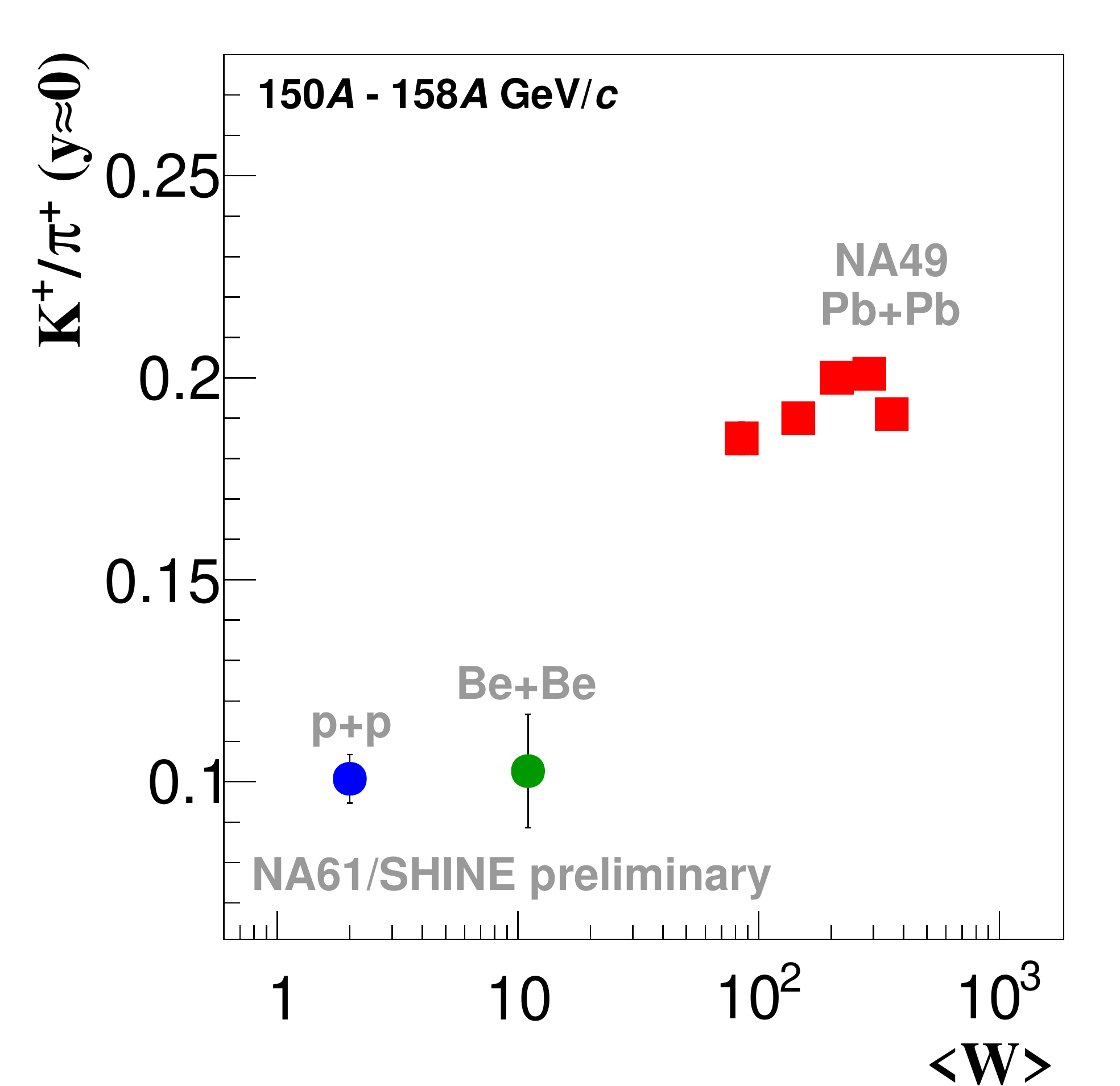}
\includegraphics[width=0.35\textwidth]{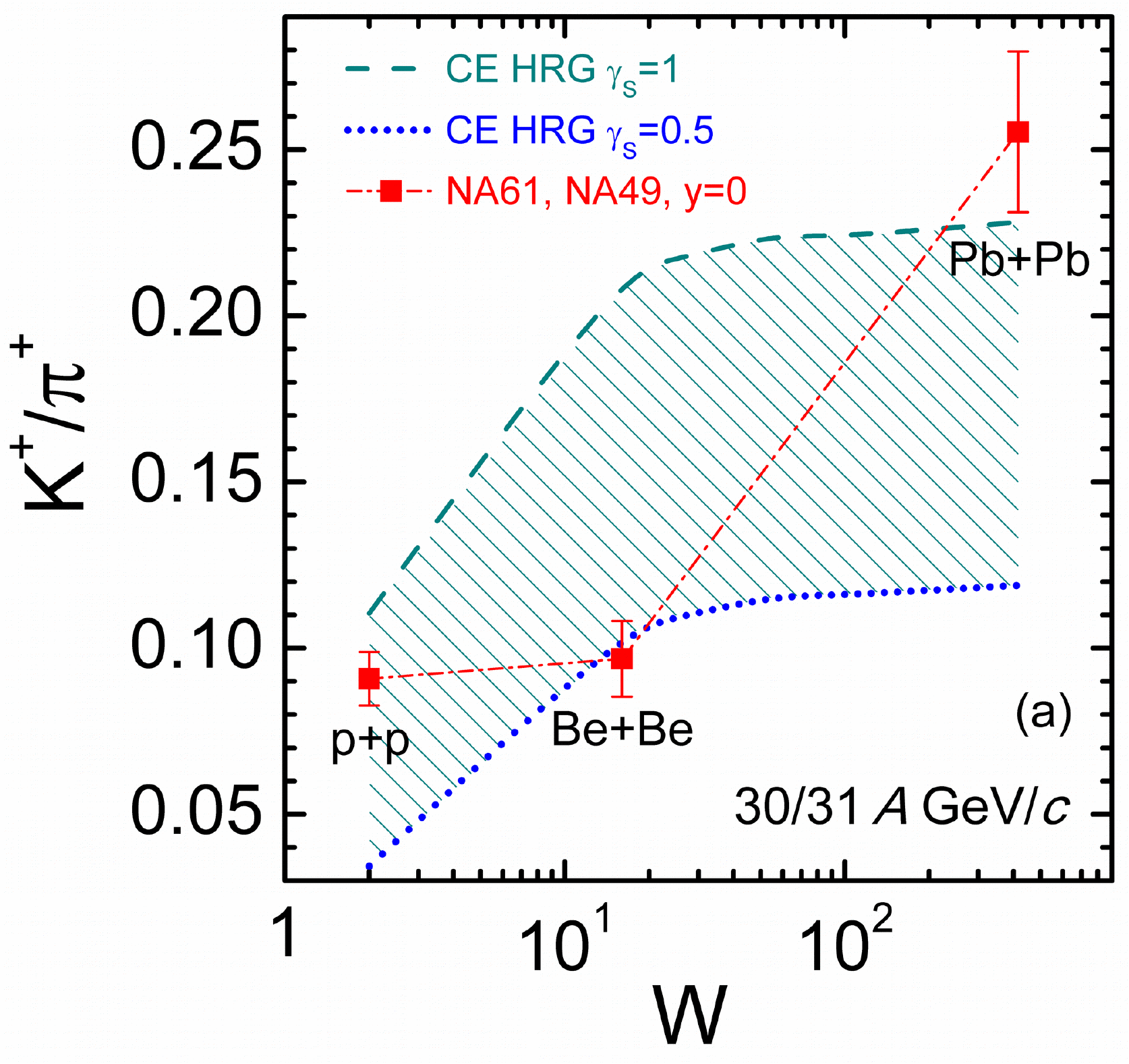}
\includegraphics[width=0.35\textwidth]{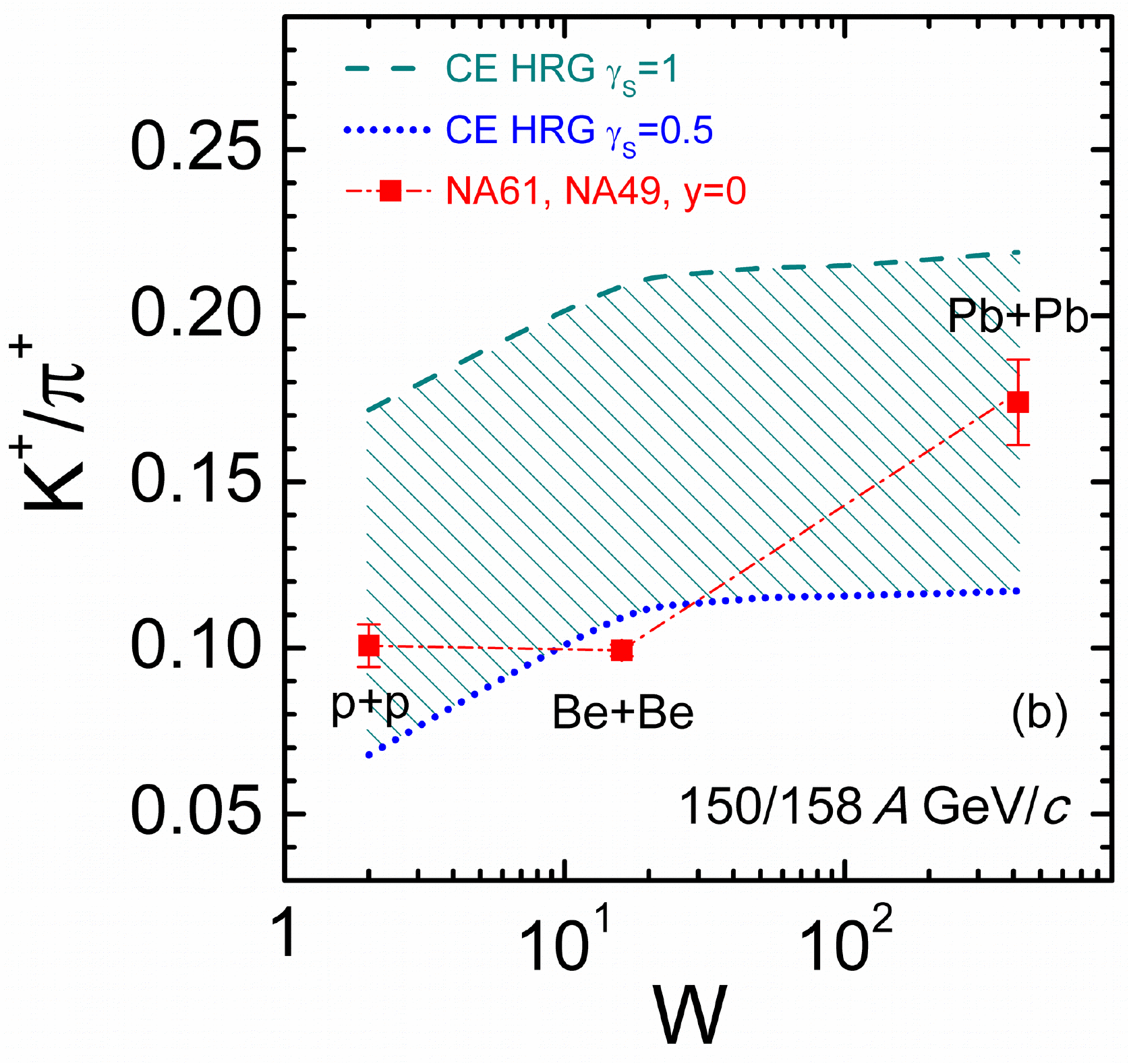}
\vspace{-0.2cm}
\caption[]{\footnotesize Upper panel: $K^{+}/\pi^{+}$ ratio versus mean number of wounded nucleons at mid-rapidity. Results for 30$A$/31$A$ GeV/c and 150$A$/158$A$ GeV/c beam momentum. NA61/SHINE p+p data are taken from Ref.~\cite{Aduszkiewicz:2017sei}, 0-20\% most central Be+Be results are preliminary. NA49 Pb+Pb data are taken from Refs.\cite{Alt:2007aa, Anticic:2012ay}.
Lower panel: comparison of NA61/SHINE results to predictions of the hadron resonance gas model in canonical ensemble formulation~\cite{Motornenko:2018gdc} (here $W=A_{1}+A_{2}$ represents the total number of nucleons in the colliding nuclei).}
\label{kpi_Nw}
\end{figure}

Figure~\ref{kpi_Nw} shows the system size (and centrality) dependence of the $K^{+}/\pi^{+}$ ratio at two SPS energies. The plot confirms the similarities of p+p and Be+Be collisions and clearly shows a rapid jump from p+p and Be+Be collisions (small clusters) to the heavy Pb+Pb system (fireball). The lower panel of Fig.~\ref{kpi_Nw} demonstrates that the hadron resonance gas model in canonical ensemble (CE HRG) formulation ($\gamma_{S}=1$) cannot describe NA61/SHINE data~\cite{Motornenko:2018gdc}.

\begin{figure}
\centering
\includegraphics[width=0.4\textwidth]{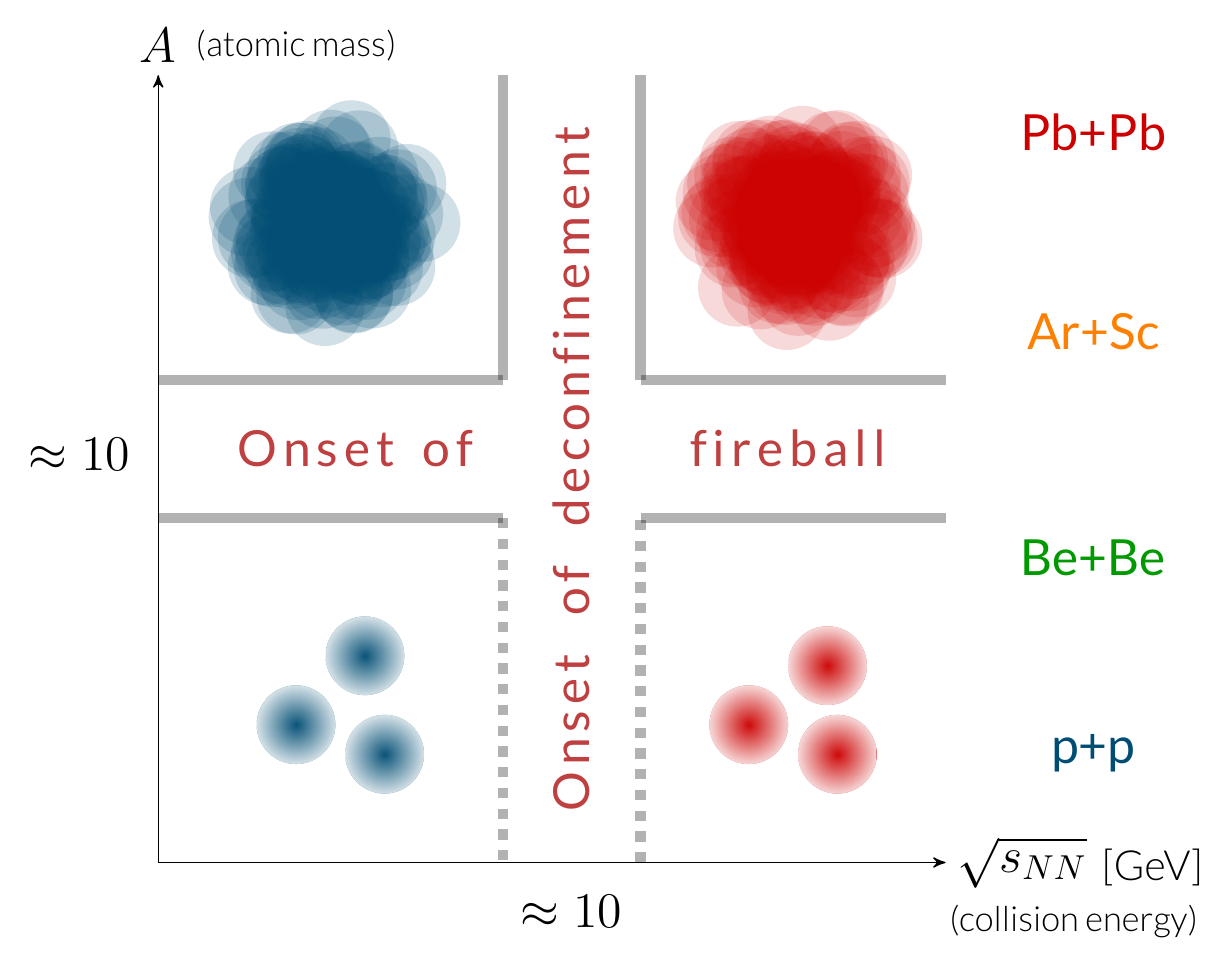}
\caption[]{\footnotesize Two-dimensional scan performed by NA61/SHINE by varying collision energy and nuclear mass number of colliding nuclei. The suggested four domains of hadron production properties are separated by two thresholds: the \textit{onset of deconfinement} and the \textit{onset of fireball}.}
\label{four_domains}
\end{figure}

The small clusters (such as p+p or Be+Be collisions) can be understood as independent, non-statistical small objects for which the Wounded Nucleon Model works. In contrast, a fireball can be treated as a large object that decays statistically. Based on the recent NA61/SHINE p+p and Be+Be results a new sketch of the energy -- system size plot has been proposed (Fig.~\ref{four_domains}). Two onsets
are expected in nucleus-nucleus collisions. Namely, the old and known \textit{onset of deconfinement},
where we observe the beginning of QGP formation with increasing collision energy, and the \textit{onset of fireball}, where we see the beginning of formation of a large cluster which decays statistically. According to NA61/SHINE results the onset of fireball should be located somewhere between Be+Be and Ar+Sc collisions.

\begin{figure}
\centering
\includegraphics[width=0.19\textwidth]{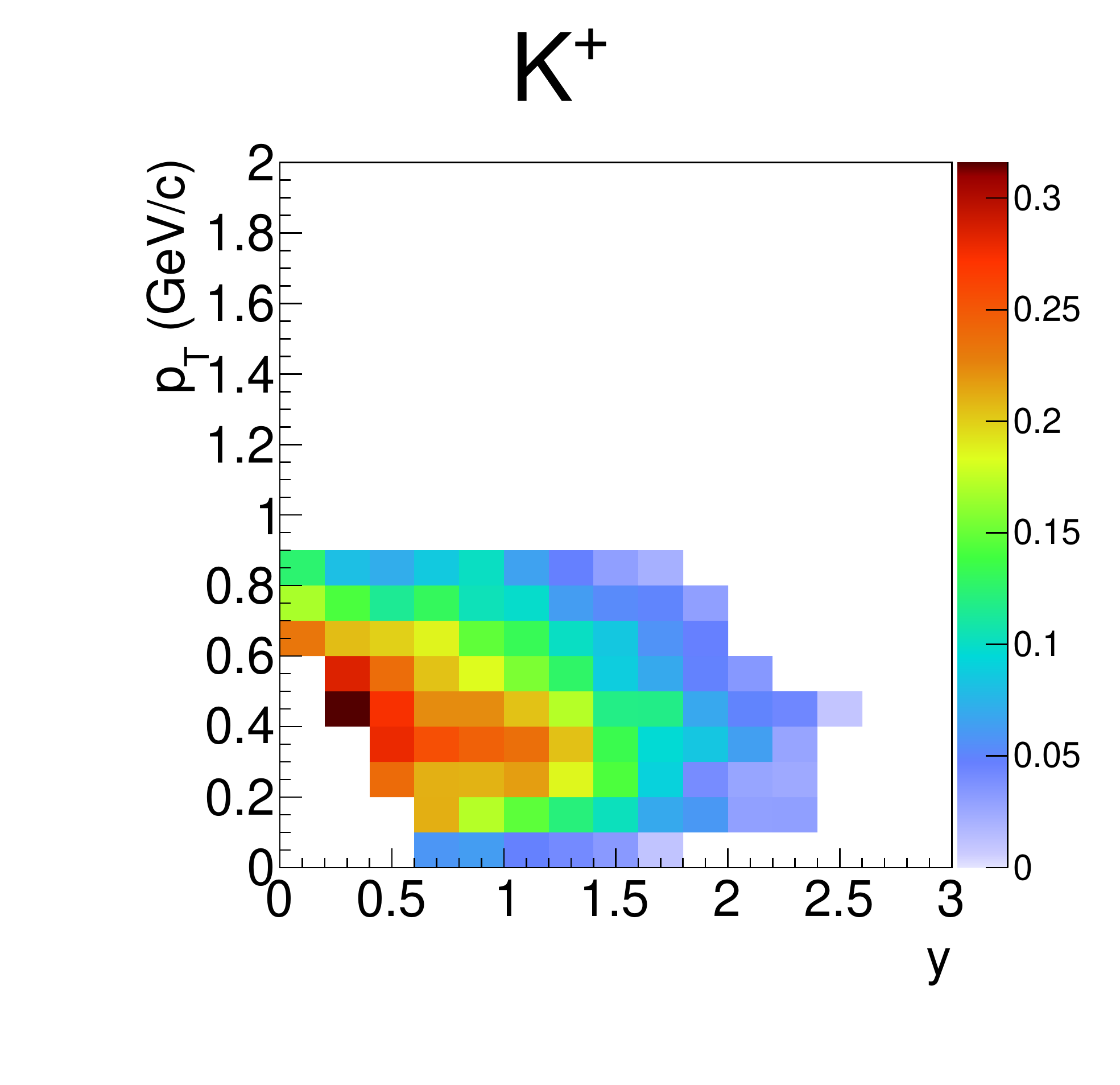}
\includegraphics[width=0.19\textwidth]{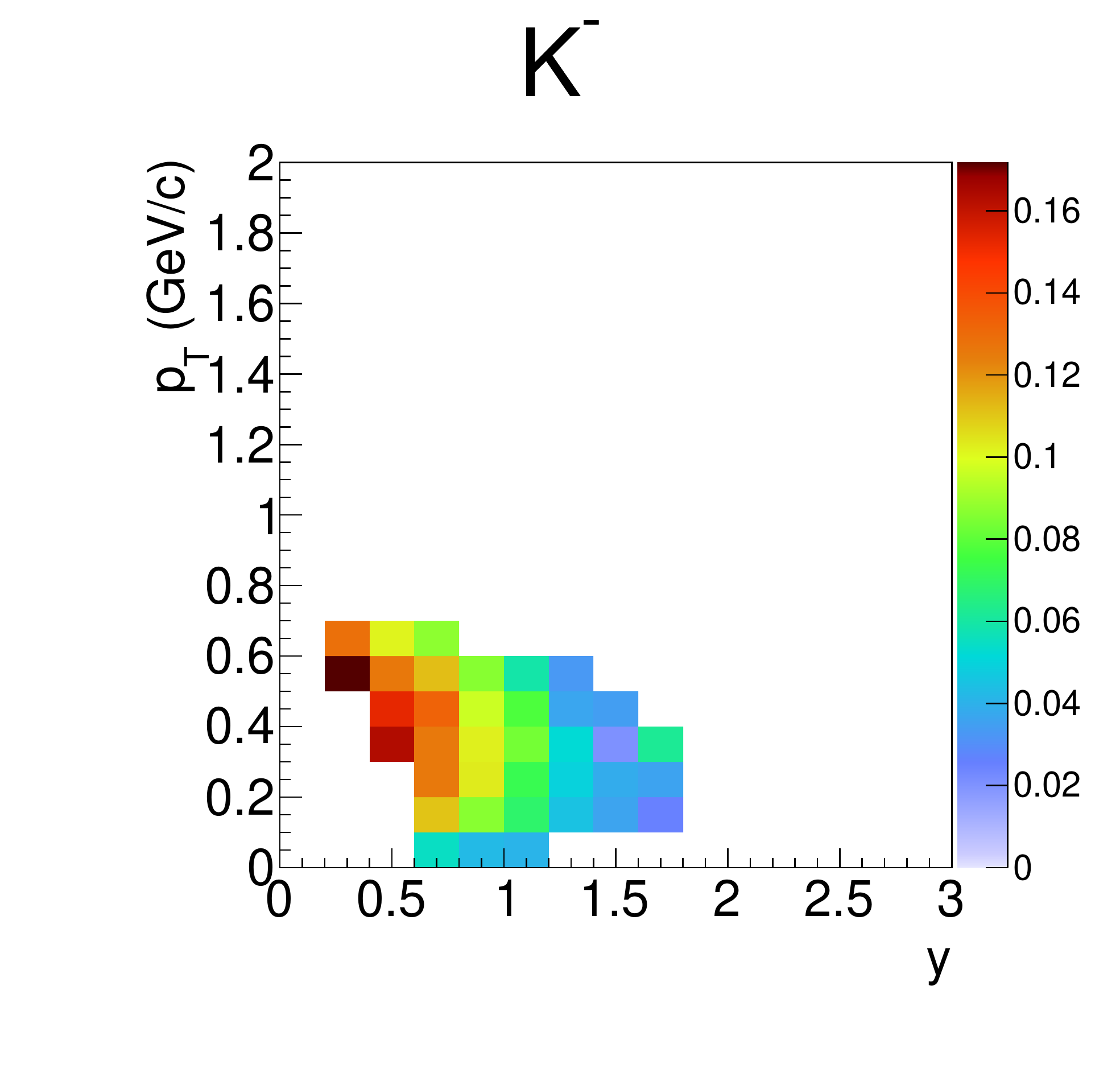}
\includegraphics[width=0.19\textwidth]{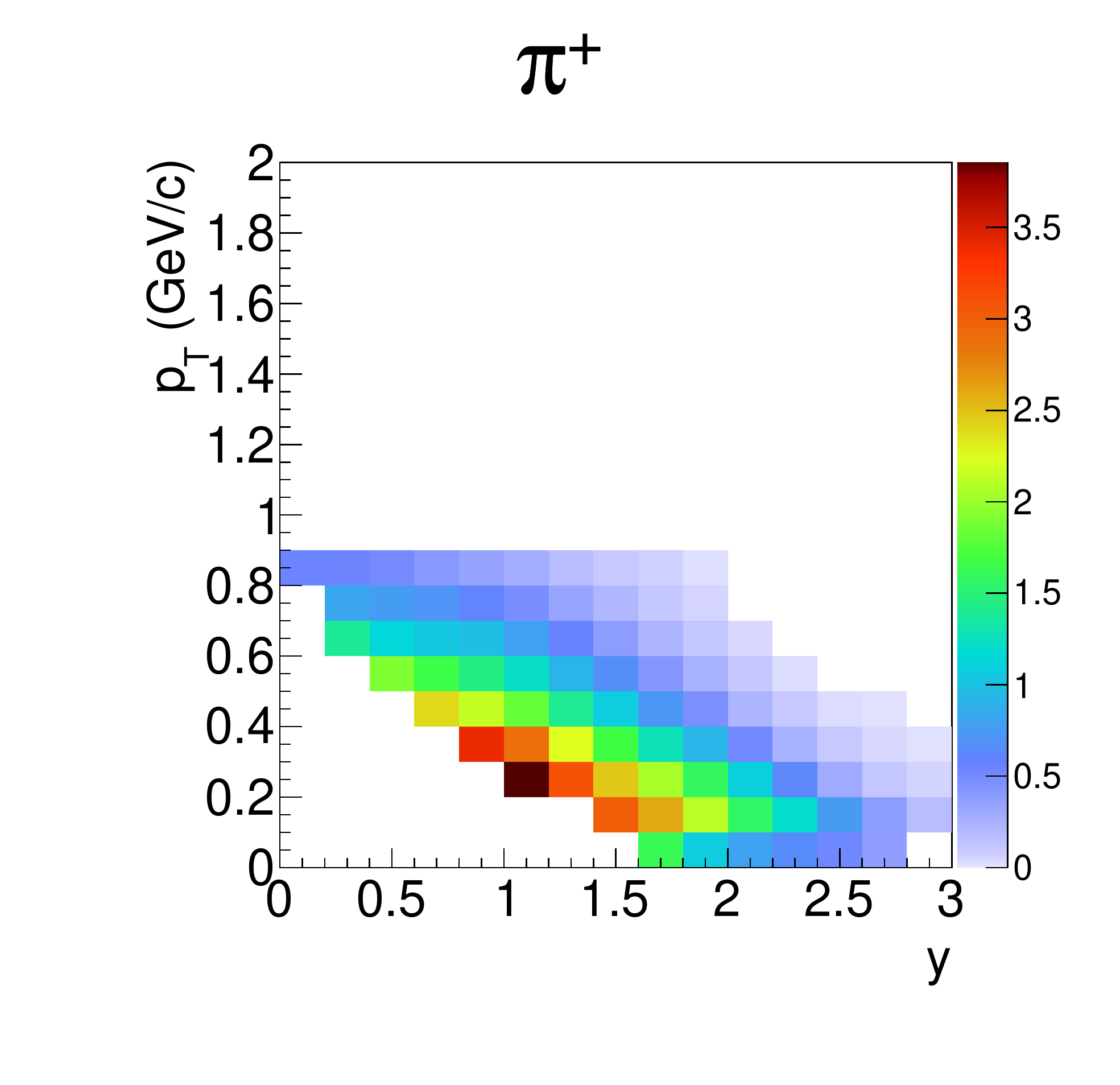}
\includegraphics[width=0.19\textwidth]{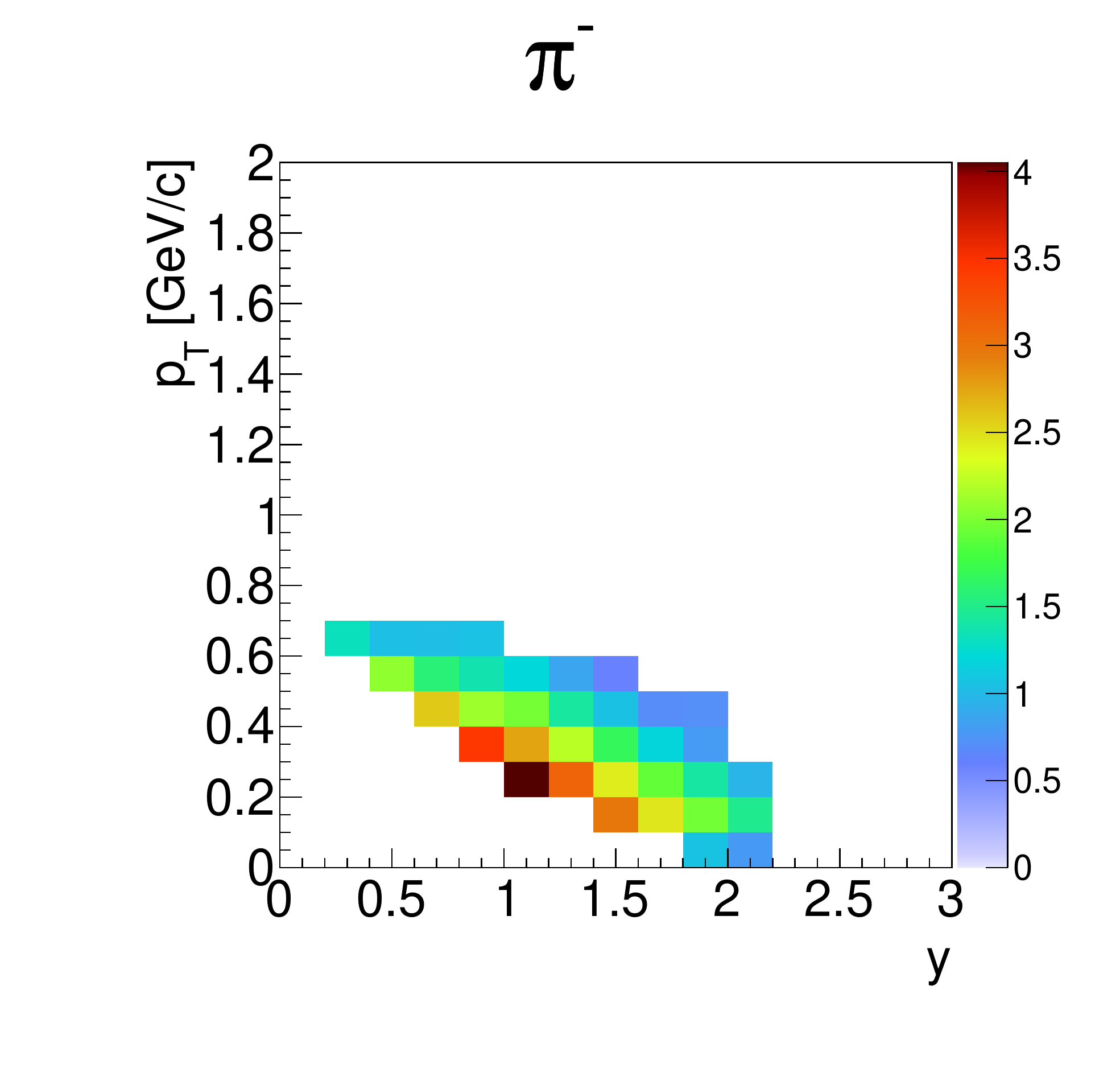}
\includegraphics[width=0.19\textwidth]{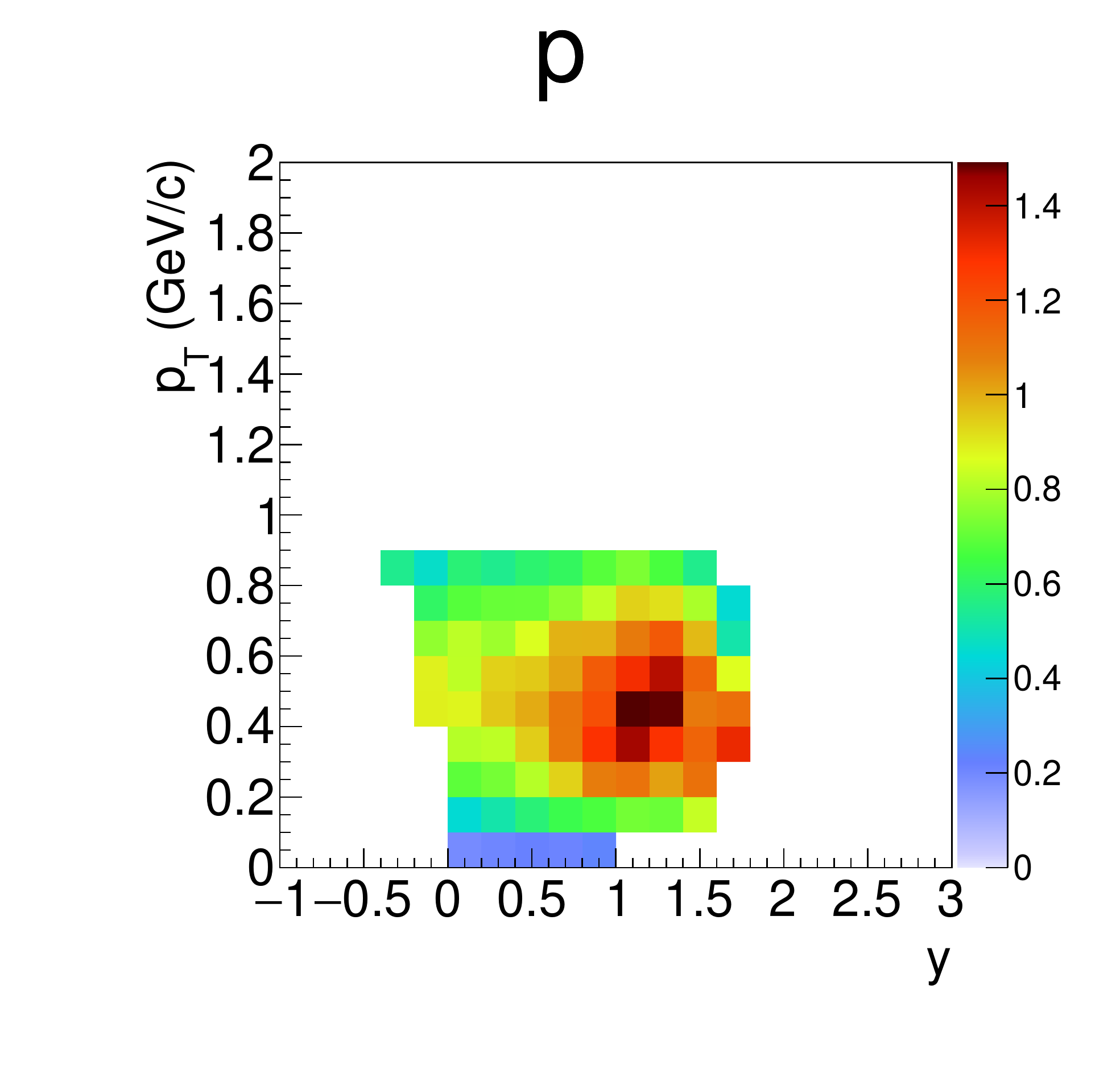}
\includegraphics[width=0.19\textwidth]{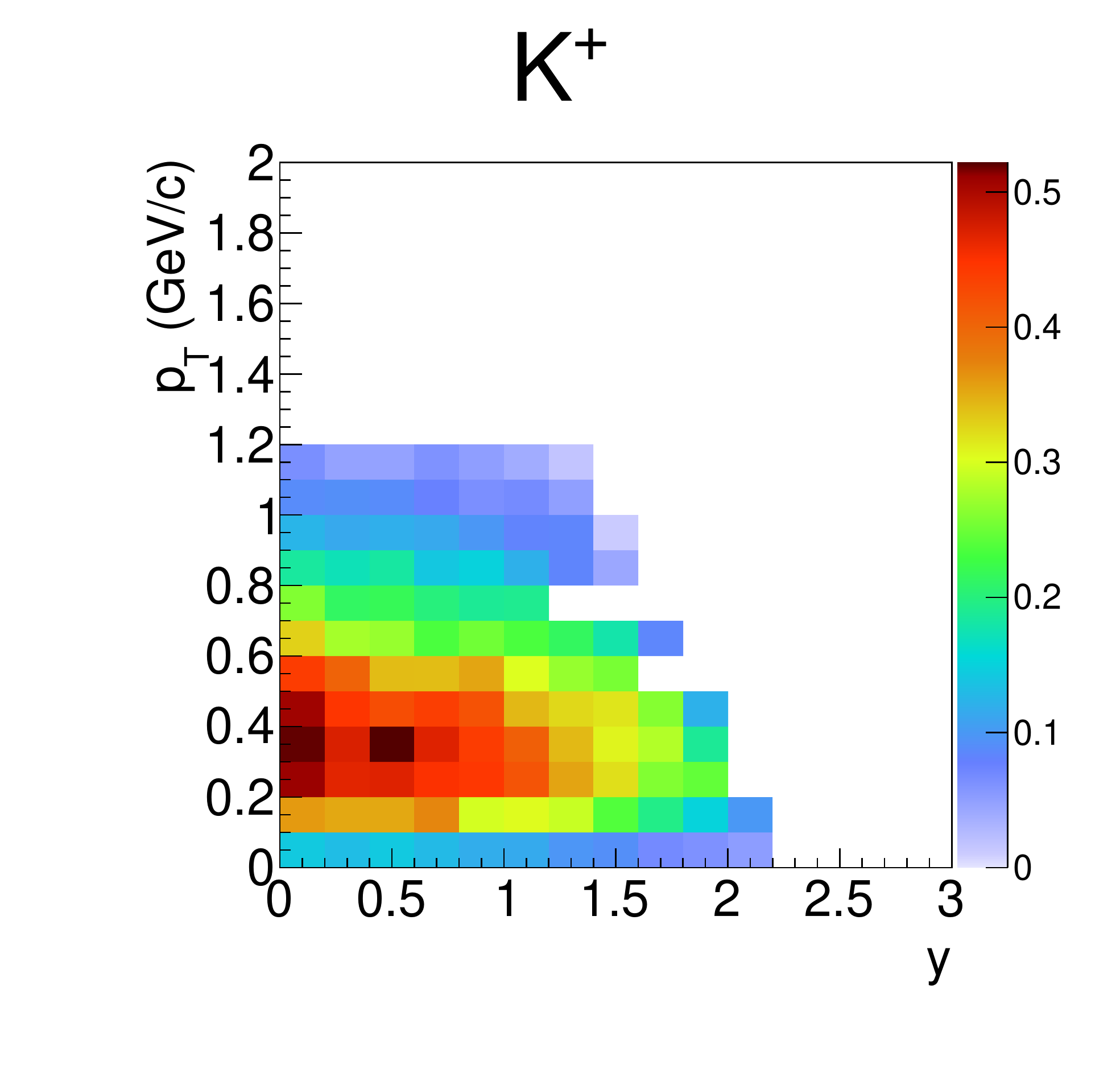}
\includegraphics[width=0.19\textwidth]{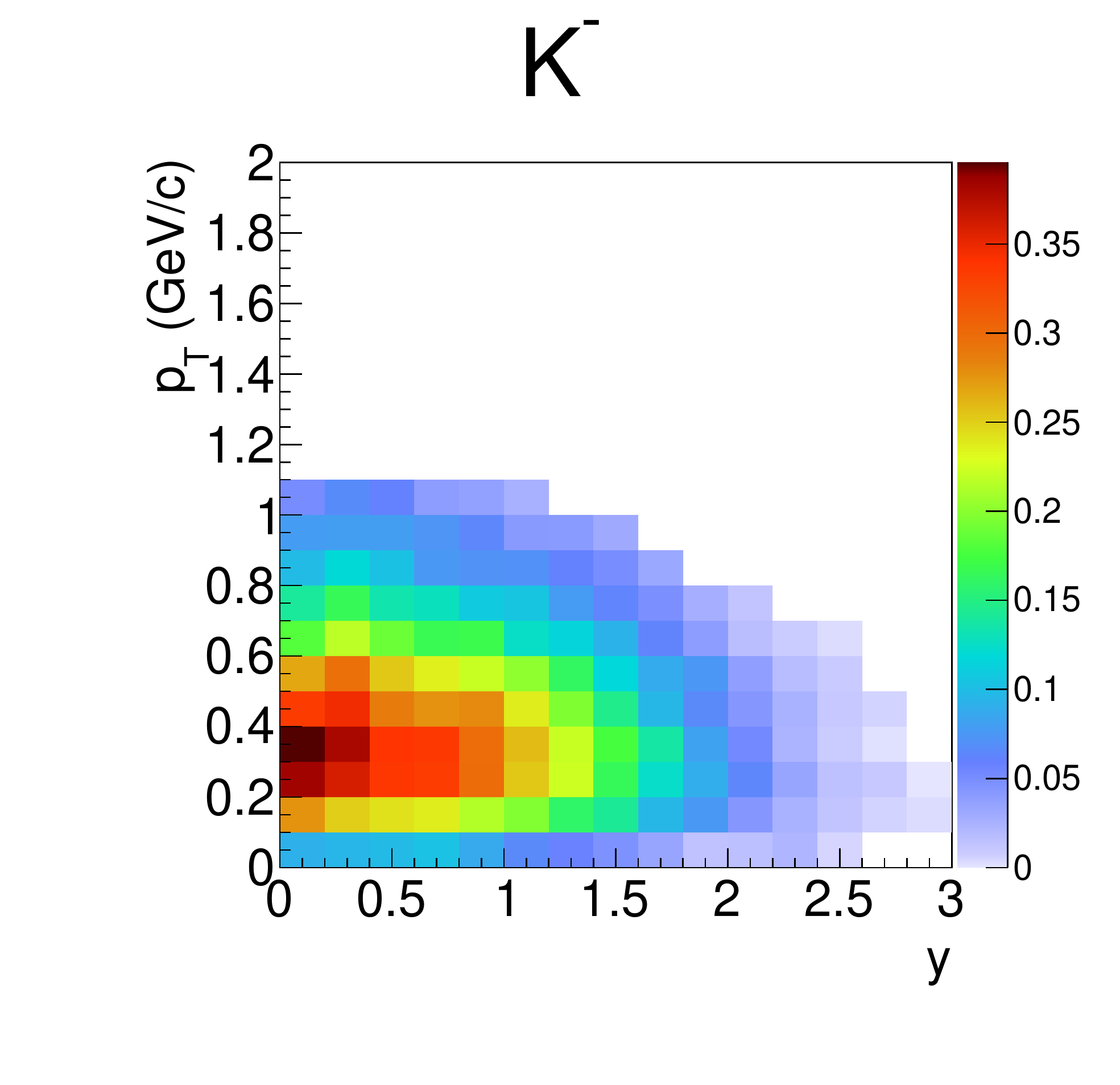}
\includegraphics[width=0.19\textwidth]{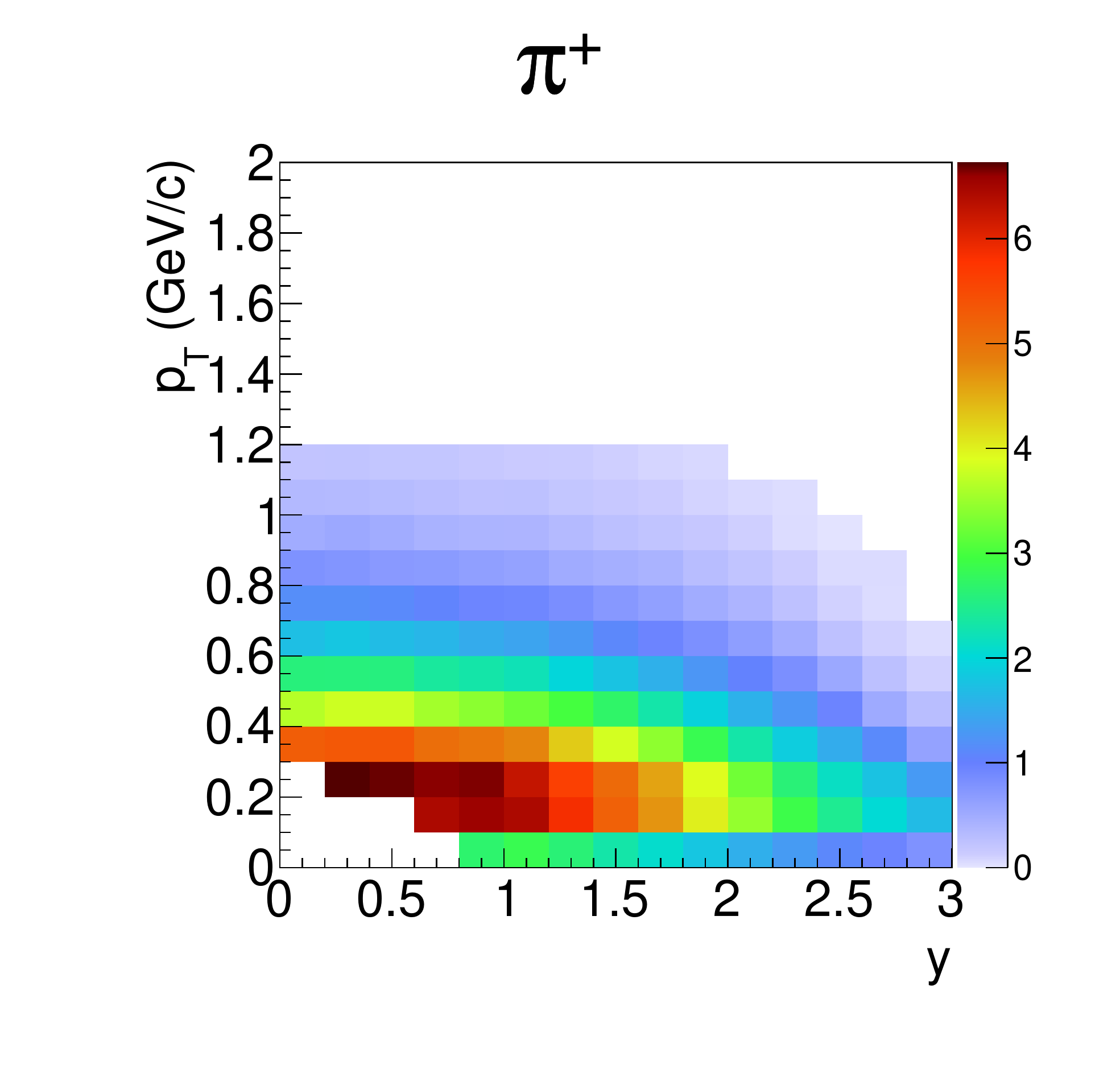}
\includegraphics[width=0.19\textwidth]{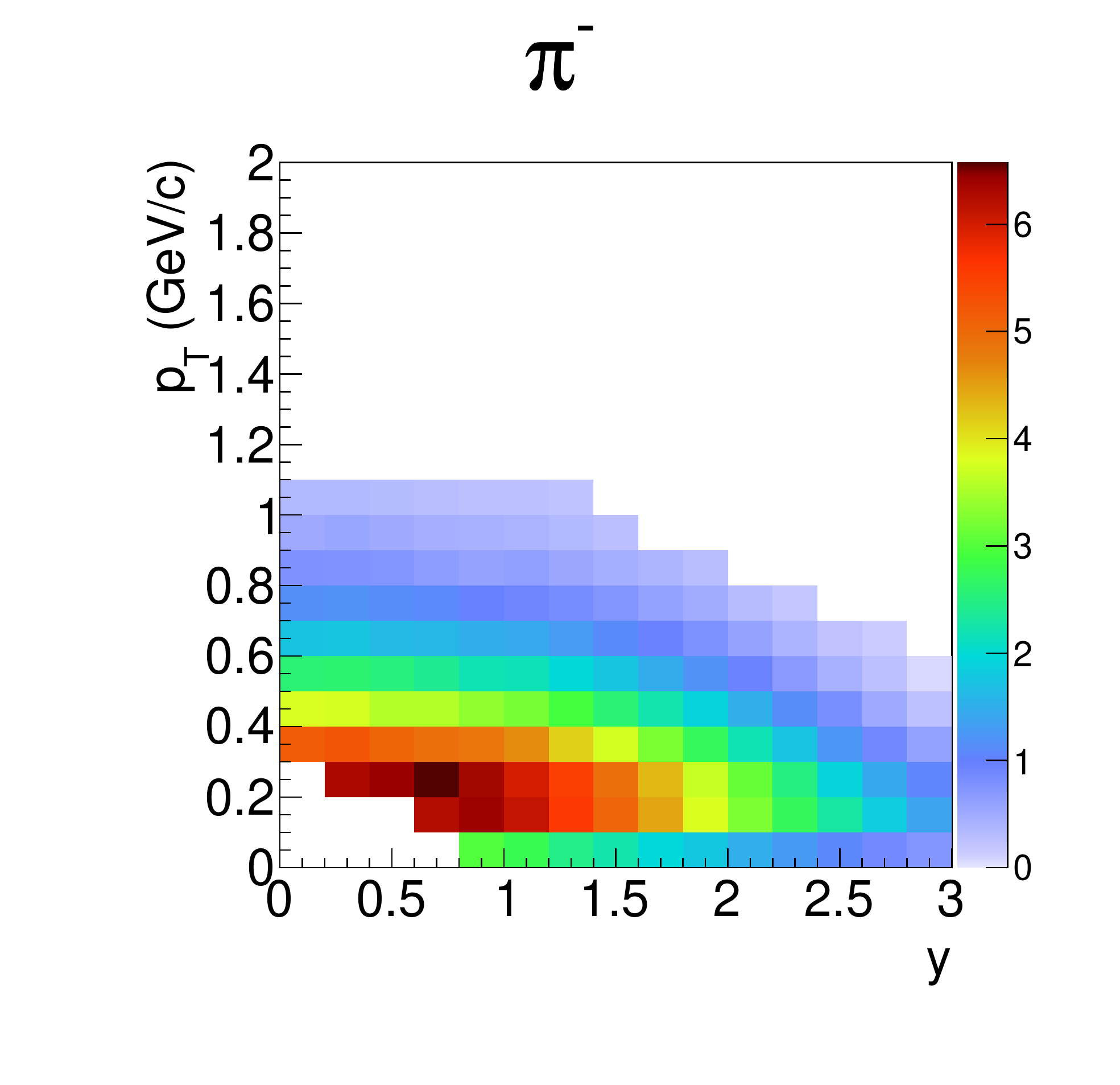}
\includegraphics[width=0.19\textwidth]{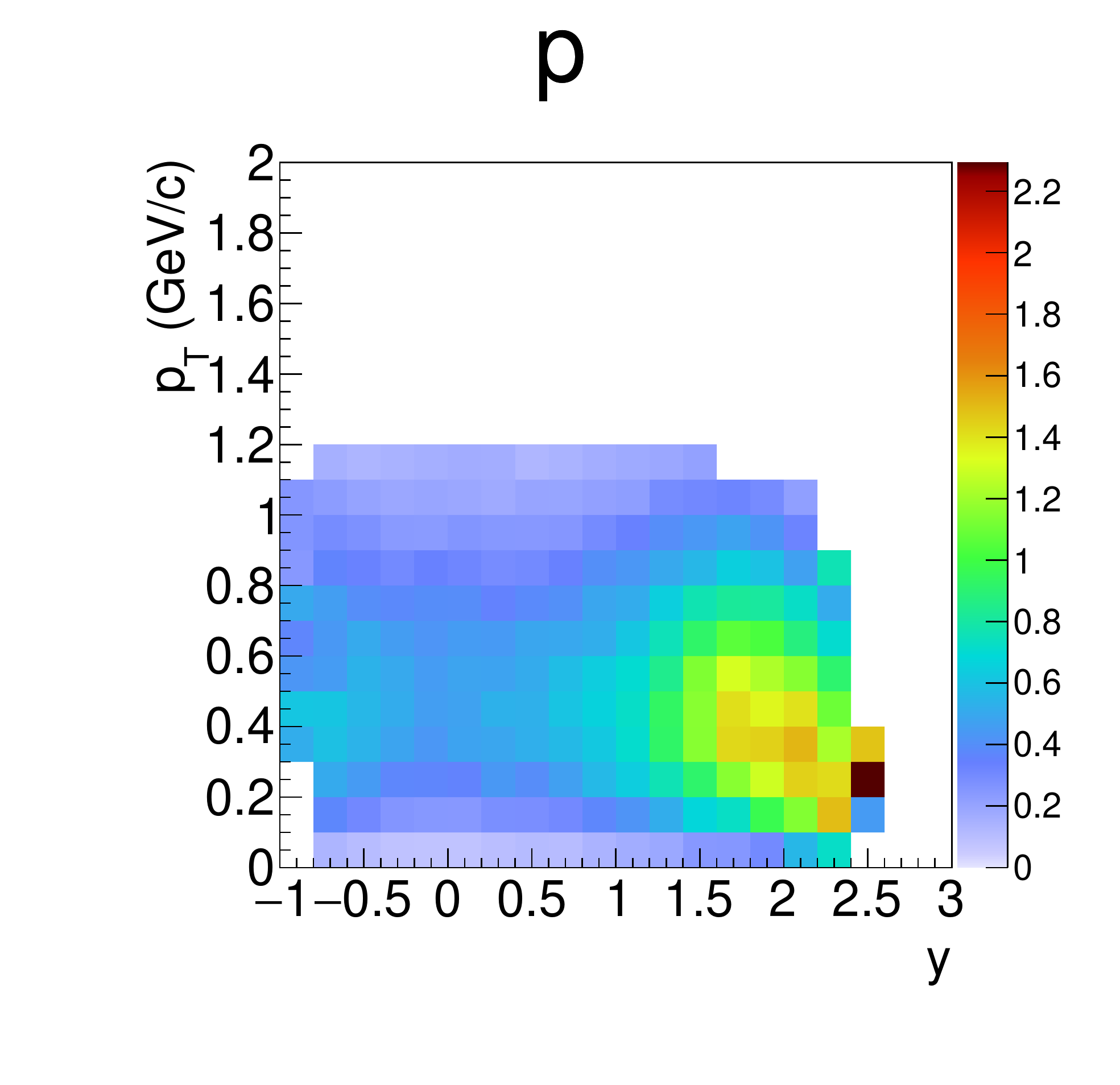}
\vspace{-0.2cm}
\caption[]{\footnotesize Preliminary results on double differential spectra $\frac{d^{2}n}{dydp_{T}}$ of $K^{\pm}$, $\pi^{\pm}$ and protons produced in strong and electromagnetic processes in the 20\% most central Be+Be collisions at 30$A$ (upper panel) and 150$A$ (lower panel) GeV/c. Results obtained with the $dE/dx$ identification method.}
\label{2D_szymon_bebe}
\end{figure}

For this conference the new set of results on identified particles in Be+Be collisions have been obtained (see Ref.~\cite{Lewicki_CPOD2018} for more details). Figure~\ref{2D_szymon_bebe} shows preliminary $\pi^{\pm}$, $K^{\pm}$ and proton spectra from the 20\% most central Be+Be collisions at two selected SPS energies. Results were obtained via the $dE/dx$ identification method. 
The new results extend the acceptance of the $tof-dE/dx$ identification method to the forward hemisphere (the older $tof-dE/dx$ results for charged kaons can be found in Ref.~\cite{Aduszkiewicz:2017mei}).

\begin{figure}
\centering
\includegraphics[width=0.8\textwidth]{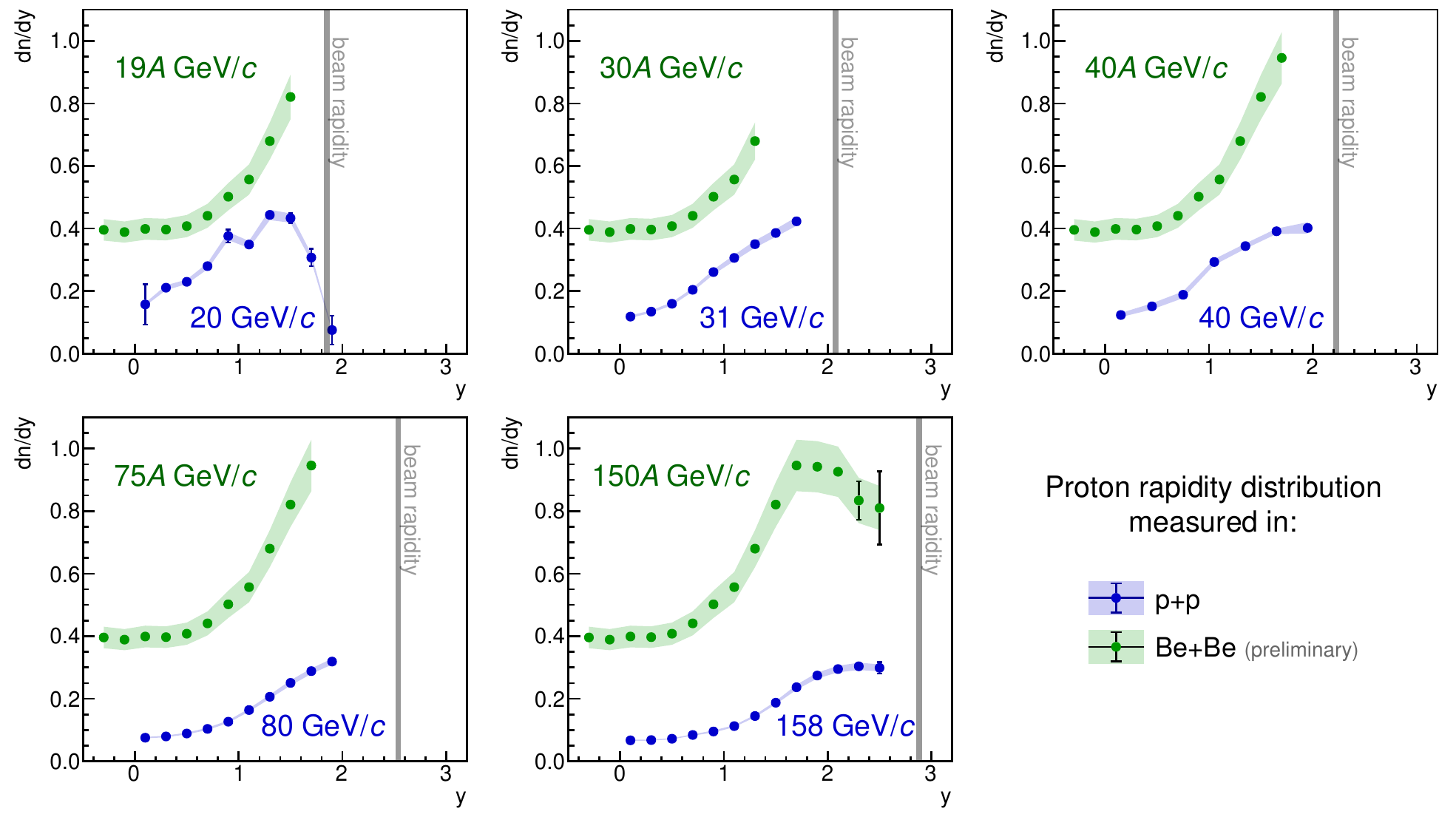}
\vspace{-0.2cm}
\caption[]{\footnotesize Rapidity distributions of protons measured in inelastic p+p interactions~\cite{Aduszkiewicz:2017sei} and in the 20\% most central Be+Be collisions. Be+Be results are preliminary.}
\label{rap_protons_pp_bebe}
\end{figure}

The $p_T$-extrapolated and $p_T$-integrated rapidity spectra of protons are presented in Fig.~\ref{rap_protons_pp_bebe} and compared to published results~\cite{Aduszkiewicz:2017sei} from p+p collisions. The plot is another indication of the similarity between p+p and Be+Be collision systems. The shape of the distributions is qualitatively similar for Be+Be and p+p reactions. This is contrary to the results for proton, net-proton or net-baryon distributions in Pb+Pb collisions~\cite{Blume:2008zza} which exhibit a significant evolution of shape between AGS energies (mid-rapidity maximum due to stopping) and top SPS energies (double maximum due to partial transparency).

\begin{figure}
\centering
\includegraphics[width=0.35\textwidth]{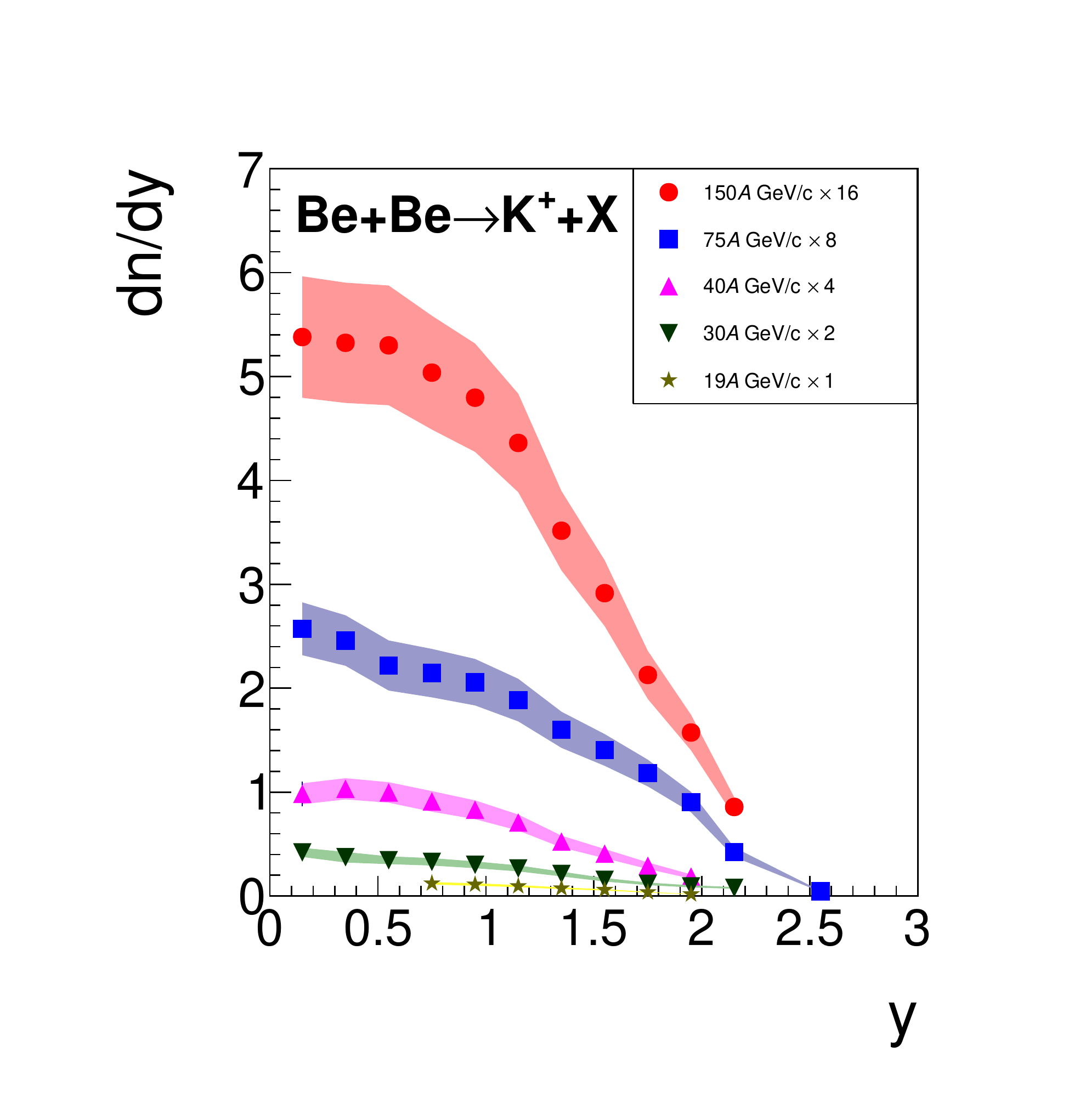}
\includegraphics[width=0.35\textwidth]{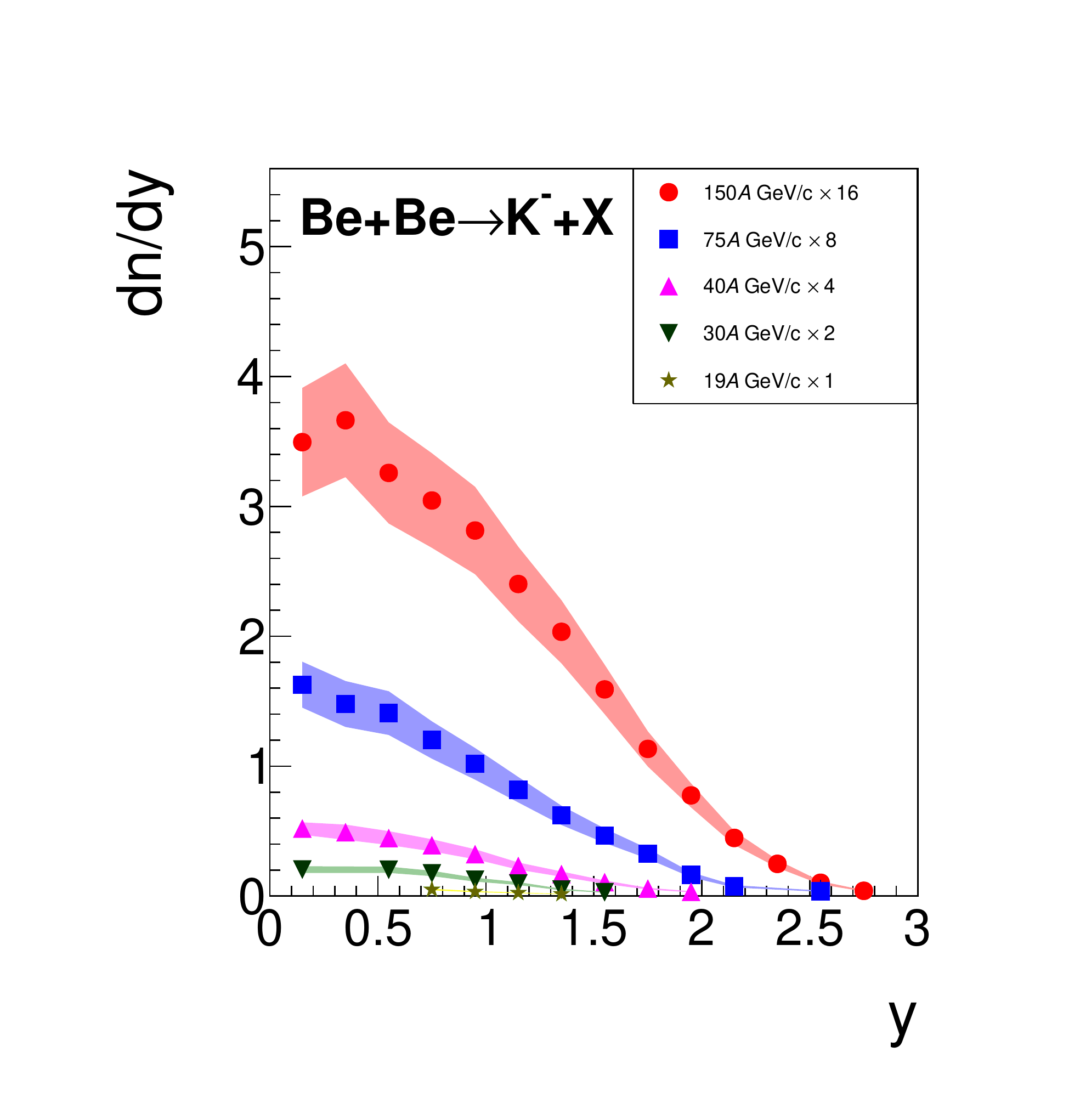}
\vspace{-0.6cm}
\caption[]{\footnotesize Preliminary results on rapidity distributions of charged kaons produced in the 20\% most central Be+Be collisions at 19$A$--150$A$ GeV/c using the $dE/dx$ and $tof-dE/dx$ identification methods. Statistical uncertainties are plotted as vertical bars and systematic ones as shaded bands.}
\label{rap_kaons_bebe}
\end{figure}

The rapidity spectra of charged kaons are presented in Fig.~\ref{rap_kaons_bebe}. They were obtained by combining the current results from the $dE/dx$ identification method with the previously obtained mid-rapidity results from the $tof-dE/dx$ identification method~\cite{Aduszkiewicz:2017mei}. 
The results were reflected around mid-rapidity and fitted by two Gaussian functions (symmetrically displaced around mid-rapidity) to obtain the $4\pi$ mean multiplicities of charged kaons. They are needed to draw the \textit{horn} plot. Finally, symmetry between production of $\pi^{+}$ and $\pi^{-}$ was assumed when calculating the $\langle K^{+} \rangle / \langle \pi^{+} \rangle$ ratio. Namely, the mean multiplicity $\langle \pi^{+} \rangle$ was taken to be the same as $\langle \pi^{-} \rangle$, the latter obtained from the $h^{-}$ analysis method~\cite{Kaptur:2015xzu}. Such an assumption is justified in the limited acceptance of the $dE/dx$ identification method in Be+Be collisions.

\begin{figure}
\centering
\includegraphics[width=0.35\textwidth]{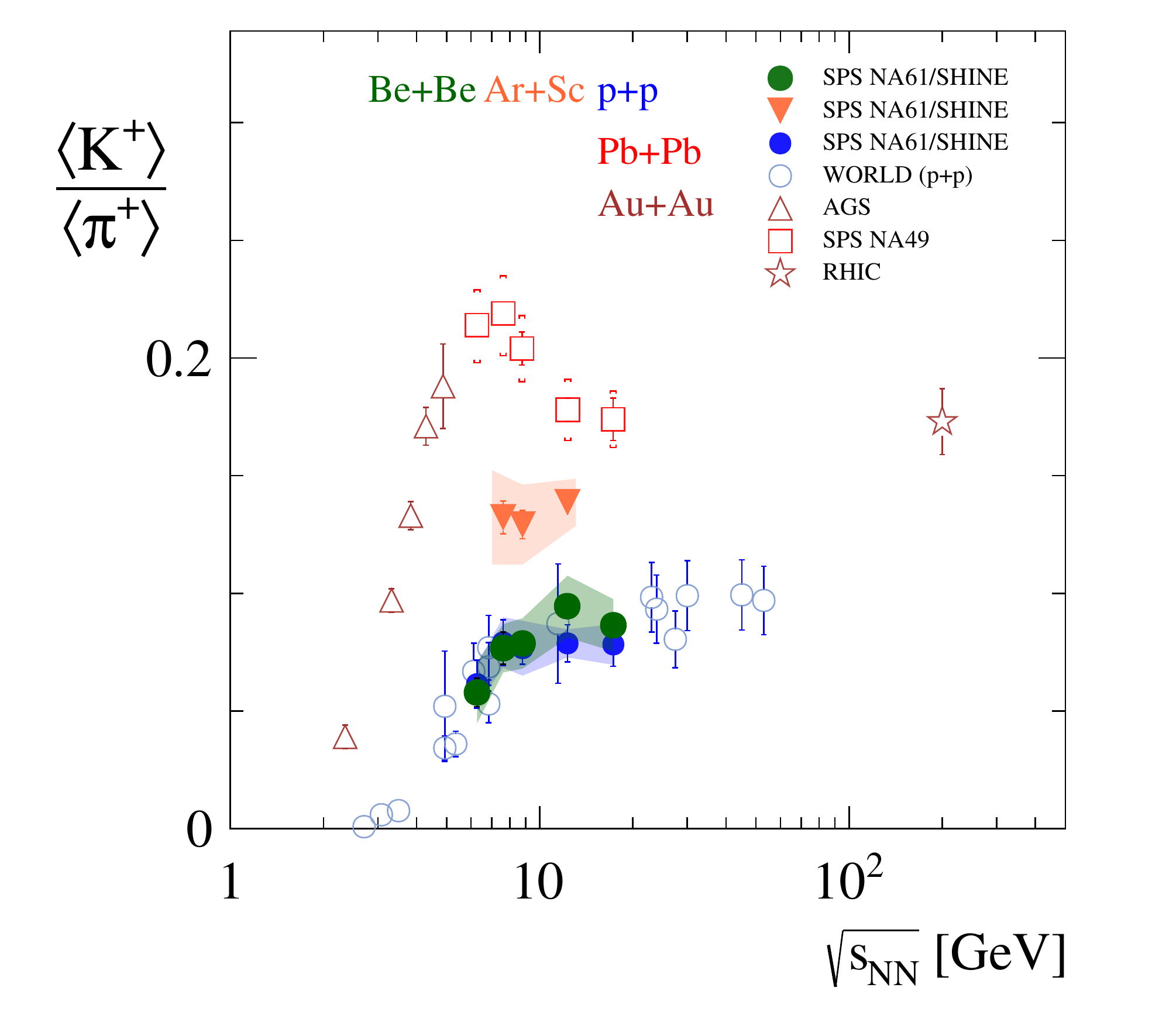}
\includegraphics[width=0.35\textwidth]{./FIGURES/hornKP.pdf}
\includegraphics[width=0.25\textwidth]{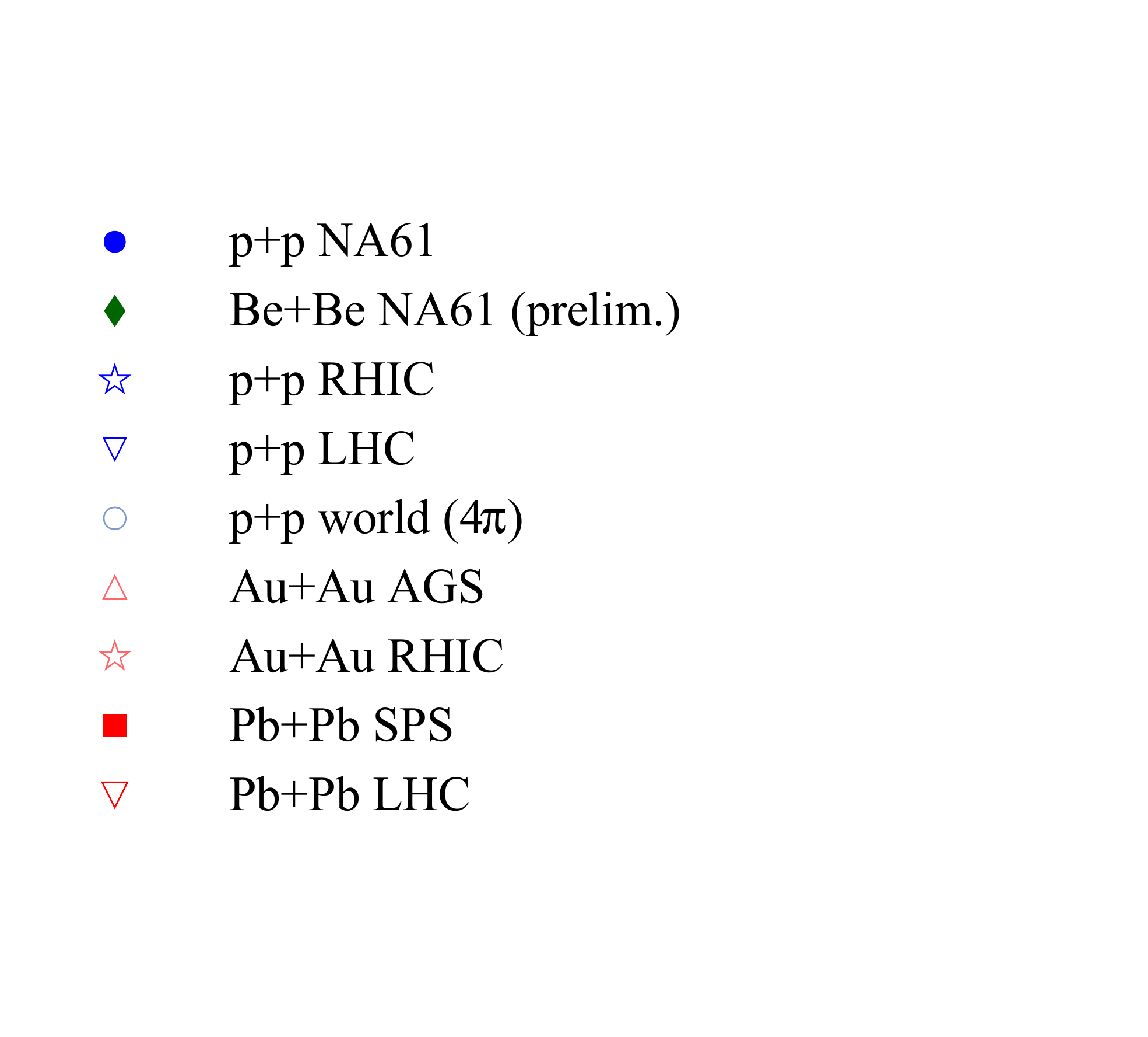}
\vspace{-0.2cm}
\caption[]{\footnotesize \textit{Horn} plots -- $K^{+}/\pi^{+}$ ratio versus energy in $4\pi$ phase space (left) and at mid-rapidity (right) (status of plots as available for the CPOD 2018 conference). Kaon and pion multiplicities in p+p data are taken from Ref.~\cite{Aduszkiewicz:2017sei}, the rest of NA61/SHINE results are preliminary. See Ref.~\cite{Pulawski:2015tka} for references to world data.}
\label{horn_2018}
\end{figure}

The current status of the \textit{horn} plot in full $4\pi$ phase space and at mid-rapidity is presented in Fig.~\ref{horn_2018}. In both cases, the ratios in the 20\% most central Be+Be collisions show similar values to those observed in p+p interactions. No horn structure is observed, in contrast to the behaviour seen in heavy Pb+Pb or Au+Au systems. It suggests that the \textit{onset of fireball} should be present for collisions of nuclei heavier than beryllium.

\section{Can we study the properties of the onset of deconfinement using anisotropic flow?}

Directed flow $v_1$ was considered to be sensitive to the first order phase transition (strong softening of the Equation of State)~\cite{Csernai:1999nf, Stoecker:2004qu, Brachmann:1999xt}. The expected effect was a non-monotonic behaviour (change from positive to negative and again to positive values) of proton $dv_1/dy$ as a function of beam energy. This effect is usually referred to as \textit{collapse of proton flow}. Numerical predictions for $v_1$ within the hydrodynamical model with and without the first order phase transition can be found for example in Ref.~\cite{Csernai:1999nf}. Quite a long time ago the NA49 experiment measured \textit{anti-flow of protons} at mid-rapidity~\cite{Alt:2003ab}. A negative value of $dv_1/dy$ was observed in peripheral Pb+Pb collisions at 40$A$ GeV/c beam momentum ($\sqrt{s_{NN}}=8.8$ GeV). 

More recently the STAR experiment at RHIC published the energy dependence of $dv_1/dy$ at mid-rapidity. The analysis was done for Au+Au collisions at $\sqrt{s_{NN}}=7.7-200$ GeV and for different particle species~\cite{Adamczyk:2014ipa, Adamczyk:2017nxg} (see also Refs.~\cite{Meehan:2017cum, Shanmuganathan:2015qxb} for additional STAR results on this subject).
The $v_1$ slopes are always negative for pions, kaons, anti-protons, anti-lambdas, and $\phi$ mesons, whereas $v_1$ slopes for protons, $\Lambda$ baryons, net-protons and net-$\Lambda's$ change signs at lower energies and show a minimum at $\sqrt{s_{NN}}=10-20$ GeV (14.5 GeV for net-protons). In fact, net-protons and net-$\Lambda's$ exhibit a double change of sign, from positive to negative and again to positive $dv_1/dy$ values. Such behaviour is consistent with hydro models with a 1st order phase transition, however, some calculations suggest that the scenario with a cross-over type transition may lead to similar structures.      

In 2018 the NA61/SHINE experiment reported the first results on anisotropic flow, measured in centrality selected Pb+Pb collisions at 30$A$ GeV/c beam momentum ($\sqrt{s_{NN}}=7.6$ GeV). According to the \textit{horn} structure in the energy dependence of the $K^{+}/\pi^{+}$ ratio in Pb+Pb, this is the the energy of the onset of deconfinement. Therefore, studying the centrality dependence of flow at this specific energy may allow to better understand the properties of the onset of deconfinement.

\begin{figure}
\centering
\includegraphics[width=0.5\textwidth]{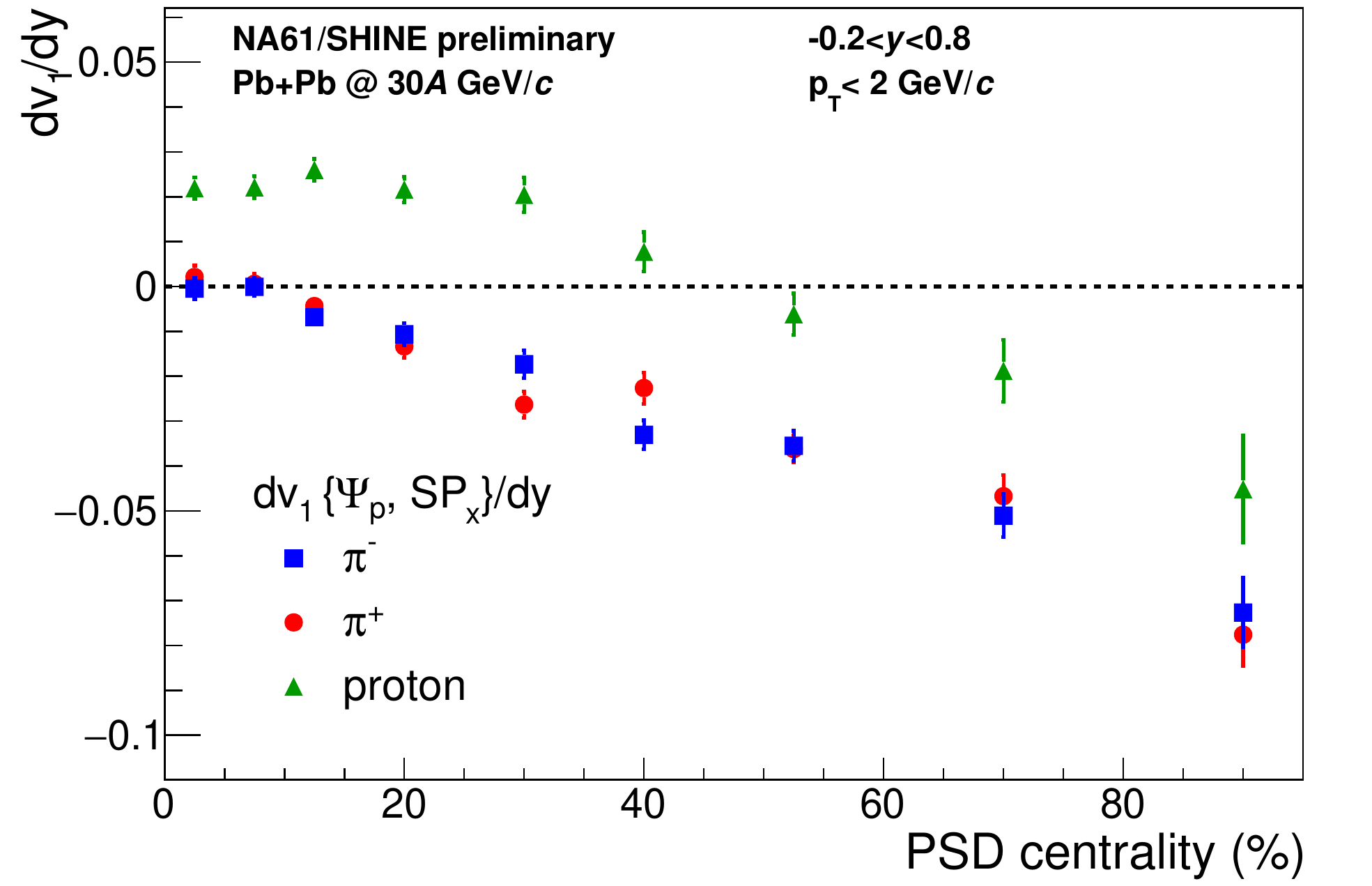}
\vspace{-0.2cm}
\caption[]{\footnotesize Preliminary results on centrality dependence of $dv_1/dy$ at mid-rapidity measured in Pb+Pb collisions at 30$A$ GeV/c.}
\label{fig_dv1dy}
\end{figure}

The NA61/SHINE fixed target setup allows tracking and particle identification over a wide rapidity range. Flow coefficients were measured relative to the spectator plane estimated with the Projectile
Spectator Detector\footnote{The Projectile Spectator Detector is located on the beam axis and measures the forward energy related to the non-interacting nucleons of the beam nucleus. In NA61/SHINE it is used to determine the centrality of $A+A$ collisions, as well as the spectator plane in the analysis of anisotropic flow.} (PSD), which is unique for NA61. Preliminary results on centrality dependence of $dv_1/dy$ at mid-rapidity, measured in Pb+Pb collisions at 30$A$ GeV/c, are presented in Fig.~\ref{fig_dv1dy}. One sees that the slope of pion $v_1$ is always negative. In contrast, the slope of proton $v_1$ changes sign for centrality of about 50\%. Results for Pb+Pb at 13$A$ and 150$A$ GeV/c as well as six energies of Xe+La are expected soon.

\begin{figure}
\centering
\vspace{0.3cm}
\includegraphics[width=0.4\textwidth]{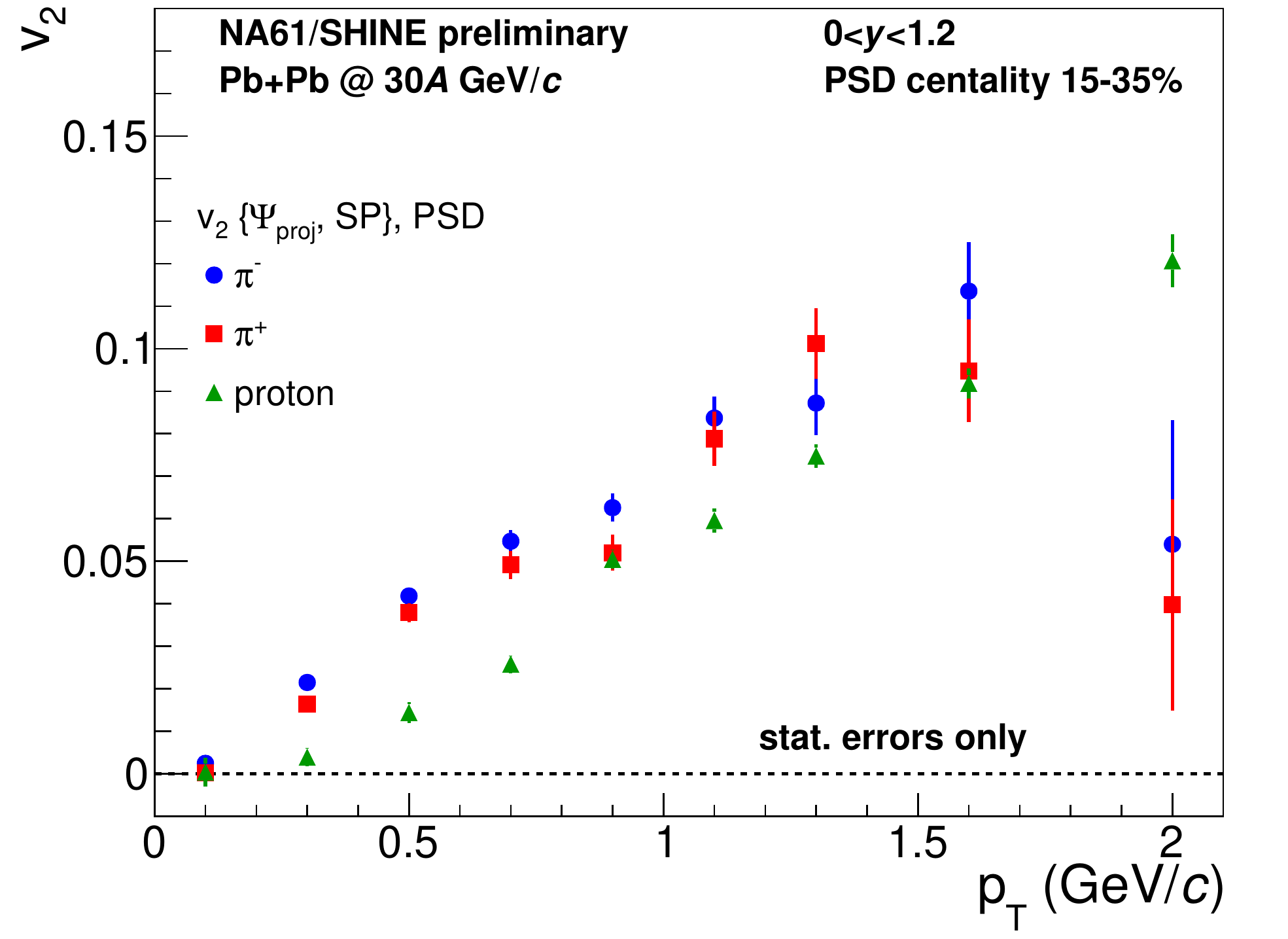}
\includegraphics[width=0.4\textwidth]{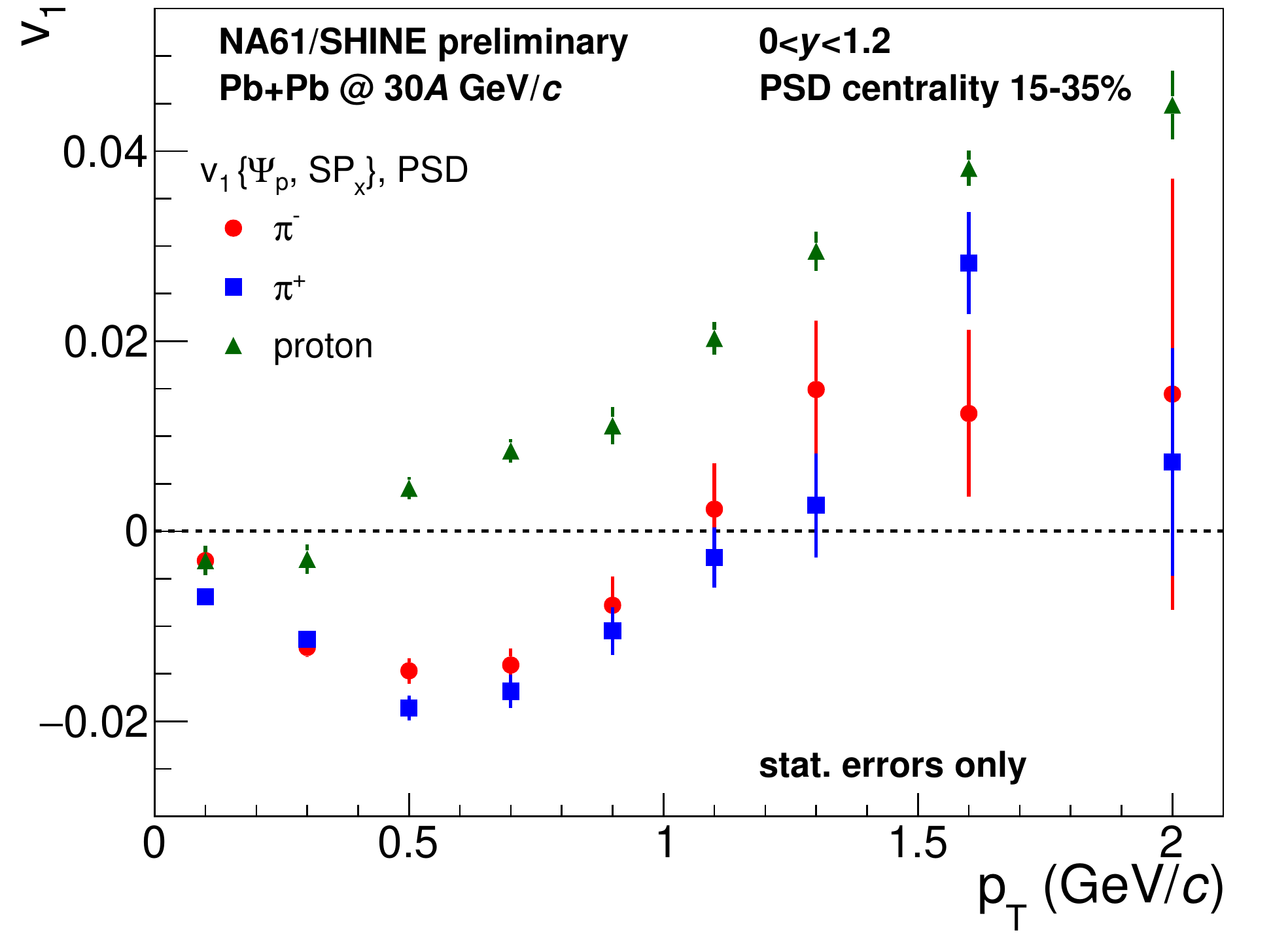}
\vspace{-0.2cm}
\caption[]{\footnotesize Preliminary results on transverse momentum dependence of elliptic (left) and directed (right) flow close to mid-rapidity measured in 15--35\% central Pb+Pb collisions at 30$A$ GeV/c.}
\label{fig_v2_v1}
\end{figure}

Figure~\ref{fig_v2_v1} shows the particle type dependence of elliptic ($v_2$) and directed flow. A clear mass hierarchy of $v_2$ is observed which is interpreted as due to radial flow\footnote{Hierarchy (mass ordering) of $v_2$ (for a given $p_T$ $v_2$ is lower for a higher-mass particle) is qualitatively reproduced (at lower $p_T$) by hydrodynamical models and understood as due to radial flow. In hydro models the elliptic flow follows: $v_2 \sim (p_T - \langle v_T \rangle m_T)/T$, where $v_T$ is the transverse expansion velocity (radial flow), $m_T$ is the particle transverse mass, and $T$ temperature.}. The difference between $v_2$ of $\pi^{+}$ and $\pi^{-}$ is small. A significant mass dependence can be also seen for $v_1$ (Fig.~\ref{fig_v2_v1}, right). The difference between $v_1$ of $\pi^{+}$ and $\pi^{-}$ is sensitive to electromagnetic effects (see charged pions yield asymmetry in the next section). More results and details on flow analysis, as well as the new analysis of NA49 data with respect to the spectator plane, can be found in Refs.~\cite{Klochkov:2018xvw, Selyuzhenkov_CPOD2018}.

\section{Can we see the electromagnetic interactions between charged pions and spectators?}

\begin{figure}
\centering
\includegraphics[width=0.5\textwidth]{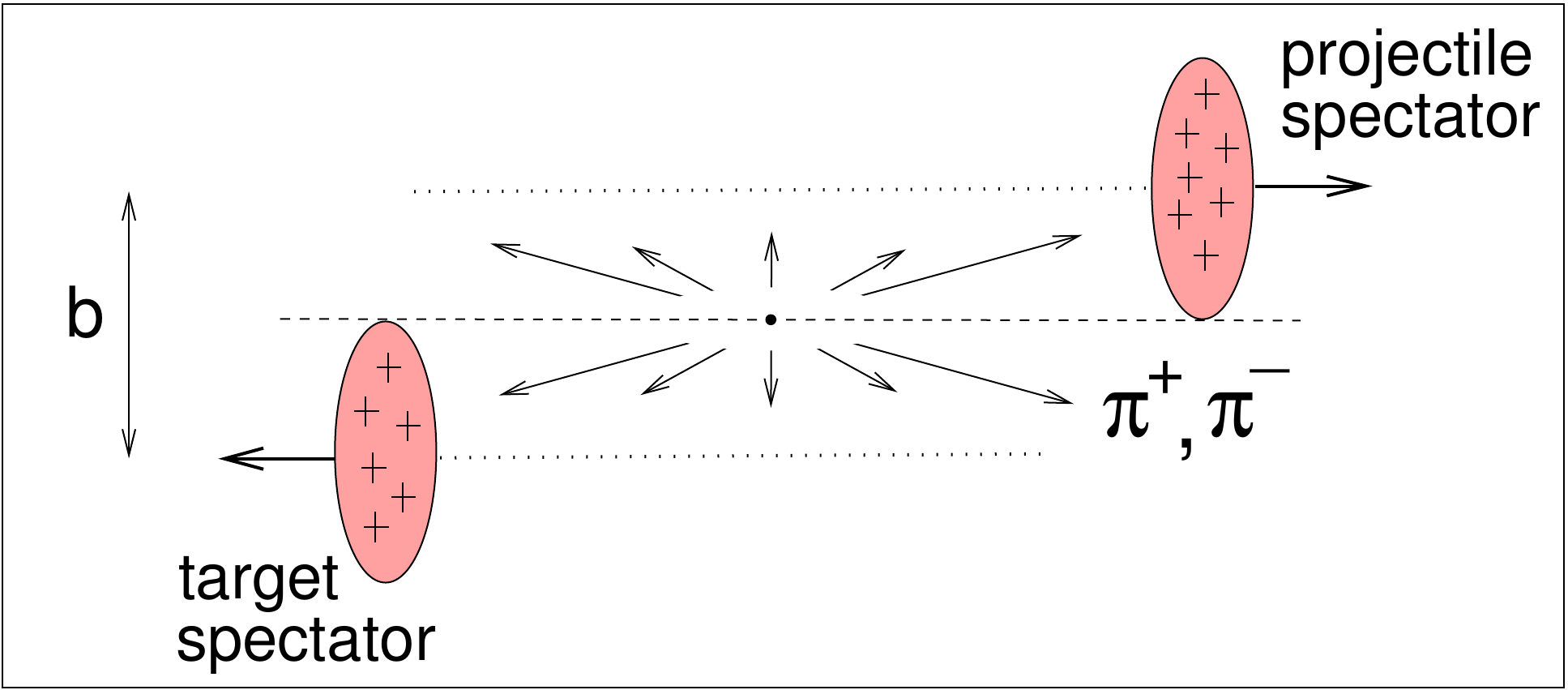}
\caption[]{\footnotesize Simplified view of a Pb+Pb collision where the trajectories of spectators and produced charged pions are schematically shown. Figure taken from Ref.~\cite{Rybicki:2006qm}.}
\label{pic_Rybicki}
\end{figure}

The measurement of the $\pi^{+}/\pi^{-}$ ratio helps to study spectator-induced electromagnetic effects. Spectators (in non-central collisions) follow their initial path with unchanged momenta (Fig.~\ref{pic_Rybicki}). Charged spectators generate electromagnetic fields. Charged pion trajectories can be modified by electromagnetic interactions (repulsion for $\pi^{+}$ and attraction for $\pi^{-}$) with the spectators. This effect is sensitive to the space-time evolution the system. Therefore, the $\pi^{+}/\pi^{-}$ ratio allows to study spectator-induced electromagnetic effects and brings new information on the space and time evolution of the particle production process~\cite{Rybicki:2006qm, Rybicki:2013qla}.

\begin{figure}
\begin{tikzpicture}
	\begin{scope} [xshift=-3.5cm]
	\node {\includegraphics[width=0.32\textwidth]{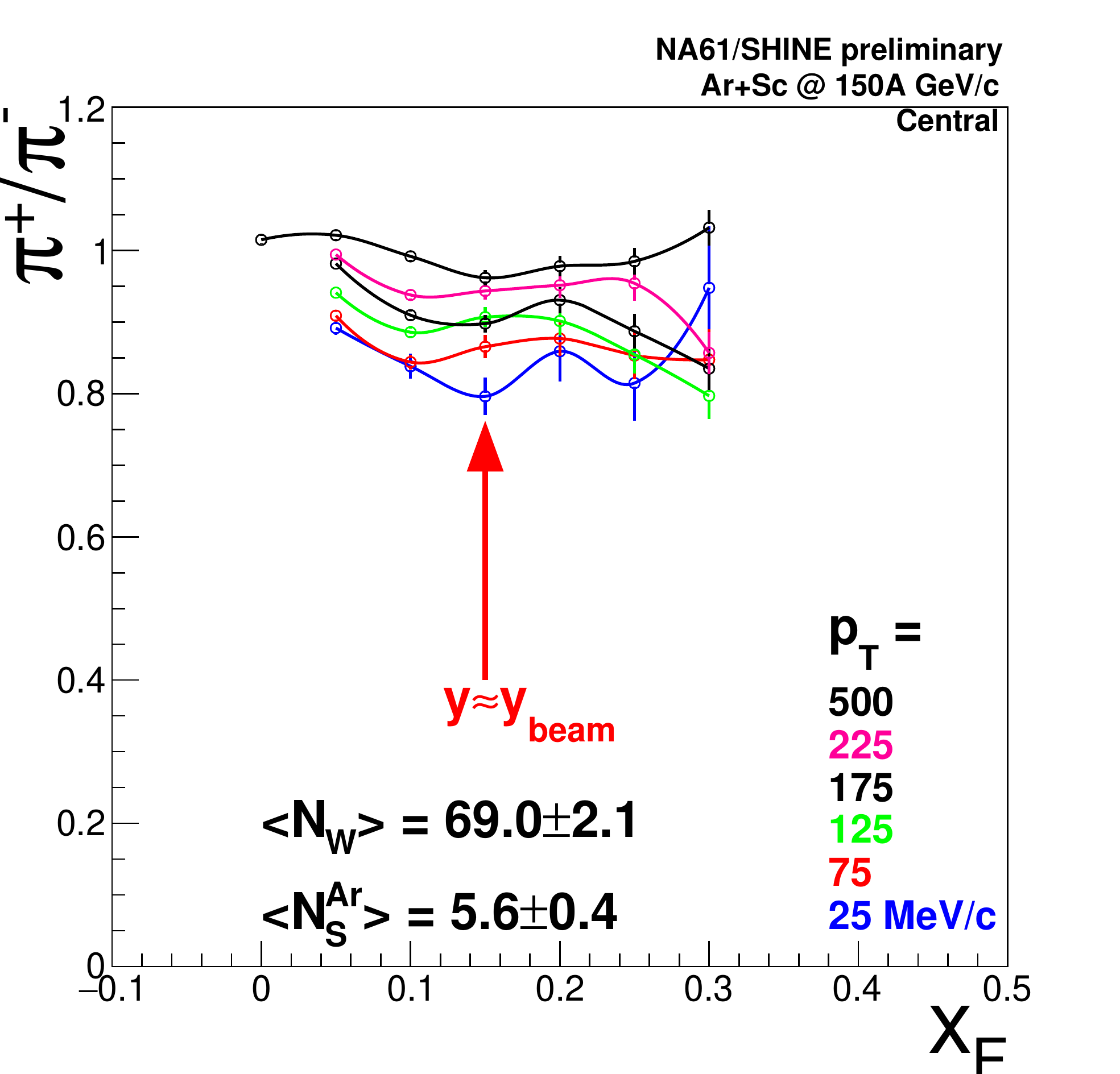}};
	\end{scope}
	\begin{scope} [xshift=1.8cm]	
	\node {\includegraphics[width=0.32\textwidth]{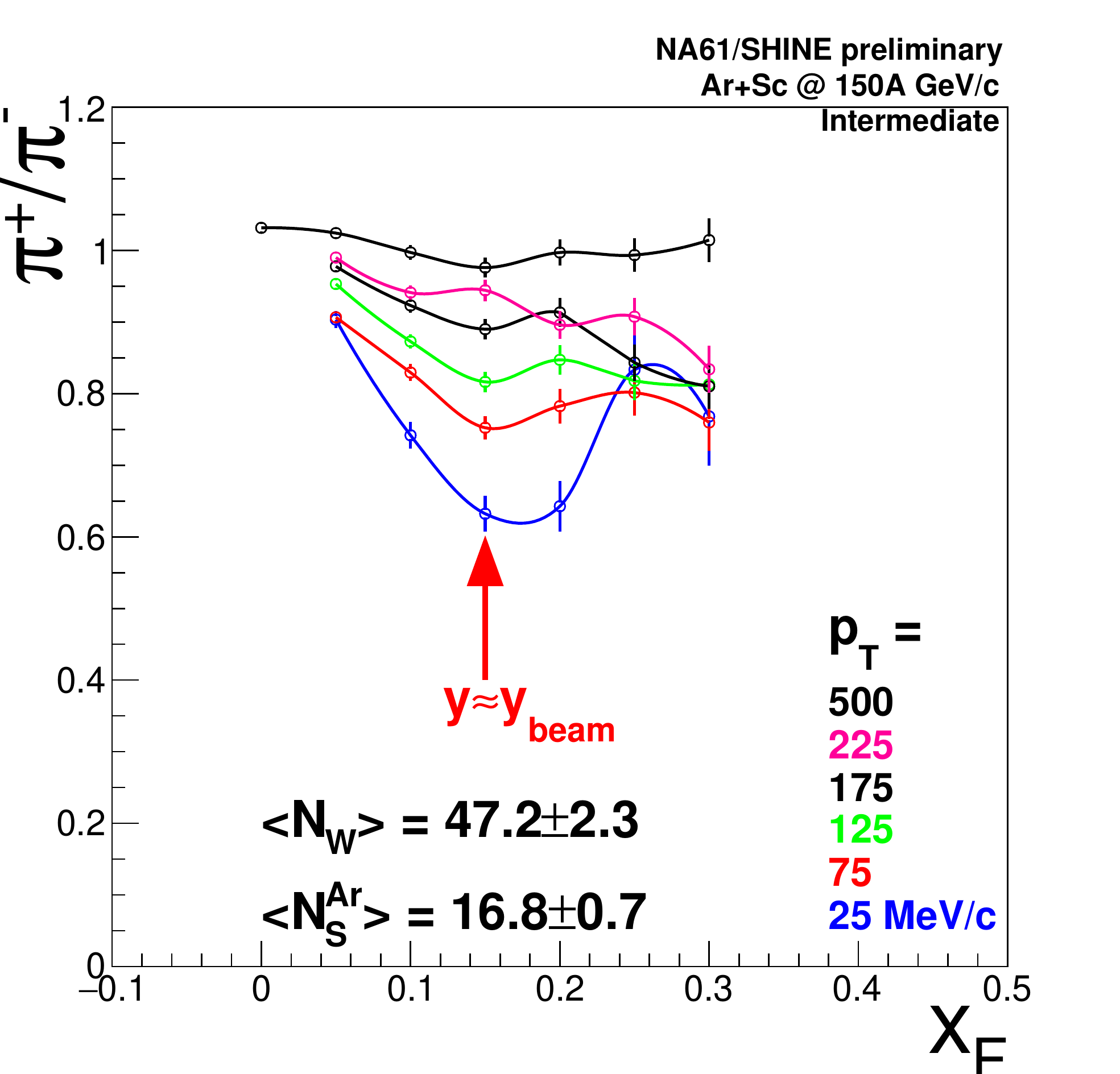}};
	\end{scope}
	\begin{scope} [xshift=6.5cm,  yshift=-0.25cm]
	\node {\includegraphics[width=0.3\textwidth]{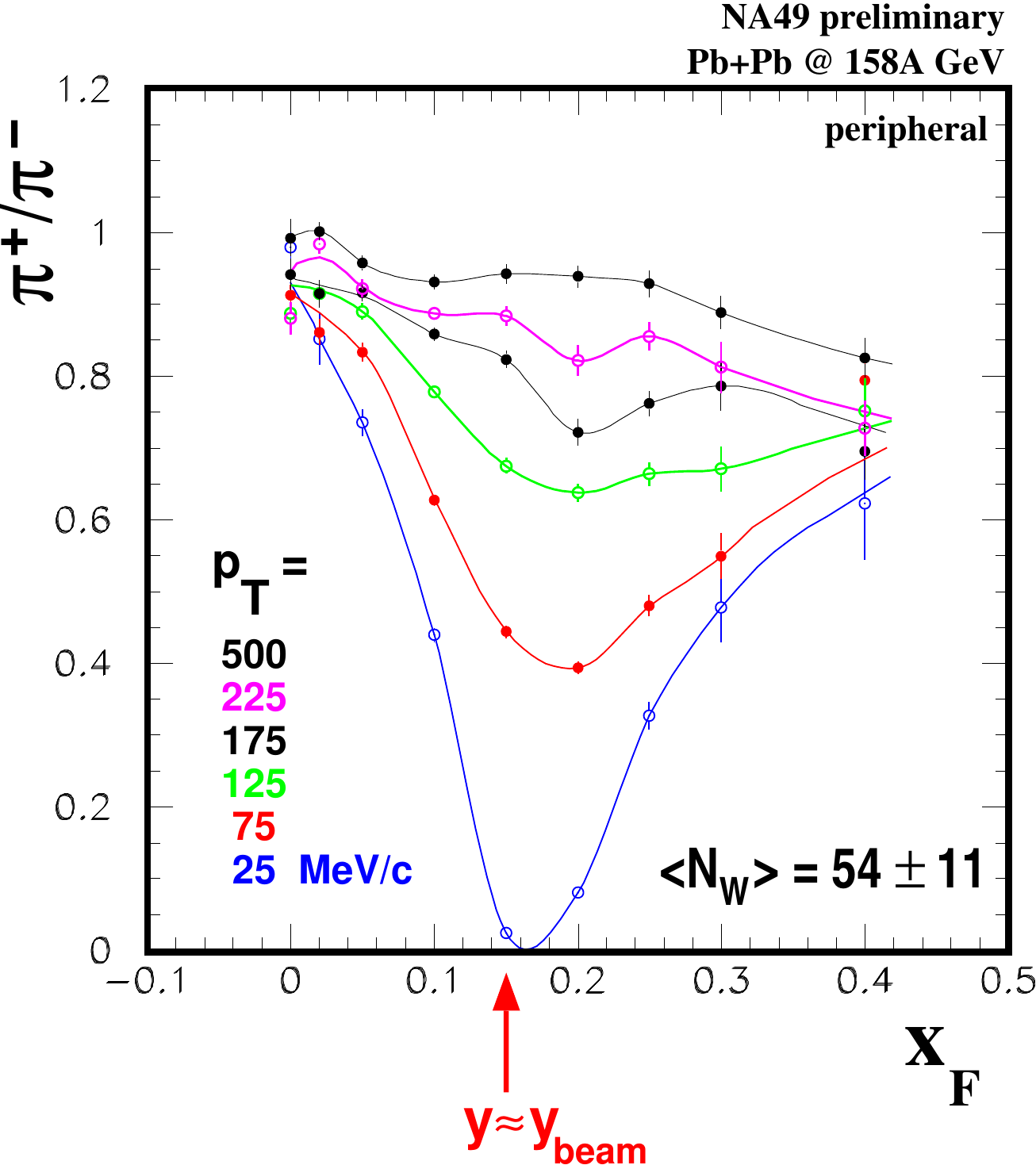}};
	\end{scope}
\end{tikzpicture}
\vspace{-0.4cm}
\caption[]{\footnotesize Preliminary results on Feynman's $x$ dependence of the $\pi^{+}/\pi^{-}$ ratio in central Ar+Sc (left), intermediate centrality Ar+Sc (middle) and peripheral Pb+Pb (right) collisions at 150$A$/158$A$ GeV/c. $\langle N_W \rangle$ represents the mean number of wounded nucleons and $\langle N_S^{~Ar} \rangle$ the mean number of projectile spectator nucleons. Only statistical uncertainties are shown. The NA49 Pb+Pb plot is taken from Ref.~\cite{Rybicki:2009zz}.}
\label{pip_pim_arsc_pbpb}
\end{figure}

Figure~\ref{pip_pim_arsc_pbpb} shows preliminary NA61/SHINE results on Feynman's $x$ ($x_F=p_L/p_L^{beam}$ measured in the center-of-mass reference system) dependence of the $\pi^{+}/\pi^{-}$ ratio in central Ar+Sc and intermediate centrality Ar+Sc collisions at 150$A$ GeV/c. For a comparison the older NA49 results~\cite{Rybicki:2009zz} on peripheral Pb+Pb interactions at 158$A$ GeV/c are also shown. It can be seen that the repulsion of $\pi^{+}$ is the strongest for pions with rapidities close to beam rapidity (spectators) and with low $p_T$ (the minimum at the blue curve in Fig.~\ref{pip_pim_arsc_pbpb}, middle). This is the first observation of spectator-induced electromagnetic effects in small systems at the SPS energies. Qualitatively, a similar effect is seen both in intermediate centrality Ar+Sc and in NA49 peripheral Pb+Pb collisions. A comparison of NA61/SHINE results with models, and resulting conclusions for the space-time evolution of the system, can be found in Ref.~\cite{Davis_CPOD2018}.

\section{Can we contribute to understanding of the time evolution of the fireball?}

\begin{figure}
\centering
\vspace{0.5cm}
\includegraphics[width=0.3\textwidth]{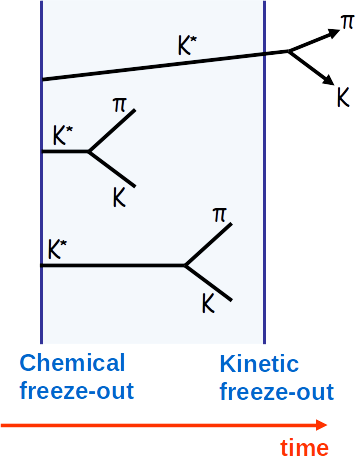}
\includegraphics[width=0.5\textwidth]{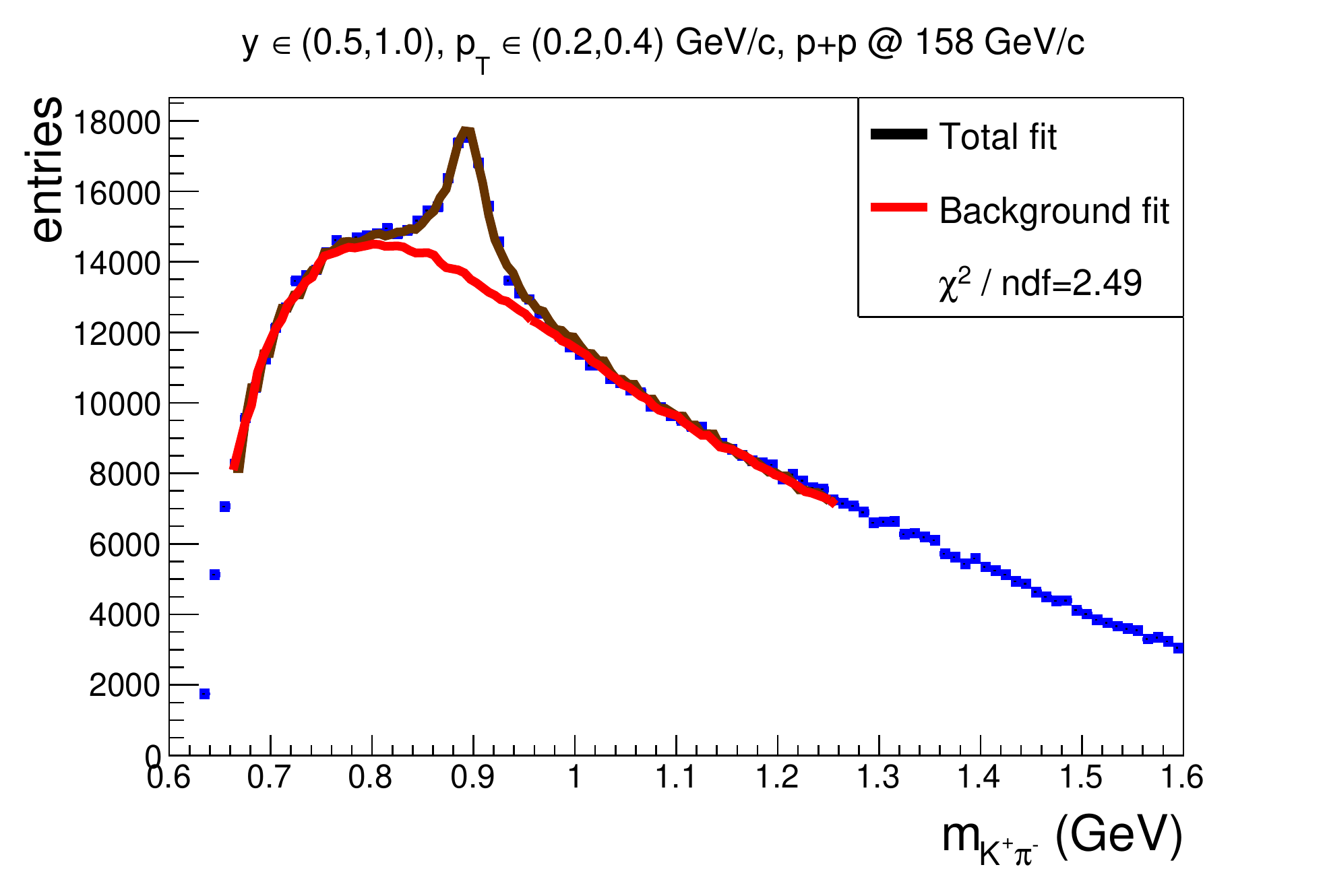}
\vspace{-0.2cm}
\caption[]{\footnotesize Left: schematic picture of the time evolution of the fireball, where the decay of a certain fraction of $K^{*}(892)^0$ resonances (without regeneration processes) is shown. Right: an example $K^{+}\pi^{-}$ invariant mass distribution, where the background is a template of $K^{+}\pi^{-}$ pairs from mixed events and from Monte Carlo.}
\label{Kstar_pic_inv}
\end{figure}

The analysis of short-lifetime resonances can contribute to our understanding of the time evolution of high-energy nucleus-nucleus collision. It may help to distinguish between two possible freeze-out scenarios: the sudden and the gradual one~\cite{Markert:2002rw}. Namely, the ratio of $K^{*}(892)^0$ to charged kaon production, may allow to study the length of the time interval between chemical 
and kinetic 
freeze-out. The lifetime of the $K^{*}(892)^0$ meson ($\sim 4$ fm/$c$) is comparable to the expected duration of the rescattering hadronic gas phase between freeze-outs. Consequently, some $K^{*}(892)^0$ resonances may decay inside the fireball. The momenta of their decay products can be modified due to elastic scatterings, preventing the experimental reconstruction of the resonance via an invariant mass analysis. In such a case a suppression of the observed $K^{*}(892)^0$ yield is expected. Assuming no regeneration processes (Fig.~\ref{Kstar_pic_inv}, left), and that conditions at chemical freeze-out of p+p and Pb+Pb collisions are the same, the time between freeze-outs can be determined~\cite{Adams:2004ep} from the formula:

\begin{equation}
\frac{K^{*}}{K} \mid_{kinetic} = \frac{K^{*}}{K} \mid_{chemical} \cdot e^{- \frac{\Delta t}{\tau}},
\label{eq:time_freezeout}
\end{equation}
where the $\langle K^{*}(892)^0 \rangle/ \langle K^{+/-} \rangle$ ratio in inelastic p+p interactions can be treated as the one at chemical freeze-out, the $\langle K^{*}(892)^0 \rangle/ \langle K^{+/-} \rangle$ ratio for central Pb+Pb or Au+Au interactions can be used as the one at kinetic freeze-out, $\tau$ is the $K^{*}(892)^0$ lifetime of 4.17 fm/c~\cite{Tanabashi:2018oca}, and $\Delta t$ is the time between chemical and kinetic freeze-outs.

NA61/SHINE has measured $K^{*}(892)^0$ meson production via its $K^{+}\pi^{-}$ decay mode in inelastic p+p collisions at beam momentum of 158 GeV/c ($\sqrt{s_{NN}}=17.3$ GeV). The \textit{template} method was used to extract raw $K^{*}(892)^0$ signals (see Ref.~\cite{Tefelska_CPOD2018} for details). In this method the background (Fig.~\ref{Kstar_pic_inv}, right) is described as a sum of two components: mixed events and Monte Carlo generated templates which describe the contribution of $K^{+}\pi^{-}$ pairs coming from sources other than the $K^{*}(892)^0$. For the studied resonance, the \textit{template} method was found to be much more effective in estimating the background than the \textit{standard} procedure relying on mixed events only.

\begin{figure}
\centering
\includegraphics[width=0.45\textwidth]{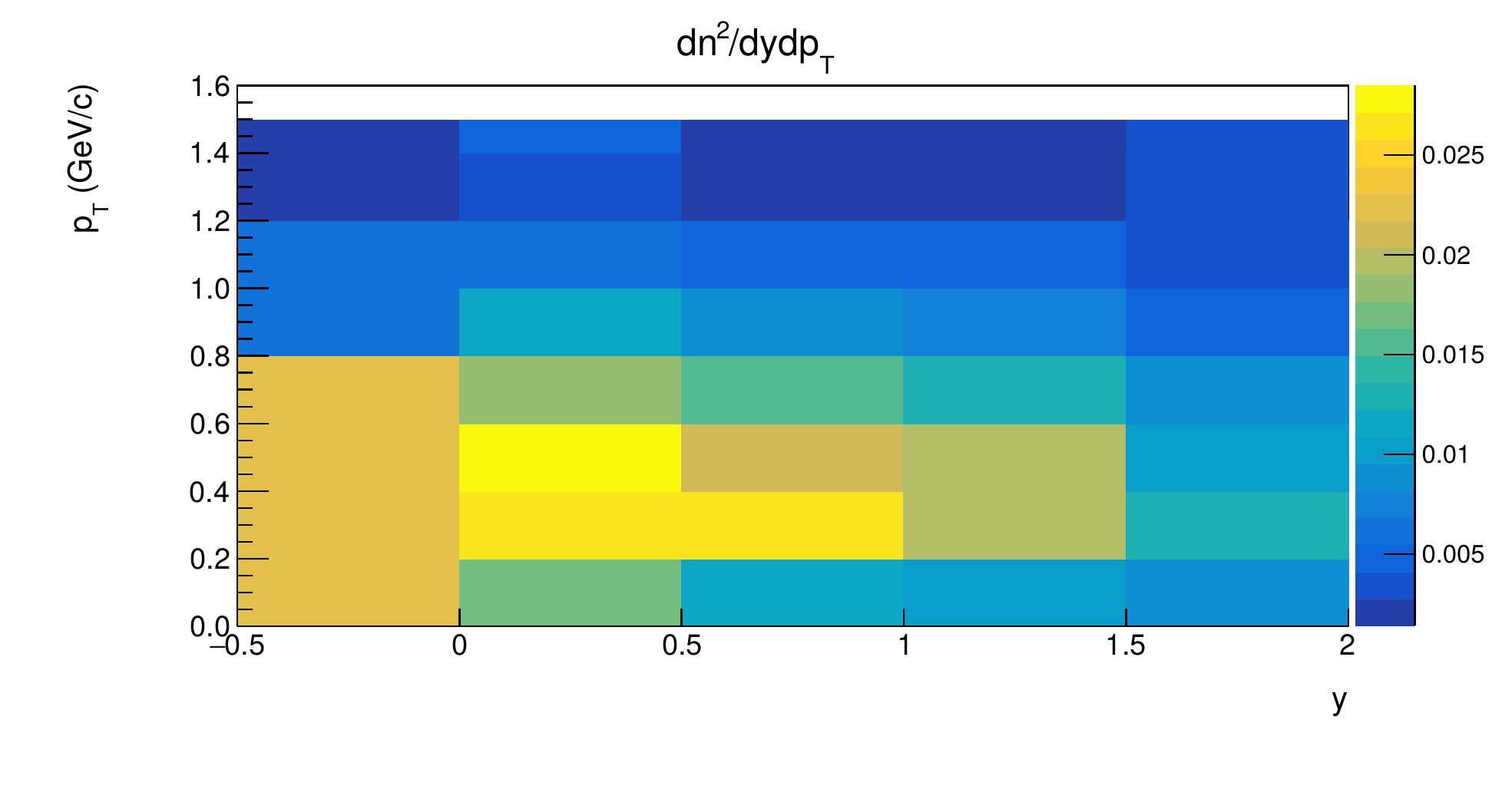}
\includegraphics[width=0.45\textwidth]{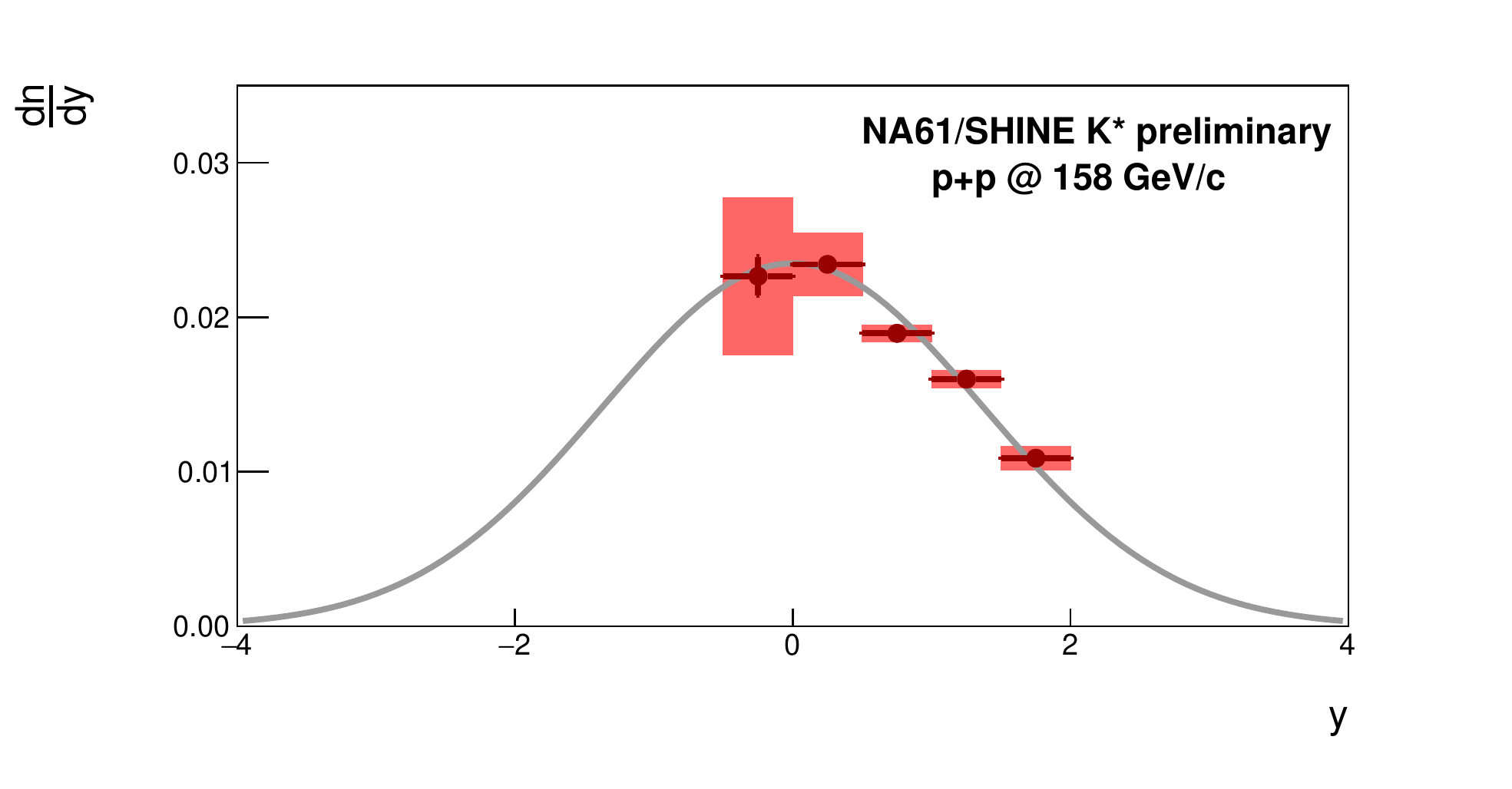}
\vspace{-0.4cm}
\caption[]{\footnotesize Left: preliminary results on double differential yields of $K^{*}(892)^0$, in (GeV/c)$^{-1}$, in bins of ($y$, $p_T$) for inelastic p+p collisions at 158 GeV/c. Right: $p_T$-extrapolated and $p_T$-integrated rapidity spectrum of $K^{*}(892)^0$. Statistical uncertainties are plotted as bars and systematic ones as shaded regions.}
\label{dndydpt_2D_dndy}
\end{figure}

The double differential yields of $K^{*}(892)^0$ mesons in inelastic p+p collisions at 158 GeV/c are presented in bins of ($y$, $p_T$) in the left panel of Fig.~\ref{dndydpt_2D_dndy}. The right panel shows the $p_T$-extrapolated and $p_T$-integrated rapidity spectrum of $K^{*}(892)^0$ from which the $4\pi$ multiplicity can be obtained. 
The preliminary value of the mean multiplicity of $K^{*}(892)^0$ mesons was found to be $0.08058 \pm 0.00059 \pm 0.0026$, where the first uncertainty is statistical and the second is systematic. This result is similar to the NA49 measurement ($0.0741 \pm 0.0015 \pm 0.0067$) for the same system and energy~\cite{Anticic:2011zr}, however the NA49 multiplicity was obtained for the range $0 < p_T < 1.5$ GeV/c instead of $4\pi$ acceptance as in NA61/SHINE.

\begin{figure}
\centering
\includegraphics[width=0.4\textwidth]{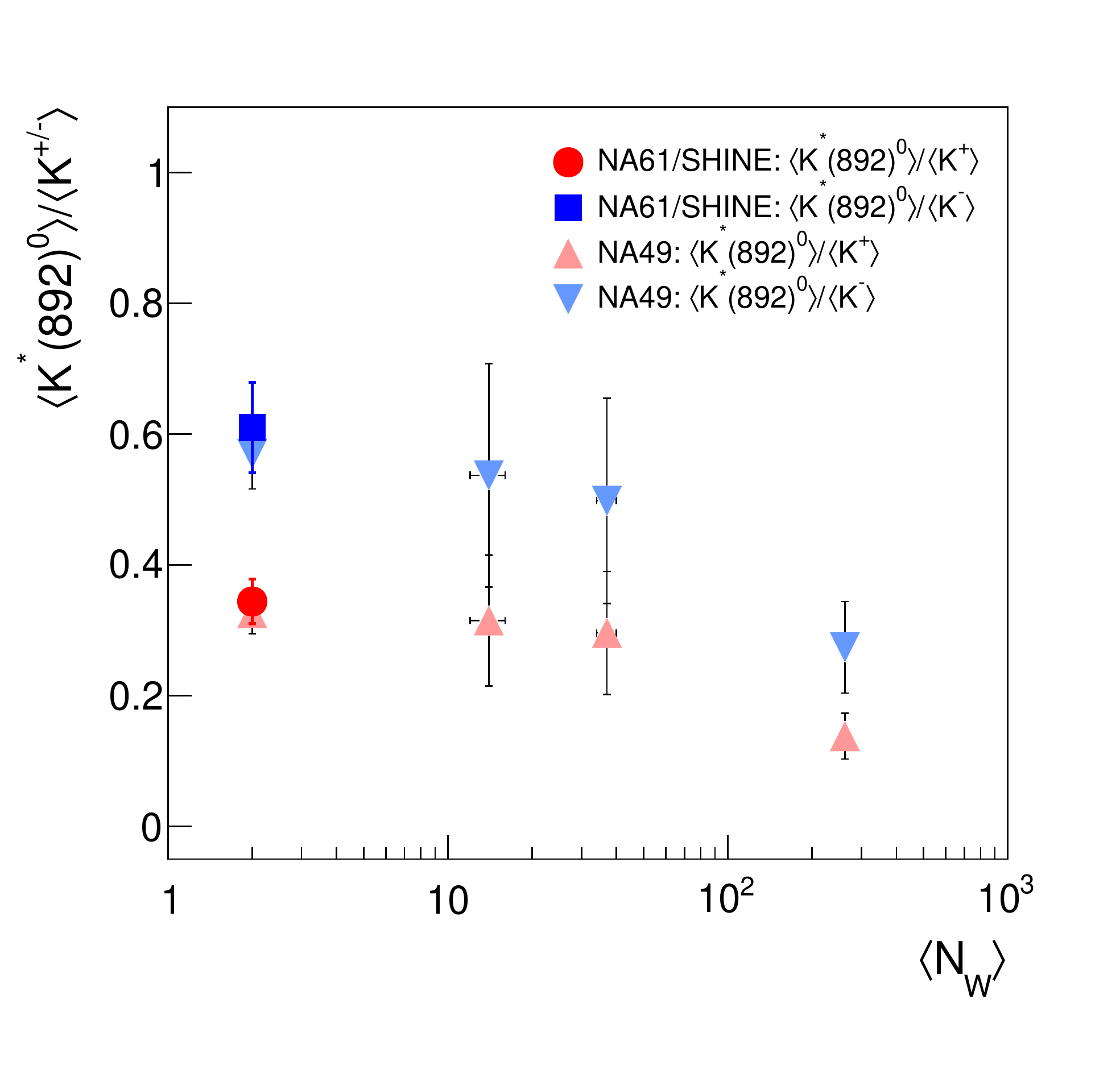}
\includegraphics[width=0.4\textwidth]{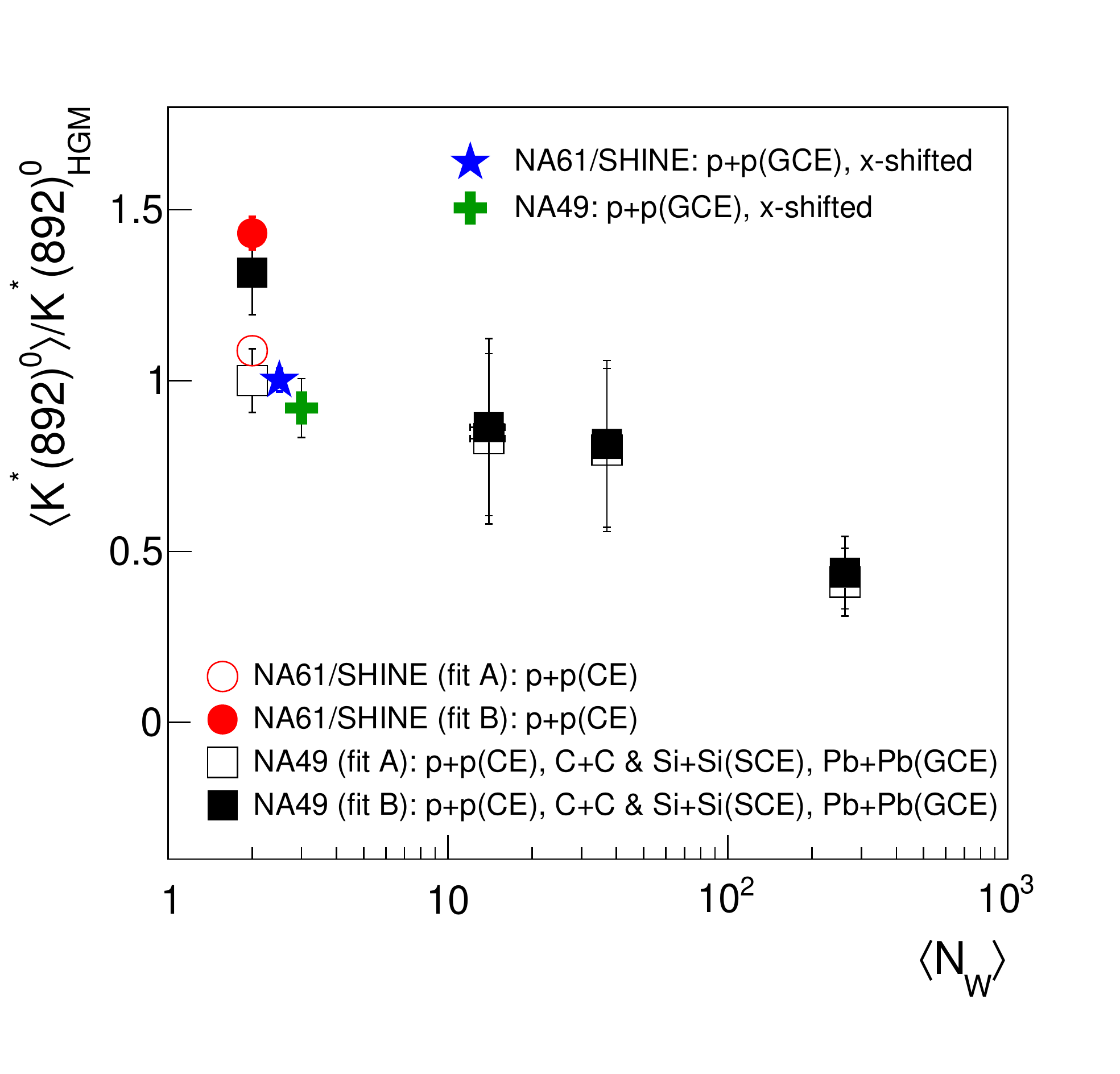}
\vspace{-0.5cm}
\caption[]{\footnotesize Left: system size dependences of $\langle K^{*}(892)^0 \rangle/ \langle K^{+} \rangle$ and $\langle K^{*}(892)^0 \rangle/ \langle K^{-} \rangle$ yield ratios in p+p, C+C, Si+Si and Pb+Pb collisions at 158$A$ GeV/c. NA49 values are taken from Ref.~\cite{Anticic:2011zr} ($K^{*}(892)^0$), and Refs.~\cite{Anticic:2010yg, Alt:2004wc, Afanasiev:2002mx} ($K^{+/-}$); NA49 Pb+Pb $K^{+/-}$ yields are rescaled from 5\%~\cite{Afanasiev:2002mx} to 23.5\% most central collisions. NA61/SHINE $K^{+/-}$ yields in p+p are taken from Ref.~\cite{Aduszkiewicz:2017sei}. Both statistical and systematic uncertainties are included. 
Right: system size dependence of $\langle K^{*}(892)^0 \rangle$ multiplicity scaled by predictions of two Hadron Resonance Gas Models~\cite{Becattini:2005xt, Begun:2018qkw}. NA49 data are taken from Ref.~\cite{Anticic:2011zr}. Experimental data include both statistical and systematic uncertainties. NA61/SHINE results on both panels are preliminary.}
\label{Kstar_ratio_HGM}
\end{figure}

Figure~\ref{Kstar_ratio_HGM}, left shows the system size dependence of $\langle K^{*}(892)^0 \rangle/ \langle K^{+} \rangle$ and $\langle K^{*}(892)^0 \rangle/ \langle K^{-} \rangle$ yield ratios in p+p, C+C, Si+Si and Pb+Pb collisions at 158$A$ GeV/c. A decrease of this ratio with increasing system size is observed, as expected due to the increasing rescattering time between chemical and kinetic freeze-out. Assuming that the losses of $K^{*}(892)^0$ before kinetic freeze-out are due to rescattering effects and there are no regeneration processes, the time between chemical and kinetic freeze-outs can be estimated as 3.8 $\pm$ 1.1 fm/c from the $\langle K^{*}(892)^0 \rangle/ \langle K^{+} \rangle$ ratio and 3.3 $\pm$ 1.2 fm/c from the $\langle K^{*}(892)^0 \rangle/ \langle K^{-} \rangle$ ratio. The value of $\Delta t$ is larger at SPS than $\Delta t = 2 \pm 1$ fm/c obtained by RHIC~\cite{Adams:2004ep}\footnote{In Ref.~\cite{Adams:2004ep} the lifetime of $K^{*0}$ was assumed to be 4 fm/c.}, suggesting that regeneration effects may start to play a significant role for higher energies. As the $K^{*}(892)^0$ regeneration may happen also at SPS energies, the obtained $\Delta t$ values can be treated rather as the lower limit of the time between chemical and kinetic freeze-outs.

Figure~\ref{Kstar_ratio_HGM}, right shows the comparison of $\langle K^{*}(892)^0 \rangle$ production to predictions of two Hadron Resonance Gas Models (HGM). In Ref.~\cite{Becattini:2005xt} the HGM predictions were performed for two versions of the fits. Fit "B" used "standard" $\gamma_S$ strangeness under-saturation factor, and for p+p data $\Xi's$ and $\Omega's$ baryons were excluded from the fit. In case of fit "A" $\gamma_S$ was replaced by the mean number of strange quark pairs $\langle s \bar{s} \rangle$, and for p+p data the $\phi$ meson was excluded from the fit. For p+p collisions the Canonical Ensemble (CE) was used. For heavier C+C and Si+Si systems the S-Canonical Ensemble (SCE) was applied (assumes exact strangeness conservation and grand-canonical treatment of electric charge and baryon number), and for Pb+Pb the Grand Canonical Ensemble (GCE) was assumed\footnote{Note that the centrality of Pb+Pb collisions used in the HGM fits was 0-5\% whereas $\langle K^{*}(892)^0 \rangle$ in NA49 was obtained for 23.5\% most central interactions. Therefore, the HGM yield had to be rescaled by a factor $262/362$ (resulting from the corresponding mean number of wounded nucleons), before comparing it to NA49 data.}. In the another HGM calculation, described in Ref.~\cite{Begun:2018qkw}, the $K^{*}(892)^0$ multiplicity in p+p collisions was obtained assuming GCE formulation and the $\phi$ meson was normally included in the fit.    

Figure~\ref{Kstar_ratio_HGM}, right shows that the deviation from the HGM increases with increasing system size. For heavier systems (including C+C and Si+Si), there is no significant difference between fit "A" and fit "B". Surprisingly, the small p+p system can be quite well described by the GCE. The CE formulation can reproduce the $K^{*}(892)^0$ multiplicities in p+p only in case of fit "A", where the $\phi$ meson is excluded from the fit.


\section{Can we see critical point from proton density fluctuations?}

\begin{figure}
\centering
\includegraphics[width=0.3\textwidth]{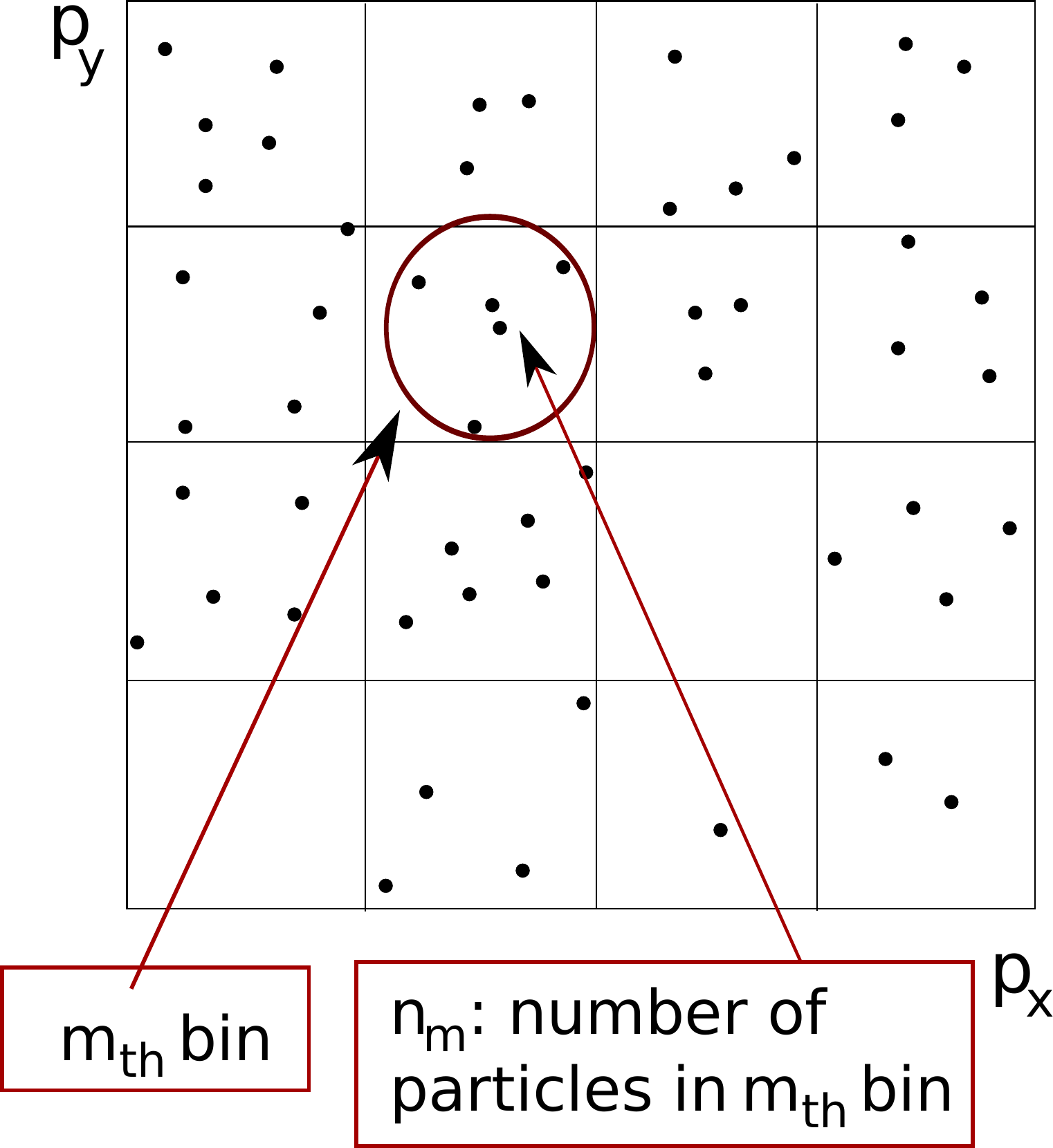}
\vspace{-0.2cm}
\caption[]{\footnotesize Schematic picture of division of $p_x$, $p_y$ space of protons into bins.}
\label{intermit_pic}
\end{figure}

An intermittency signal in protons was predicted close to the critical point. This will manifest in local power-law fluctuations of the baryon density which can be searched for by studying the scaling behaviour of second factorial moments $F_2(M)$ with the cell size (see Fig.~\ref{intermit_pic}) or, equivalently, with the number of cells in ($p_x$, $p_y$) space of protons at mid-rapidity (see Refs.~\cite{Bialas:1985jb, Turko:1989dc, Diakonos:2006zz}). The transverse momentum space is partitioned into $M \times M$ equal-size bins, and the proton distribution is quantified by multiplicities in individual bins. The second order factorial moment in transverse momentum space is expressed as:

\begin{equation}
F_2(M)=\frac{\langle \frac{1}{M^2} \sum _{m=1}^{M^2}n_m(n_m-1) \rangle} 
{{\langle \frac{1}{M^2} \sum _{m=1}^{M^2}n_m \rangle}^2},
\end{equation}
where $M^2$ is the number of bins ($M$ bins in $p_x$ and $M$ bins in $p_y$) and $n_m$ is the number of protons in the $m$-th bin. Combinatorial background subtracted (by mixed events) second factorial moments, $\Delta F_2(M)$, should scale according to a power-law (for $M \gg 1$):
\begin{equation}
\Delta F_2(M) \sim (M^2)^{\phi_2}
\end{equation}  
The resulting exponent (intermittency index $\phi_2$) can be compared to the theoretically expected~\cite{Antoniou:2006zb} critical value $\phi_2 = 5/6$. 

\begin{figure}
\centering
\includegraphics[width=0.43\textwidth]{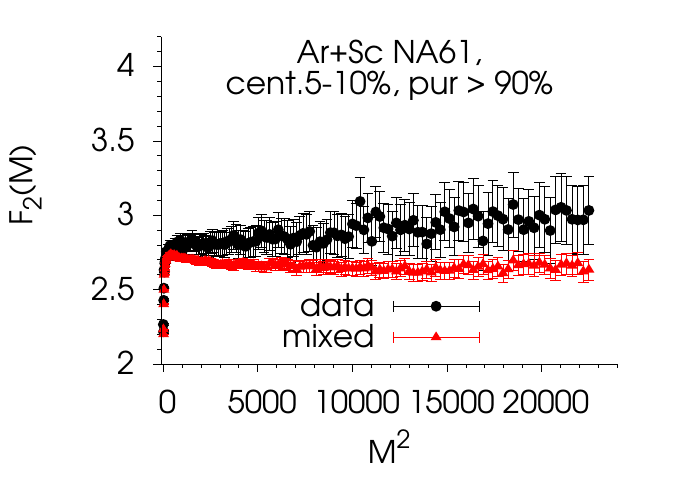}
\includegraphics[width=0.43\textwidth]{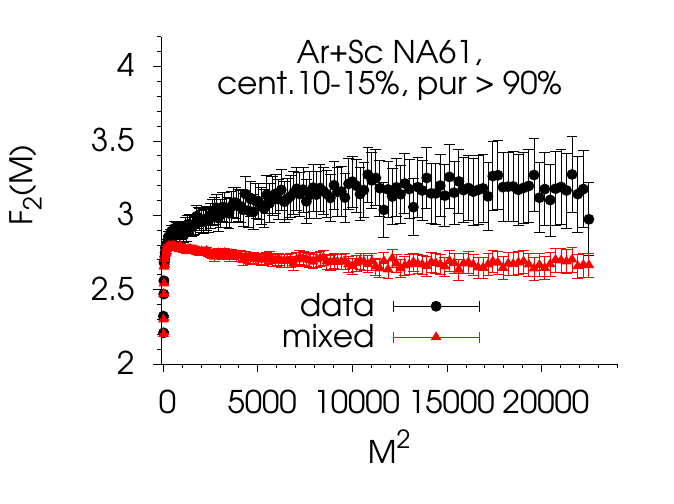}
\includegraphics[width=0.43\textwidth]{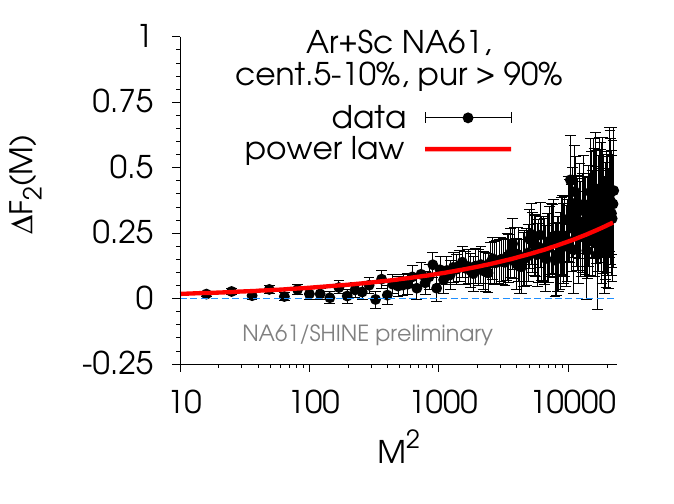}
\includegraphics[width=0.43\textwidth]{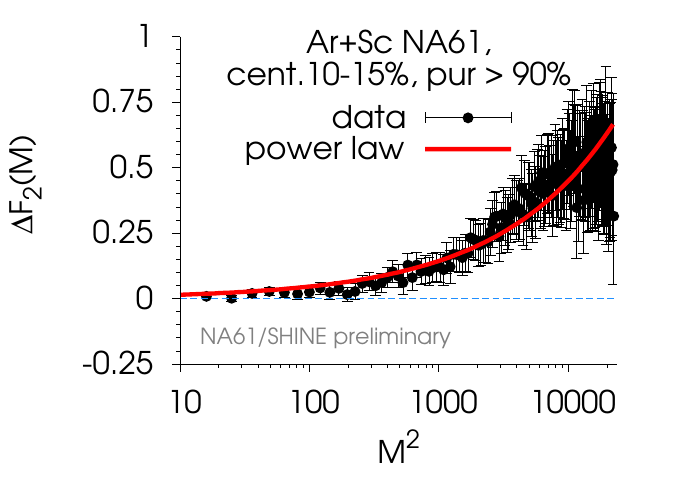}
\vspace{-0.2cm}
\caption[]{\footnotesize Preliminary results on $F_2(M)$ (top) and $\Delta F_2(M)$ (bottom) of mid-rapidity protons measured in 5-10\% (left) and 10-15\% (right) central Ar+Sc collisions at 150$A$ GeV/c. Solid red line in bottom panel represents $\Delta F_2(M) = \exp(C) \cdot (M^2)^{\phi_2}$ function with $C=-4.84$, $\phi_2=0.36$ for 5-10\% centrality, and $C=-5.40$, $\phi_2=0.49$ for 10-15\% centrality.}
\label{intermit_arsc}
\end{figure}

\begin{figure}
\centering
\includegraphics[width=0.3\textwidth]{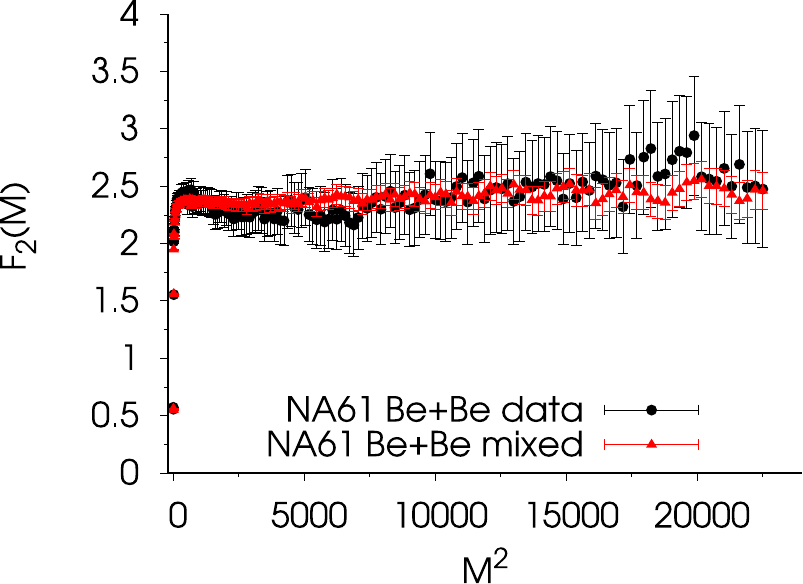}
\includegraphics[width=0.32\textwidth]{./FIGURES/NA61_ArSc_F2M_c_5_10_pur90.pdf}
\includegraphics[width=0.32\textwidth]{./FIGURES/NA61_ArSc_F2M_c_10_15_pur90.pdf}
\includegraphics[width=0.3\textwidth]{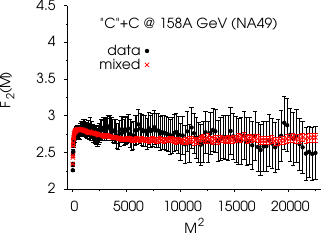}
\includegraphics[width=0.3\textwidth]{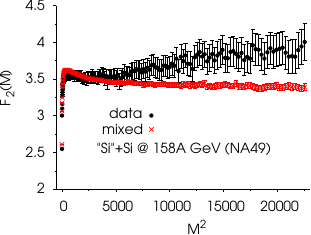}
\includegraphics[width=0.3\textwidth]{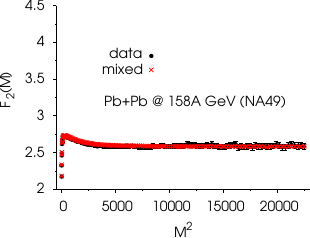}
\vspace{-0.1cm}
\caption[]{\footnotesize Preliminary NA61/SHINE results on $F_2(M)$ of protons at mid-rapidity measured in the 0-12\% most central Be+Be and 5-10\% and 10-15\% central Ar+Sc collisions at 150$A$ GeV/c (upper panel), as well as in the 0-12\% most central 'C'+C and 'Si'+Si and the most central 0-10\% Pb+Pb interactions~\cite{Anticic:2012xb} studied by the NA49 experiment (lower panel).}
\label{intermit_na49_na61}
\end{figure}

In the recent analysis of NA61/SHINE~\cite{Davis_CPOD2018_inter} the intermittency effects were studied in central Be+Be and centrality selected Ar+Sc collisions at 150$A$ GeV/c. The $dE/dx$ method was used for the identification of protons. Centrality was determined using the information on energy deposited in the PSD detector. For Ar+Sc collisions protons were selected with at least 90\% purity (see Ref.~\cite{Davis_CPOD2018_inter} for more details and other proton purities). Figure~\ref{intermit_arsc} shows preliminary results on $F_2(M)$ and $\Delta F_2(M)$ of mid-rapidity protons produced in 5-10\% and 10-15\% central Ar+Sc collisions at 150$A$ GeV/c. In Fig.~\ref{intermit_na49_na61} preliminary NA61/SHINE Be+Be and Ar+Sc results are compared to older NA49 results for 0-12\% central 'C'+C and 'Si'+Si and 0-10\% central Pb+Pb interactions~\cite{Anticic:2012xb}.

The overlap of $F_2(M)$ for data and mixed events seen in NA61/SHINE Be+Be collisions and NA49 'C'+C and Pb+Pb interactions indicates that $\Delta F_2(M)$ fluctuates around zero, and suggests that no intermittency (scaling of second order factorial moments) is observed. In contrast, an indication of an intermittency effect is seen in mid-central Ar+Sc collisions at 150$A$ GeV/c. This is similar to the effect seen by the NA49 experiment in central 'Si'+Si collisions\footnote{In NA49 intermittency analysis the fragmentation 'C' beam was a mixture of ions with $Z$=6 and 7, and fragmentation 'Si' beam was a mixture of ions
with $Z$=13, 14, 15.}
at 158$A$ GeV/c. The result seen in Ar+Sc collisions is a first possible evidence of a CP signal in the NA61/SHINE experiment. The analysis of Xe+La at 150$A$ GeV/c, as well as Ar+Sc at 75$A$ GeV/c has the highest priority in NA61/SHINE as it might strengthen the evidence for the expected non-monotonic system size and collision energy dependence of an intermittency signal from the critical point. Here, one could mention older intriguing results observed for intermediate mass systems at 158$A$ GeV/c. Namely, an intermittency signal of \textit{di-pions} was seen in 158$A$ GeV/c Si+Si interactions by NA49~\cite{Anticic:2009pe}. Also, the event-by-event transverse momentum and multiplicity fluctuations were increased in semi-central Si+Si and C+C collisions at 158$A$ GeV/c~\cite{Grebieszkow:2009jr, Anticic:2015fla} (see also next section).

\section{Can we extend the critical point search by looking at the $\Delta \eta$ dependence of fluctuations?}

NA61/SHINE uses the \textit{strongly intensive} measures $\Delta[P_T, N]$ and $\Sigma[P_T, N]$ to study transverse momentum and multiplicity fluctuations~\cite{Aduszkiewicz:2015jna}. In the Wounded Nucleon Model (WNM) they depend neither on the number of wounded nucleons ($W$) nor on fluctuations of $W$. In the Grand Canonical Ensemble they do not depend on volume and volume fluctuations. Moreover, $\Delta[P_T, N]$ and $\Sigma[P_T, N]$ have two reference values, namely they are equal to zero in case of no fluctuations and one in case of independent particle production. 

The NA61/SHINE acceptance used in this analysis is described in Ref.~\cite{Aduszkiewicz:2015jna}. Additionally, due to poor azimuthal angle acceptance and stronger electron contamination at backward rapidities, we limited the rapidity acceptance to $0 < y_{\pi} < y_{beam}$, where $y_{\pi}$ is the rapidity calculated assuming pion mass, and $y_{beam}$ is the beam rapidity. 

\begin{figure}
\centering
\includegraphics[width=0.3\textwidth]{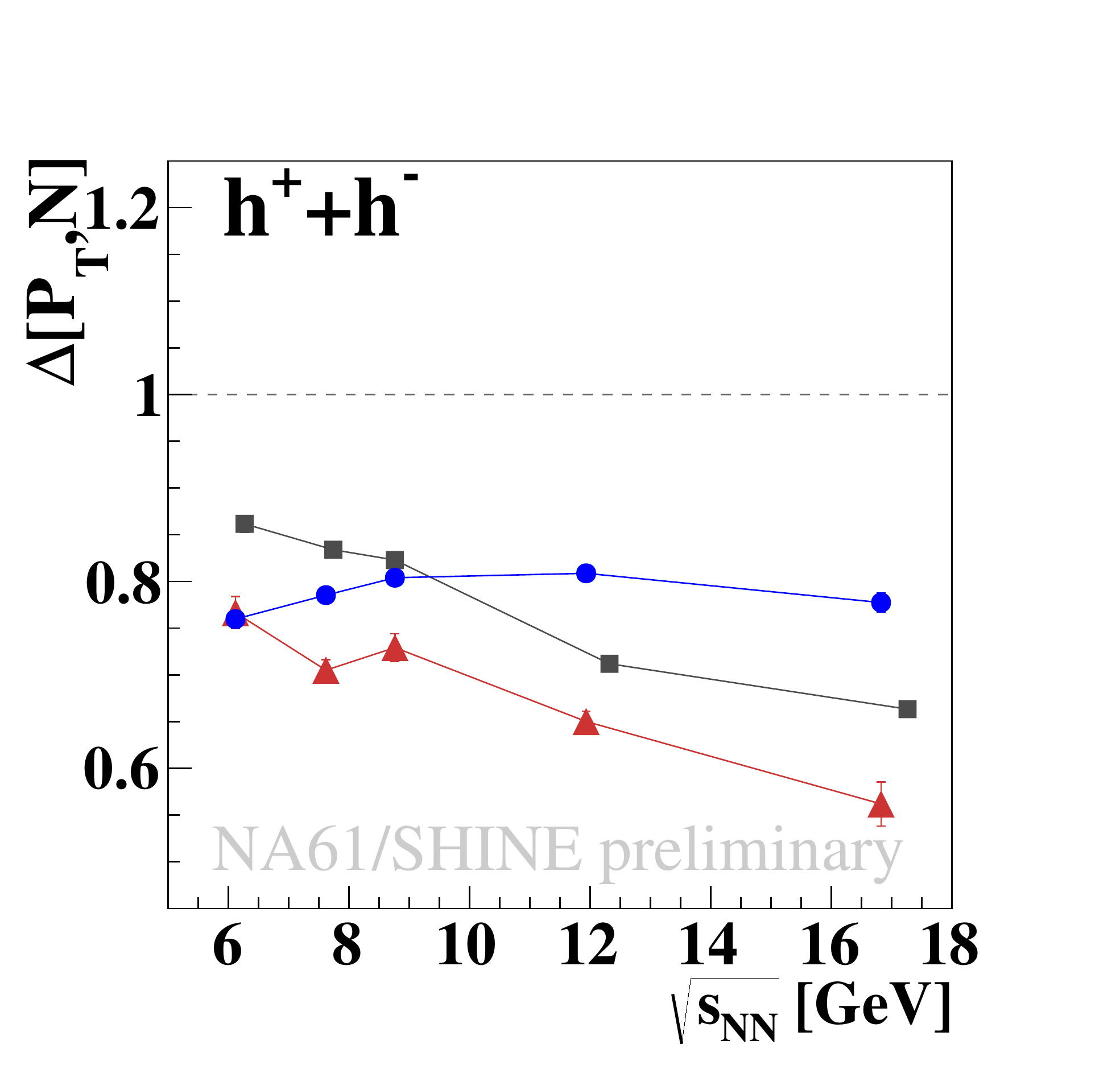}    
\includegraphics[width=0.3\textwidth]{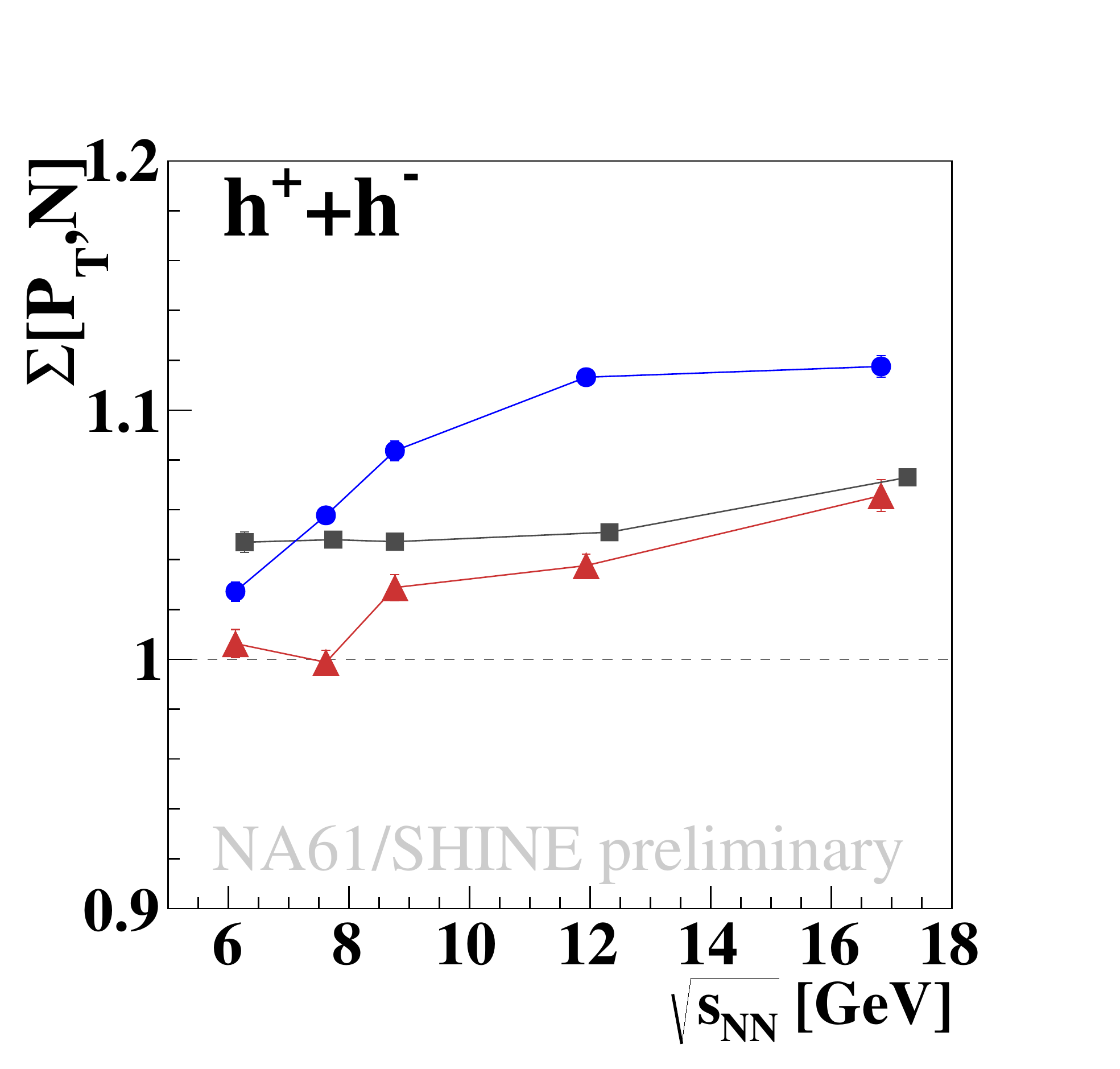} \\
\includegraphics[width=0.35\textwidth]{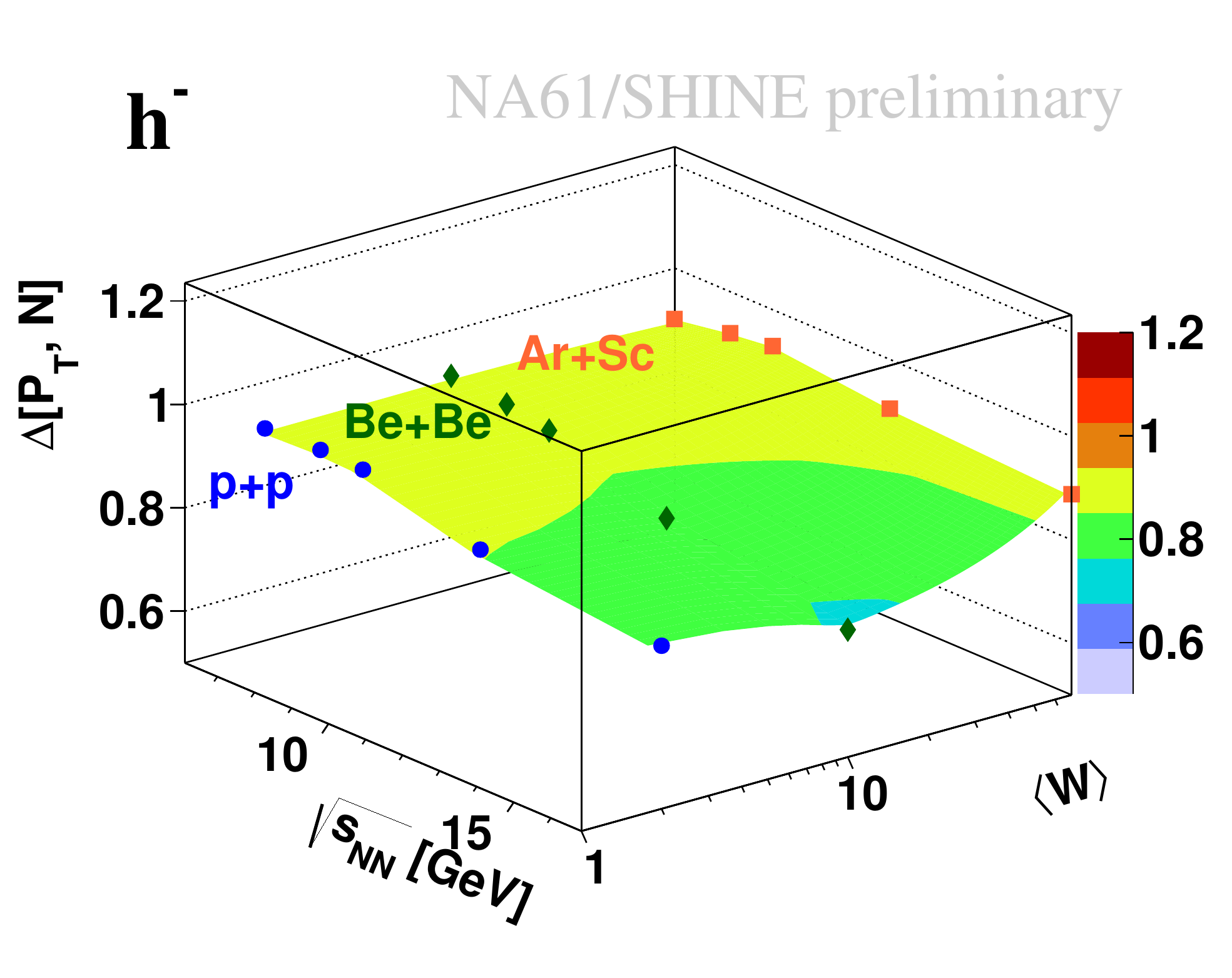}
\includegraphics[width=0.35\textwidth]{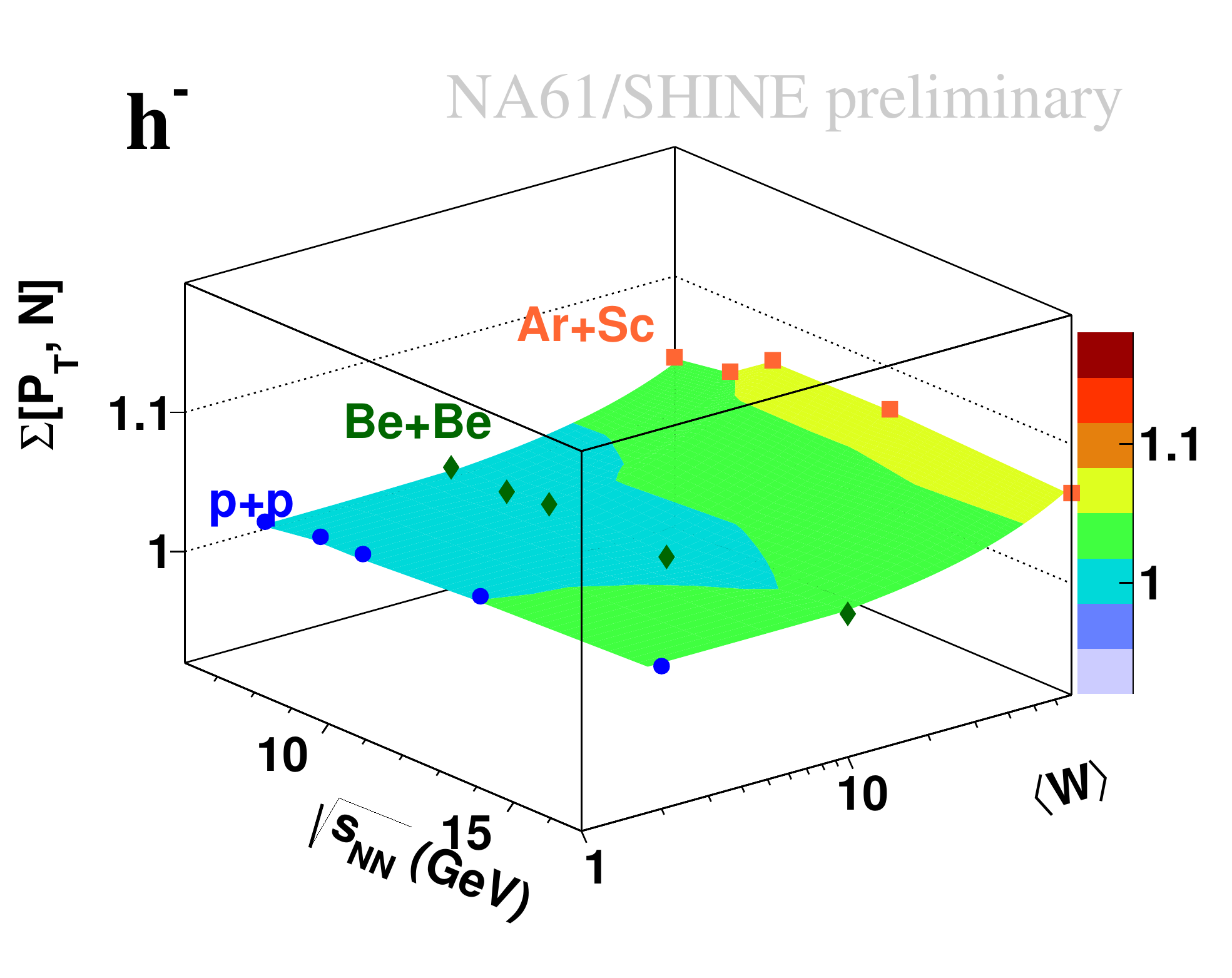}
\vspace{-0.3cm}
\caption[]{\footnotesize Upper panel: $\Delta[P_T, N]$ and $\Sigma[P_T, N]$ for all charged hadrons ($h^{+}+h^{-}$) in inelastic p+p (grey squares), 0-5\% Be+Be (red triangles), and 0-5\% Ar+Sc (blue circles) collisions obtained by NA61/SHINE in the rapidity interval $0< y_{\pi}< y_{beam}$ and with $p_T<1.5$ GeV/c. Only statistical uncertainties are shown. Bottom panel: similar as top but results for negatively charged hadrons ($h^{-}$). All NA61/SHINE results are preliminary.}
\label{evgeny_pp_bebe_arsc}
\end{figure}

Figure~\ref{evgeny_pp_bebe_arsc} shows the energy dependence of $\Delta[P_T, N]$ and $\Sigma[P_T, N]$ measured in inelastic p+p collisions, as well as in the 0-5\% most central Be+Be and Ar+Sc interactions (note that p+p and Be+Be results were already shown by NA61/SHINE \cite{Aduszkiewicz:2015jna, Czopowicz:2015mfa} but in a slightly different acceptance). Centrality was determined based on the energy measured in the PSD. The energy dependences of the studied systems show no prominent non-monotonic behaviors that could be attributed to a CP. The values of $\Delta[P_T, N]$ smaller than 1 and $\Sigma[P_T, N]$ higher than 1 may be due to Bose-Einstein statistics and/or anti-correlation between event transverse momentum and multiplicity (see Ref.~\cite{Gorenstein:2013nea} and references therein). 
A growing deviation of $\Delta[P_T, N]$ and $\Sigma[P_T, N]$ from unity (independent particle model) with energy may be due to increasing azimuthal acceptance (for a given energy azimuthal acceptance for p+p, Be+Be and Ar+Sc is the same). 

\begin{figure}
\centering
\includegraphics[width=0.3\textwidth]{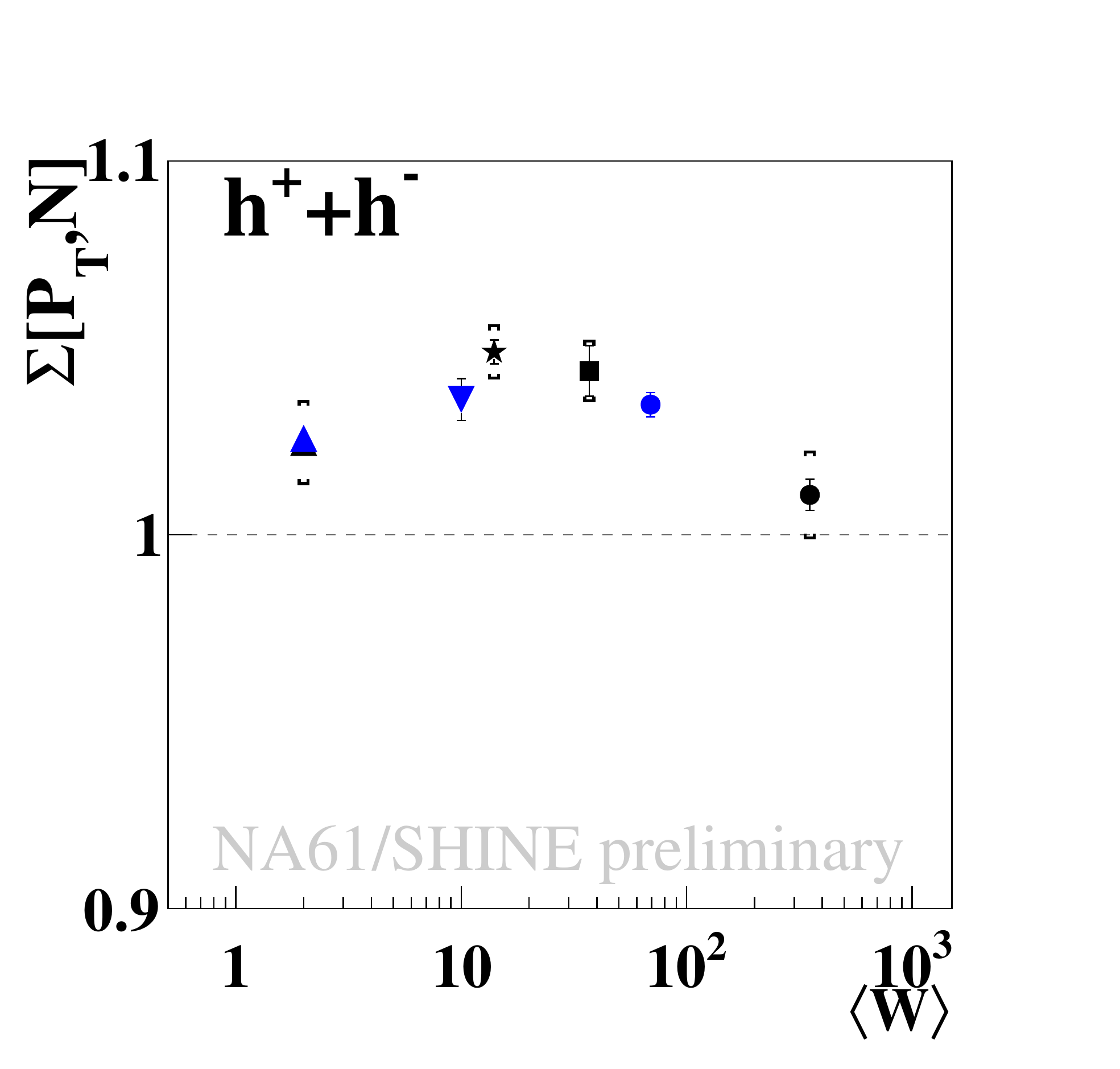}
\includegraphics[width=0.65\textwidth]{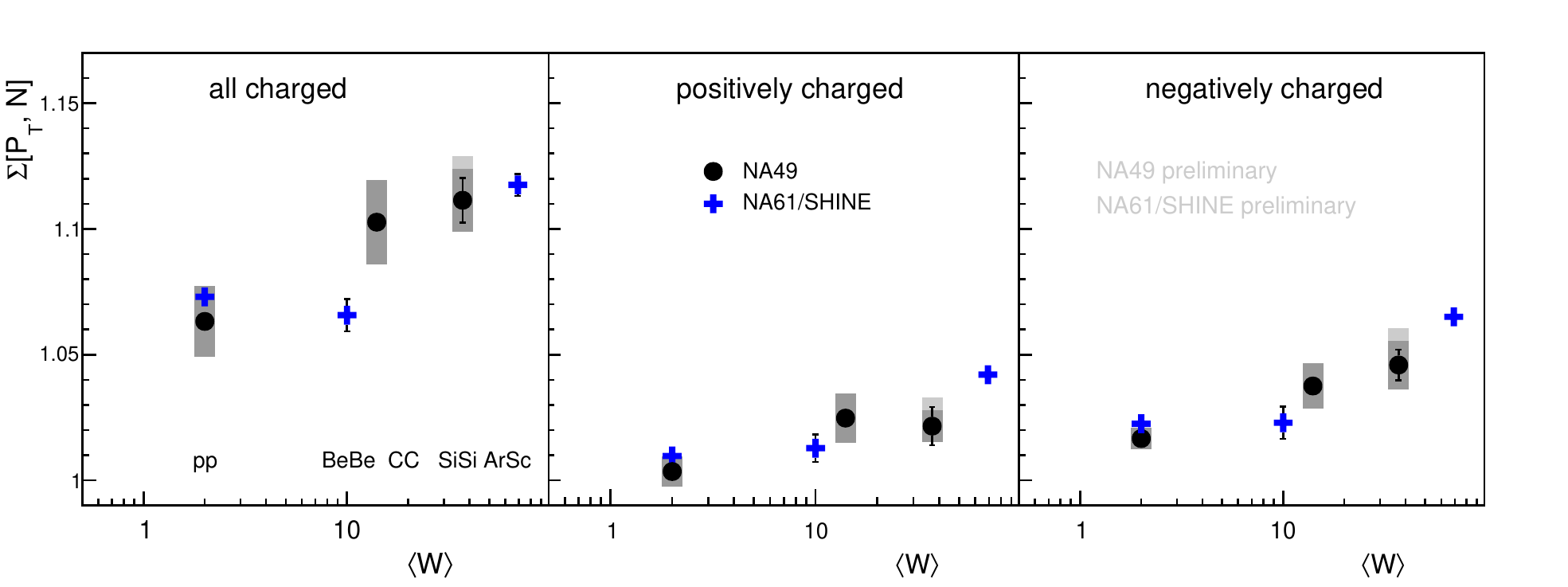}
\vspace{-0.2cm}
\caption[]{\footnotesize Left: $\Sigma[P_T, N]$ at 150$A$/158$A$ GeV/c. Black points are NA49 \cite{Anticic:2015fla} data (p+p, 0-15.3\% C+C, 0-12.2\% Si+Si, 0-5\% Pb+Pb), blue ones preliminary NA61/SHINE (p+p, 0-5\% Be+Be, 0-5\% Ar+Sc). Results in $1.1< y_{\pi}< 2.6$. Right: similar as left (no Pb+Pb point) but results in $0< y_{\pi}< y_{beam}$. Preliminary NA49~\cite{Grebieszkow:2017gqx} and NA61/SHINE data. For NA61/SHINE only statistical uncertainties are shown.}
\label{fluct_na49_na61_size}
\end{figure}

Preliminary NA61/SHINE results have been compared to NA49 data. For the system size dependence of $\Sigma[P_T, N]$ at 150$A$/158$A$ GeV/c (Fig.~\ref{fluct_na49_na61_size}, left) the NA49~\cite{Anticic:2015fla} and NA61/SHINE data points show consistent trends. $\Delta[P_T, N]$ (not shown) is more sensitive to the width of the centrality interval~\cite{Gorenstein:2013nea} and points are scattered (see Ref.~\cite{evgeny_cpod16_slides}). So far the analysis in NA49 was restricted to $1.1< y_{\pi}< 2.6$ but in 2017 p+p, C+C and Si+Si data were reanalyzed in the acceptance currently used by NA61/SHINE ($0< y_{\pi}< y_{beam}$). As expected, fluctuations are larger for the larger rapidity interval (Fig.~\ref{fluct_na49_na61_size}, right) and an increase of $\Sigma[P_T, N]$ from p+p to Ar+Sc/Si+Si can be observed. It is now very important to obtain future NA61/SHINE results in Xe+La and Pb+Pb to figure out whether a monotonic or rather non-monotonic (as seen in Fig.~\ref{fluct_na49_na61_size}, left) dependence will be observed in this larger rapidity window.

\begin{figure}
\centering
\vspace{0.1cm}
\includegraphics[width=0.35\textwidth]{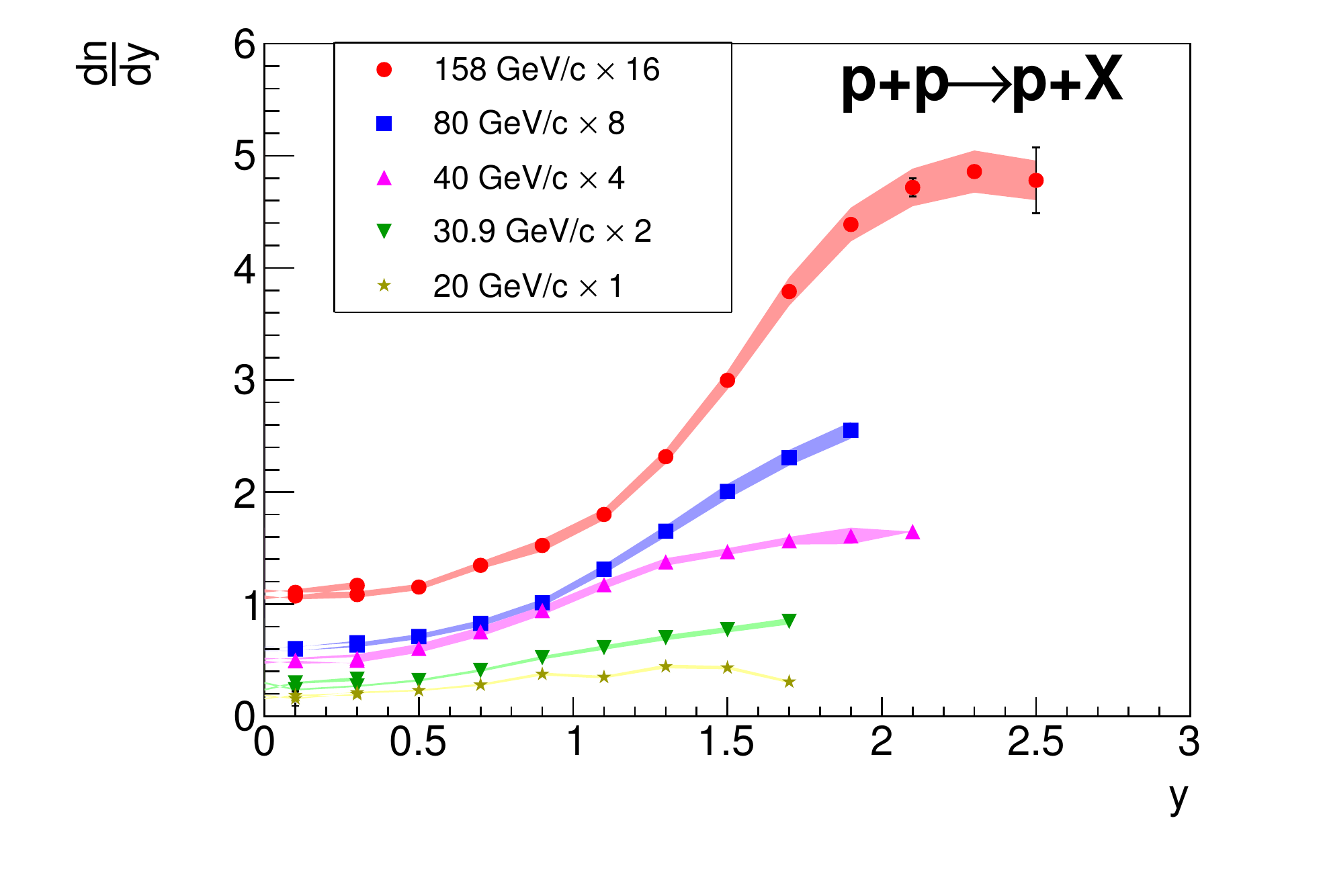}
\includegraphics[width=0.35\textwidth]{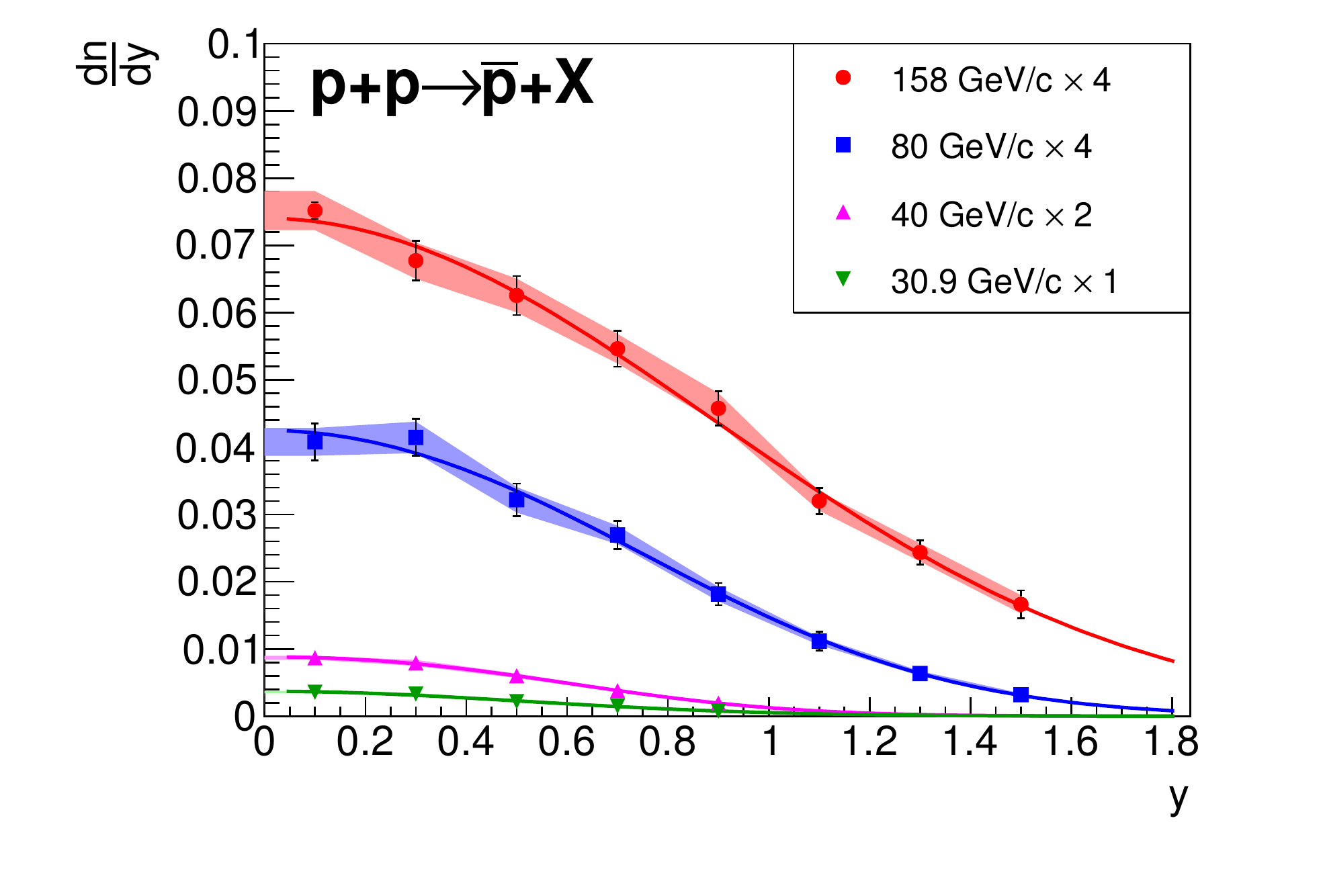}
\vspace{-0.5cm}
\caption[]{\footnotesize Rapidity spectra of protons (left) and anti-protons (right) measured by NA61/SHINE in inelastic p+p collisions at SPS energies. Figure taken from Ref.~\cite{Aduszkiewicz:2017sei}.}
\label{dndy_prot_antiprot_EPJC}
\end{figure}

Now, one may want to come back to the question asked in the title of this section, namely, can we extend the critical point search by looking at $\Delta \eta$ dependence of fluctuations? Varying the width of the rapidity acceptance window allows to study fluctuations at different baryochemical potentials. The ratio of anti-proton to proton multiplicities can be expressed as $\bar{p}/p = \exp(-(2\mu_B)/T)$, where $T$ is chemical freeze-out temperature and $\mu_B$ baryochemical potential. Rapidity spectra of protons and anti-protons in inelastic p+p collisions at SPS energies (Fig.~\ref{dndy_prot_antiprot_EPJC}) show that the anti-proton to proton ratio significantly changes with rapidity. Therefore, varying the rapidity position and/or rapidity bin width is an additional way to scan the phase diagram of strongly interacting matter.

\begin{figure}
\centering
\includegraphics[width=0.35\textwidth]{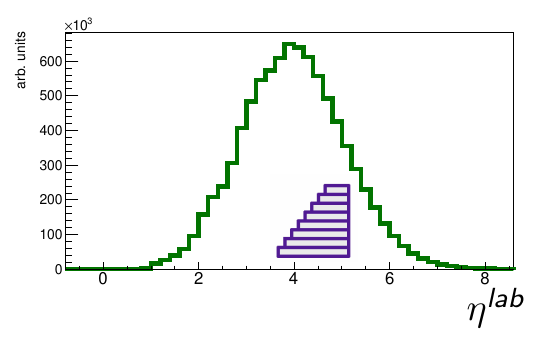}
\includegraphics[width=0.35\textwidth]{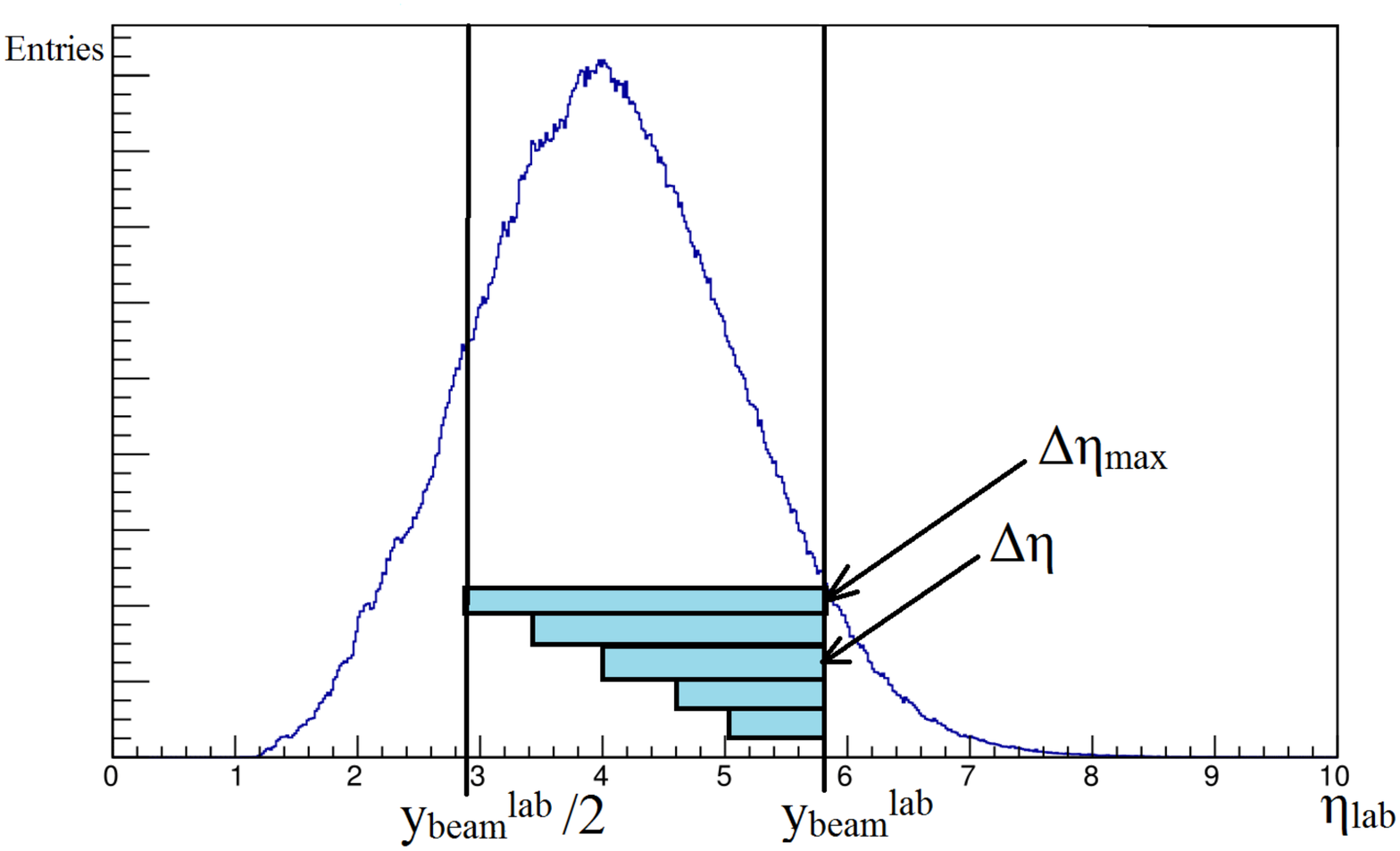}
\vspace{-0.4cm}
\caption[]{\footnotesize Sketch of pseudorapidity (in laboratory system) uncorrected spectrum of
charged hadrons with proposed analysis windows for Be+Be (left) and p+p (right) interactions at 150$A$/158 GeV/c.}
\label{pseudorap_bins_pic}
\end{figure}

\begin{figure}
\centering
\vspace{0.2cm}
\includegraphics[width=0.3\textwidth]{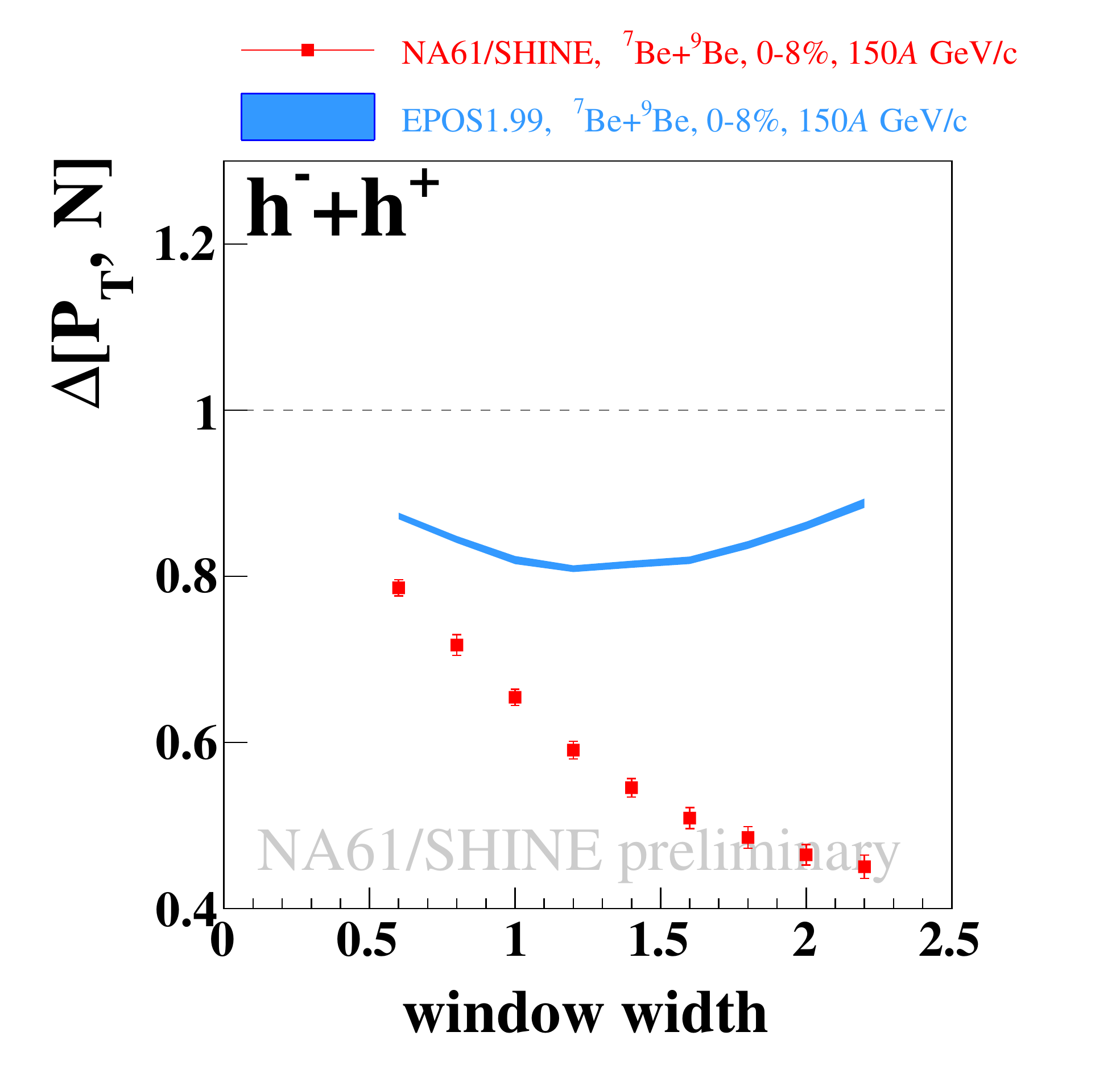}
\includegraphics[width=0.3\textwidth]{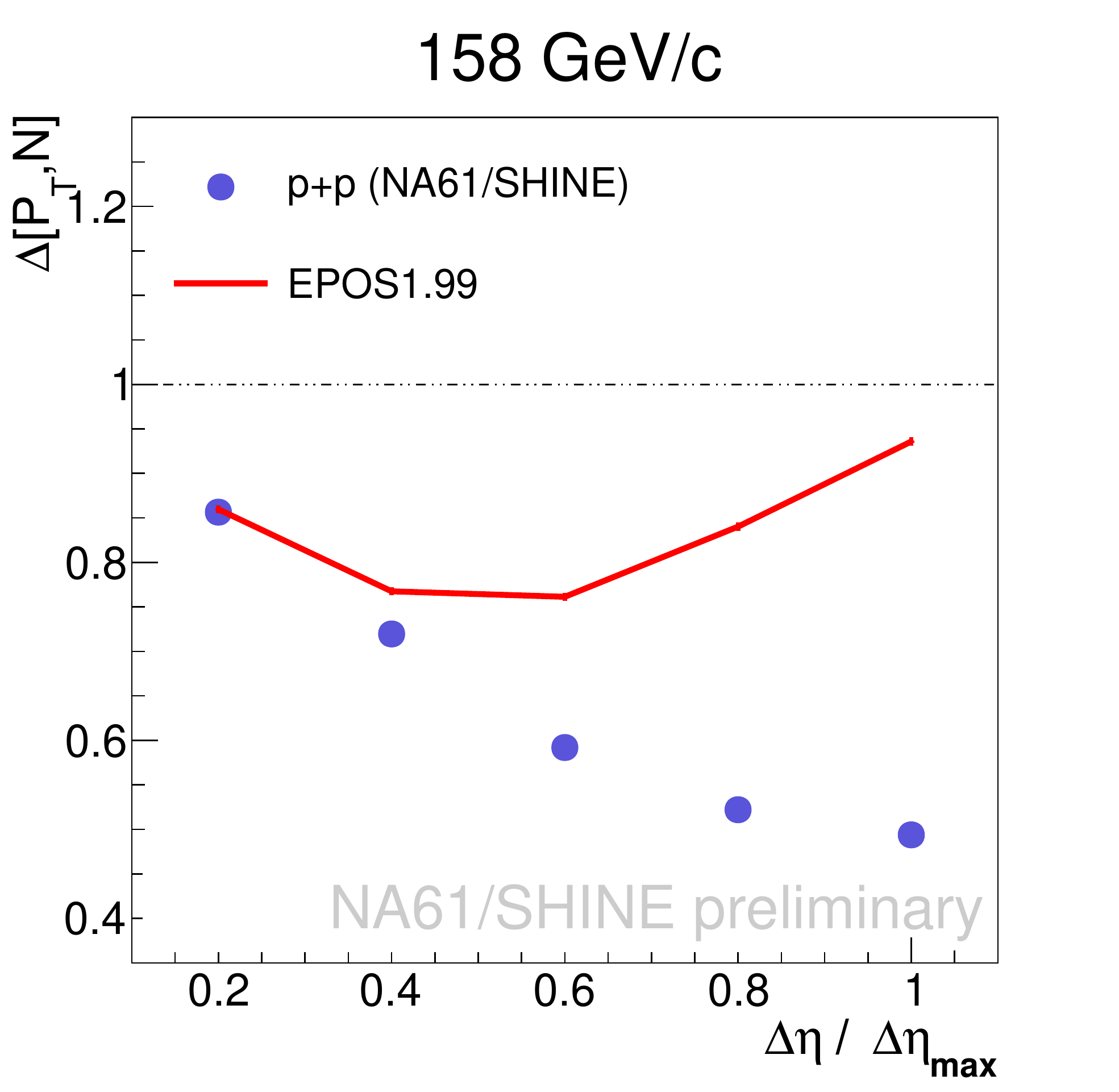}
\includegraphics[width=0.3\textwidth]{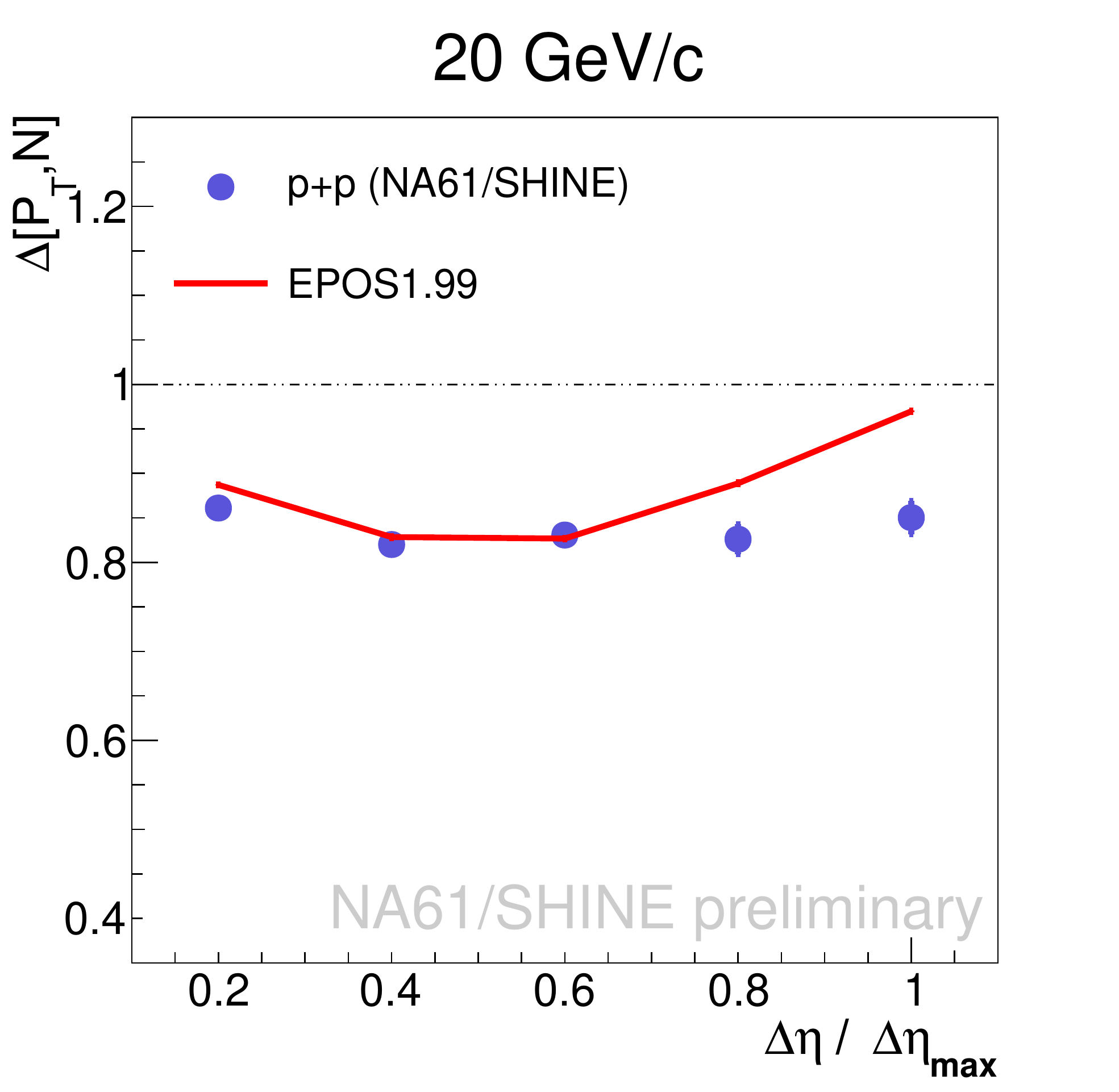}
\vspace{-0.2cm}
\caption[]{\footnotesize Preliminary results for $\Delta[P_T, N]$ dependence on the width of the pseudorapidity window measured in the 0-8\% most central Be+Be interactions at 150$A$ GeV/c (left) and in inelastic p+p collisions at 158 (middle) and 20 (right) GeV/c.}
\label{Delta_PTN_pseudorap}
\end{figure}

Figure~\ref{pseudorap_bins_pic} shows a sketch of the uncorrected psedorapidity spectrum of charged hadrons with the proposed analysis windows for Be+Be and p+p interactions at 150$A$/158 GeV/c. Note, that slightly different $\Delta \eta$ ranges were used in the analysis of Be+Be and p+p data. The corresponding $\Delta \eta$ dependences of transverse momentum and multiplicity fluctuations are presented in Fig.~\ref{Delta_PTN_pseudorap}. A non-trivial $\Delta[P_T, N]$ dependence on the width $\Delta \eta$ of the window can be observed for p+p and Be+Be interactions at 158/150$A$ GeV/c. Such a behaviour is not reproduced by the EPOS model~\cite{EPOS_CRMC_www} and the deviation from EPOS grows with the width of the pseudorapidity window. On the other hand, a disagreement of data and model is not present in p+p collisions at 20 GeV/c (Be+Be interactions at lower energies are not analyzed yet). The values and $\Delta \eta$ trends of $\Sigma[P_T, N]$ (not shown here), studied for Be+Be at 150$A$ GeV/c and p+p at 20--158 GeV/c, are fairly consistent with the EPOS model predictions. More NA61/SHINE results on this subject can be found in Refs.~\cite{Prokhorova:2018tcl, Andronov:2018ukf}.

\section{Future: Can we help to calibrate the QGP thermometer?}

\begin{figure}
\centering
\includegraphics[width=0.8\textwidth]{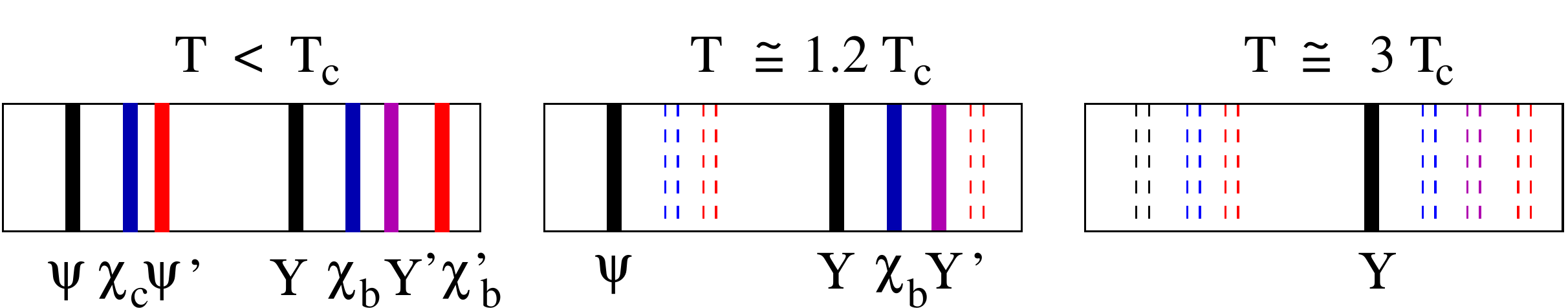}
\vspace{-0.1cm}
\caption[]{\footnotesize Surviving quarkonium states for different quark-gluon plasma temperatures. Figure taken from Ref.~\cite{Satz:2006kba}.}
\label{fig_thermometer}
\end{figure}

\begin{figure}
\centering
\includegraphics[width=0.5\textwidth]{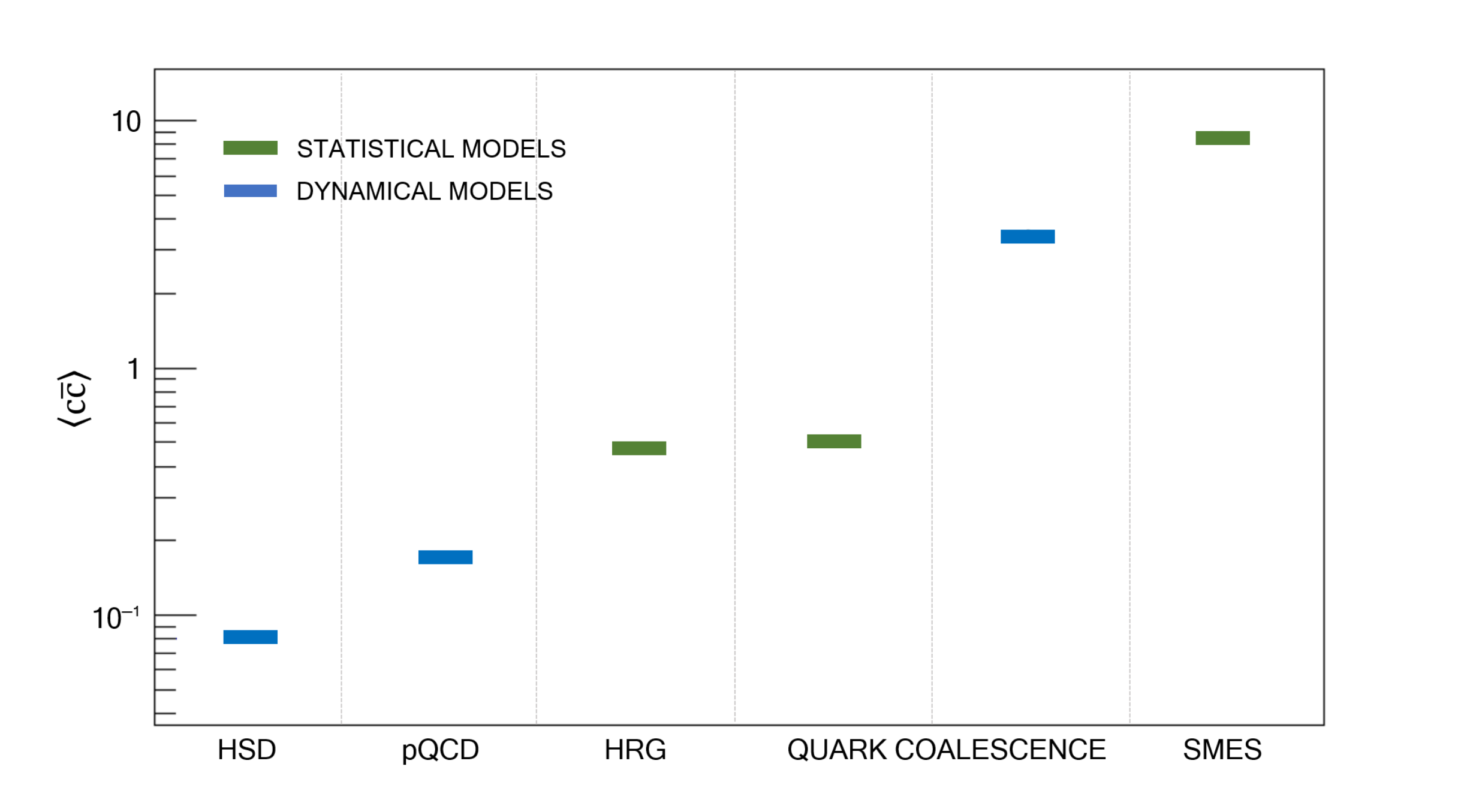}
\vspace{-0.2cm}
\caption[]{\footnotesize Statistical (HRG and Quark Coalesc. Stat.~\cite{Kostyuk:2001zd}, SMES~\cite{Gazdzicki:1998vd}) and dynamical (HSD~\cite{Linnyk:2008hp, Song_priv}, pQCD~\cite{Gavai:1994gb, BraunMunzinger:2000px}, Quark Coalesc. Dyn.~\cite{Levai:2000ne}) model predictions of average number of $\langle c \bar{c} \rangle$ quark pairs produced in central Pb+Pb collisions at 158$A$ GeV/c.}
\label{cc_models}
\end{figure}

It was suggested a long time ago that the quarkonium states may serve as a quark-gluon plasma thermometer. The melting of different states depends on the initial state temperature of the produced matter (Fig.~\ref{fig_thermometer}). The currently available models try to describe the production rates of charmonium states, by implementing sequential melting, regeneration effects, initial state effects, etc.. The cross-section for charm quark production is an important parameter for the models, but predictions of different models (Fig.~\ref{cc_models}) differ by two orders of magnitude!

\begin{figure}
\centering
\includegraphics[width=0.62\textwidth]{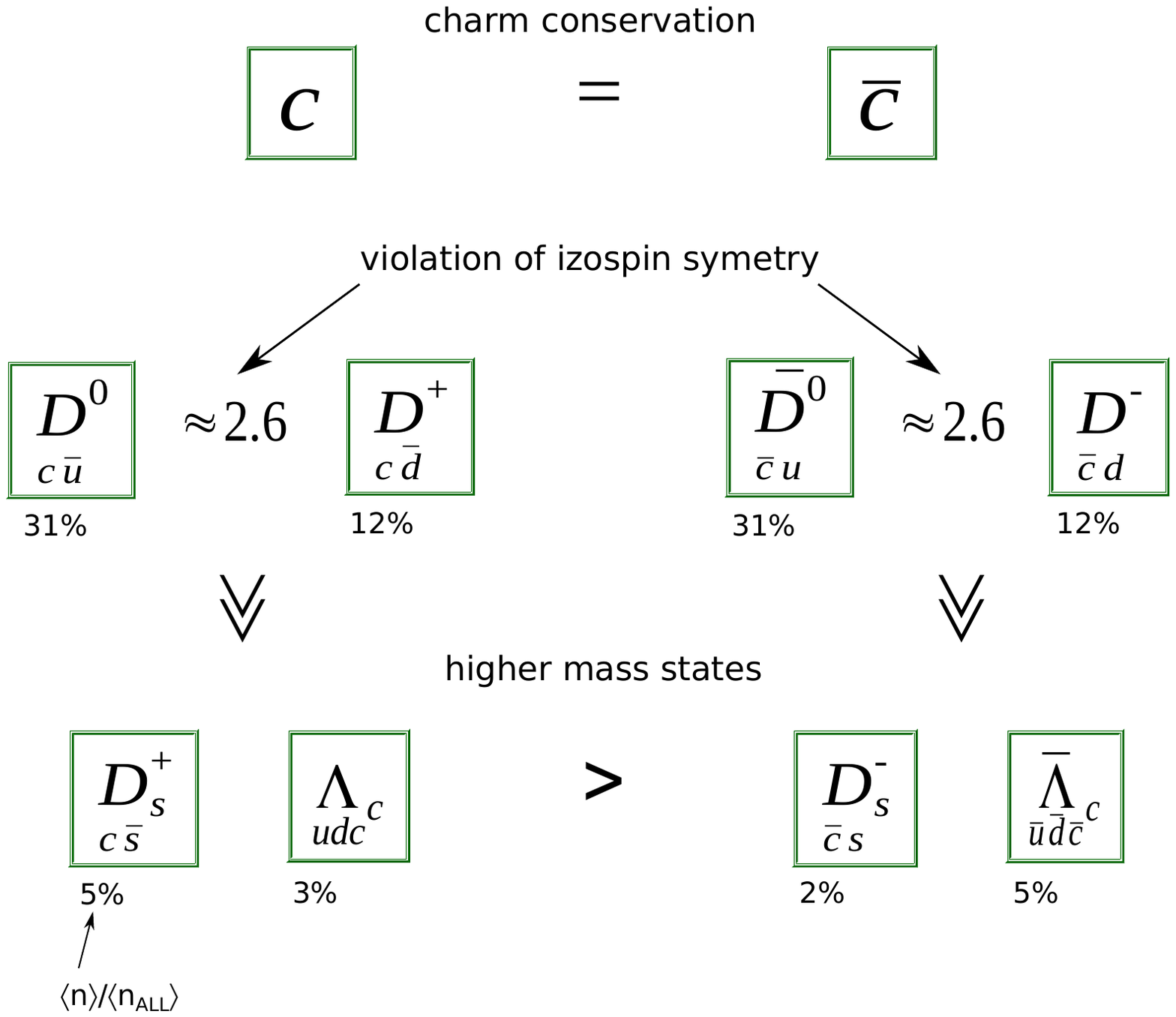}
\vspace{-6.8cm}
\caption[]{\footnotesize Distribution of open charm among charm carriers as obtained from the PHSD model~\cite{Linnyk:2008hp, Song_priv} for the 0-20\% most central Pb+Pb interactions at 150$A$ GeV/c.}
\label{charm_carriers}
\end{figure}

The NA61/SHINE fixed target experiment can measure $\langle c \bar{c} \rangle$ production in the full phase space. This will be the first time in history when such a measurement will be performed at SPS energies. In order to estimate the average number of charm quark pairs, the yields of all the most frequently produced charm carriers need to be known. Figure~\ref{charm_carriers} presents the distribution of open charm among different charm carriers, as obtained from the PHSD model~\cite{Linnyk:2008hp, Song_priv} for the 0-20\% most central Pb+Pb interactions at 150$A$ GeV/c. As a first step, NA61/SHINE will focus on $D^0$ and $\overline{D^0}$ production but after the necessary upgrades of NA61/SHINE detectors (see below) we will be able to measure the yields of all the most abundant charm hadrons allowing the first direct estimate of $\langle c \bar{c} \rangle$ production at SPS energies. Anyway, measuring $D^0$, $\overline{D^0}$, $D^{+}$, and $D^{-}$ is likely sufficient for a good $\langle c \bar{c} \rangle$ estimate.

\begin{figure}
\centering
\includegraphics[width=0.5\textwidth]{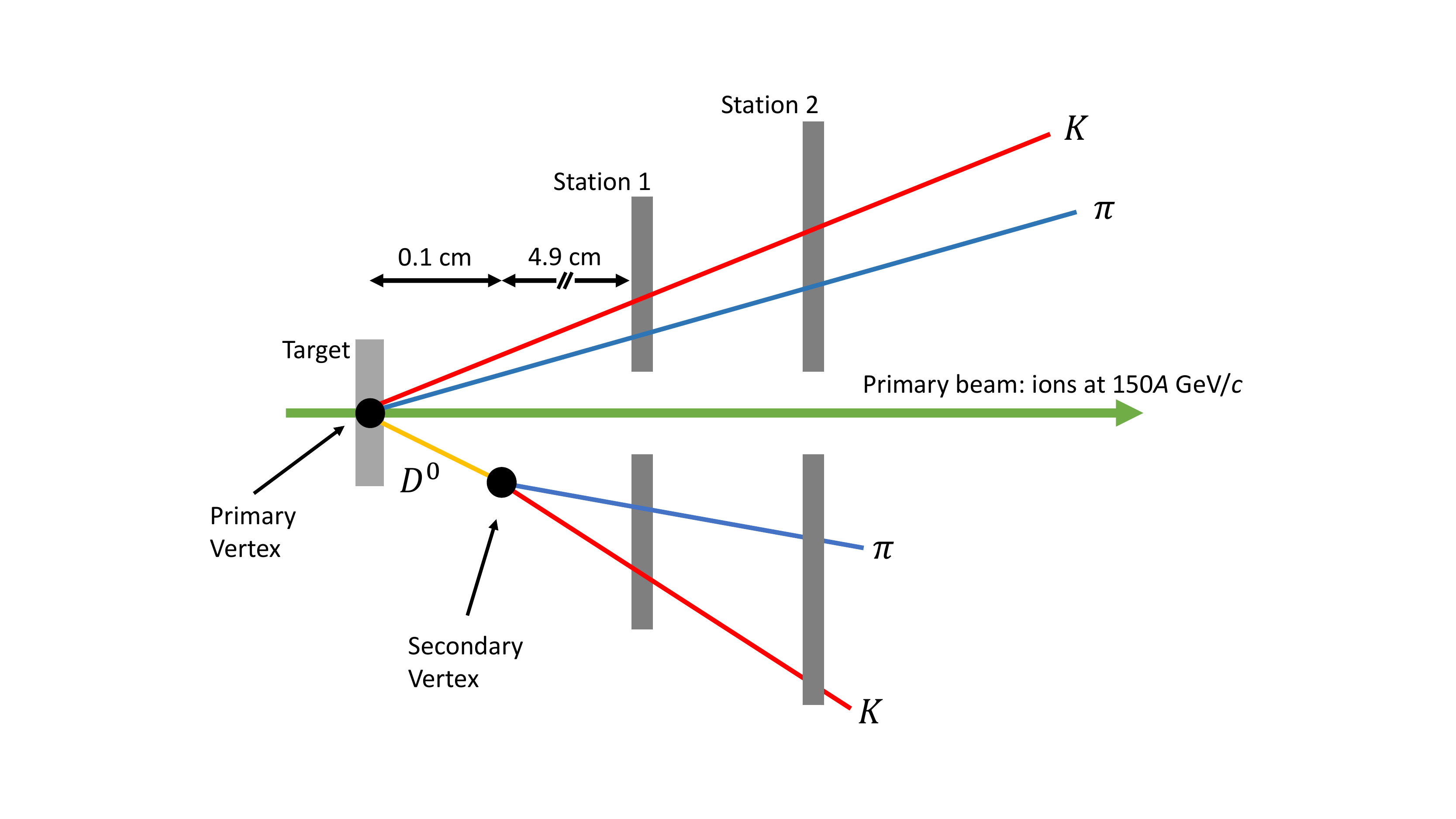}
\includegraphics[width=0.4\textwidth]{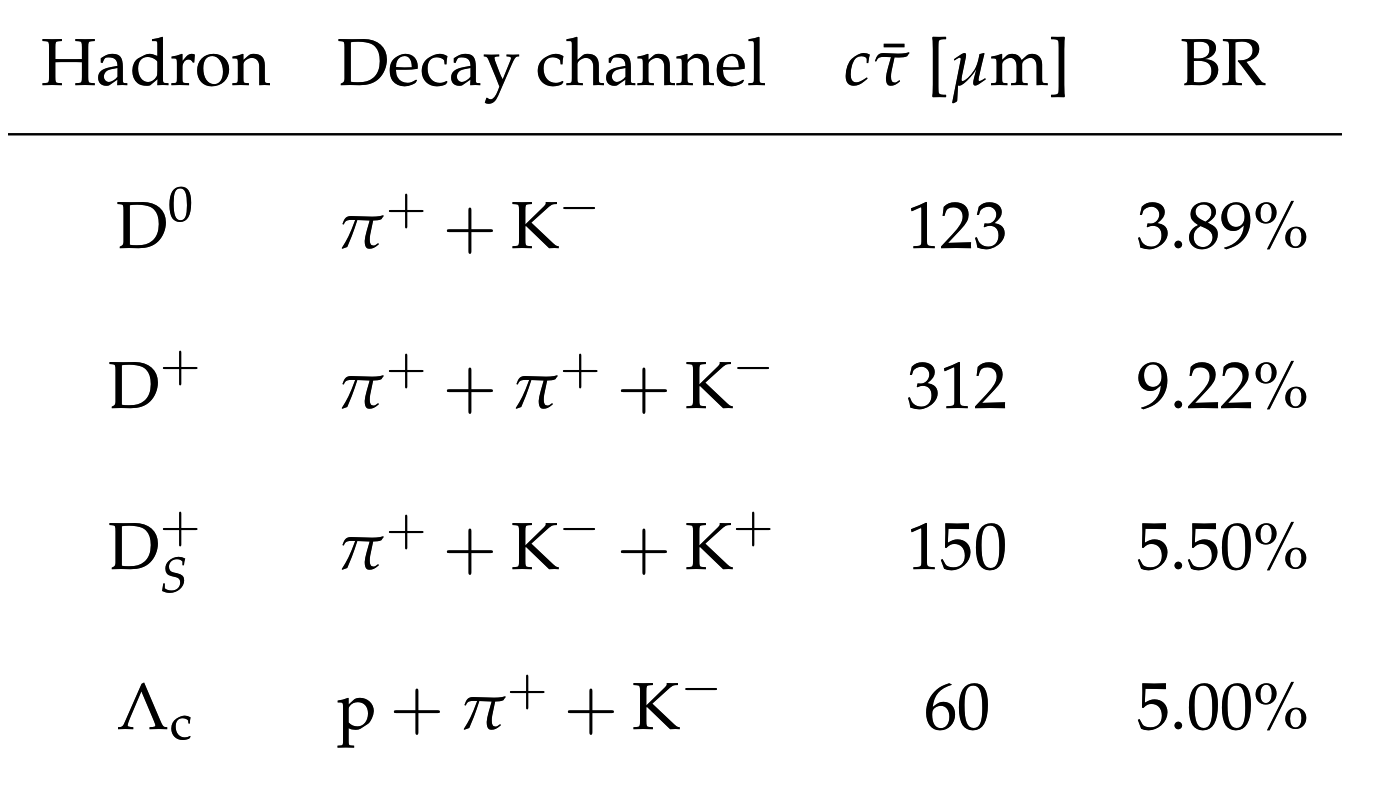}
\vspace{-0.2cm}
\caption[]{\footnotesize Left: schematic of reconstruction of $D^0 \rightarrow \pi^{+}\, K^{-}$ decay with help of the vertex detector. Right: the most frequently produced charm hadrons with their mean life times and the decay channels (with branching ratios) best suited for measurements.}
\label{V0_decay_table}
\end{figure}

Starting the open charm program in NA61/SHINE required construction of a new high-resolution
vertex detector. The schematic of the reconstruction of the $D^0 \rightarrow \pi^{+}\, K^{-}$ decay with the help of the vertex detector is shown in Fig.~\ref{V0_decay_table}, left. Daughters of $D^0$ are recognized as a pair forming a secondary vertex (so-called $V^0$) displaced from the primary
interaction vertex. The average lifetime is $c \tau (D^0) \approx 123\, \mathrm{\mu m}$ but due to Lorentz boost ($\beta \gamma \approx 10$) the secondary vertex displacement is on the level of 1 mm. Thus, the Lorentz boost makes the measurements significantly easier than in collider experiments. The list of charm hadrons, considered for measurements in NA61/SHINE, is given in Fig.~\ref{V0_decay_table}, right.

\begin{figure}
\centering
\includegraphics[width=0.3\textwidth]{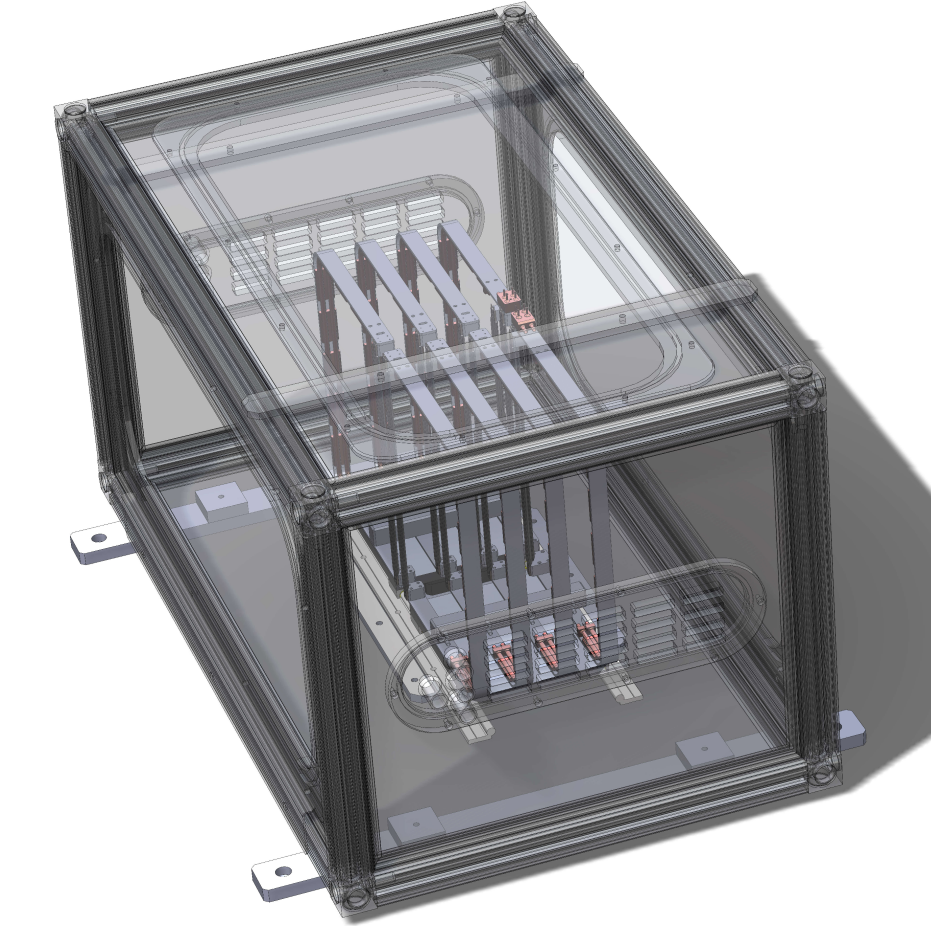}
\includegraphics[width=0.4\textwidth]{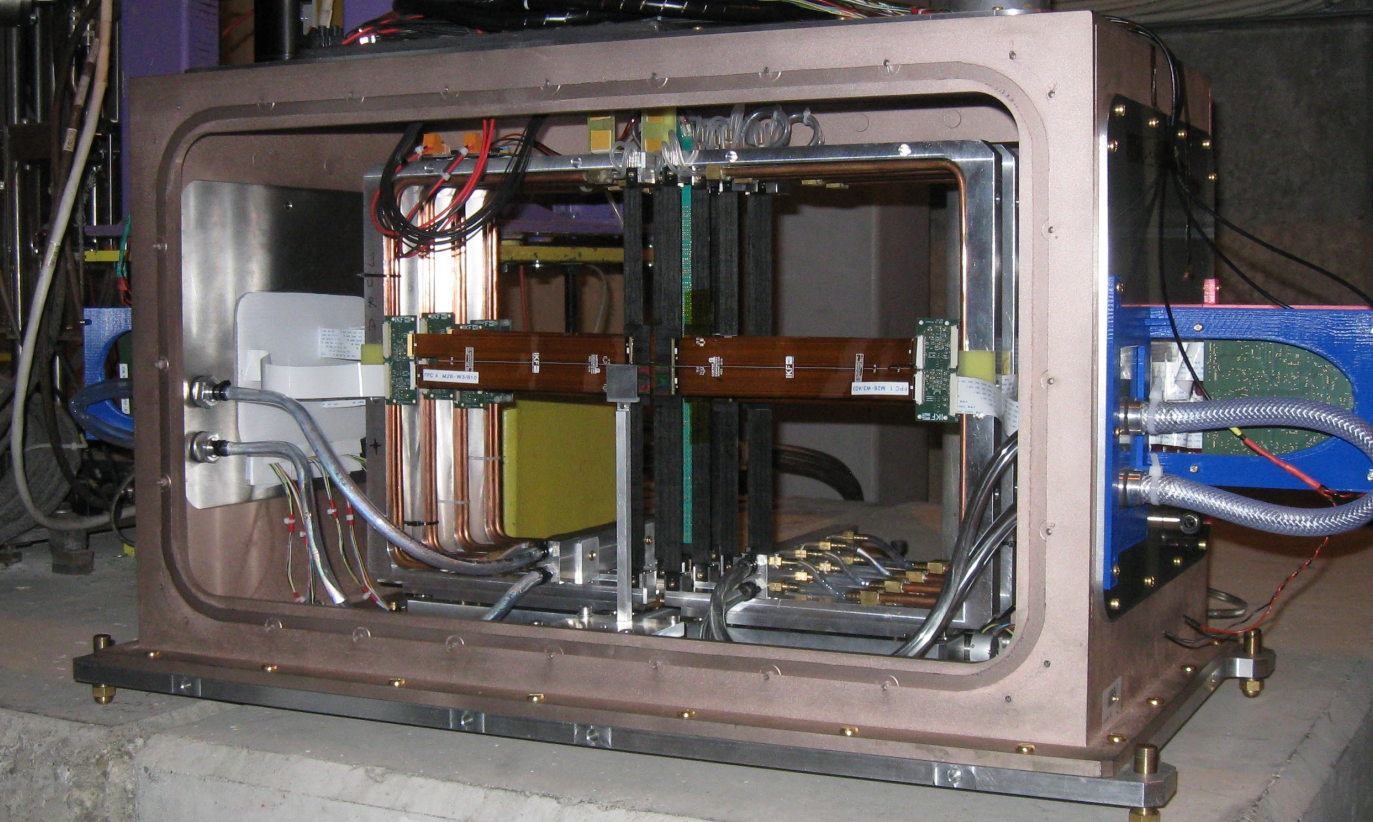}
\vspace{-0.1cm}
\caption[]{\footnotesize Project (left) and photo (right) of the Small Acceptance Vertex Detector.}
\label{SAVD_fig}
\end{figure}

The Small Acceptance Vertex Detector (SAVD) is presented in Fig.~\ref{SAVD_fig}. It consists of sixteen MIMOSA-26 silicon sensors located on two horizontally movable arms. The target holder was integrated with the SAVD base plate and the SAVD detector was commissioned in 2015. Pilot Pb+Pb data at 150$A$ GeV/c was recorded in 2016. The vertex fit resolution is on the level of 50~$\mathrm{\mu m}$.

\begin{figure}
\centering
\includegraphics[width=0.4\textwidth]{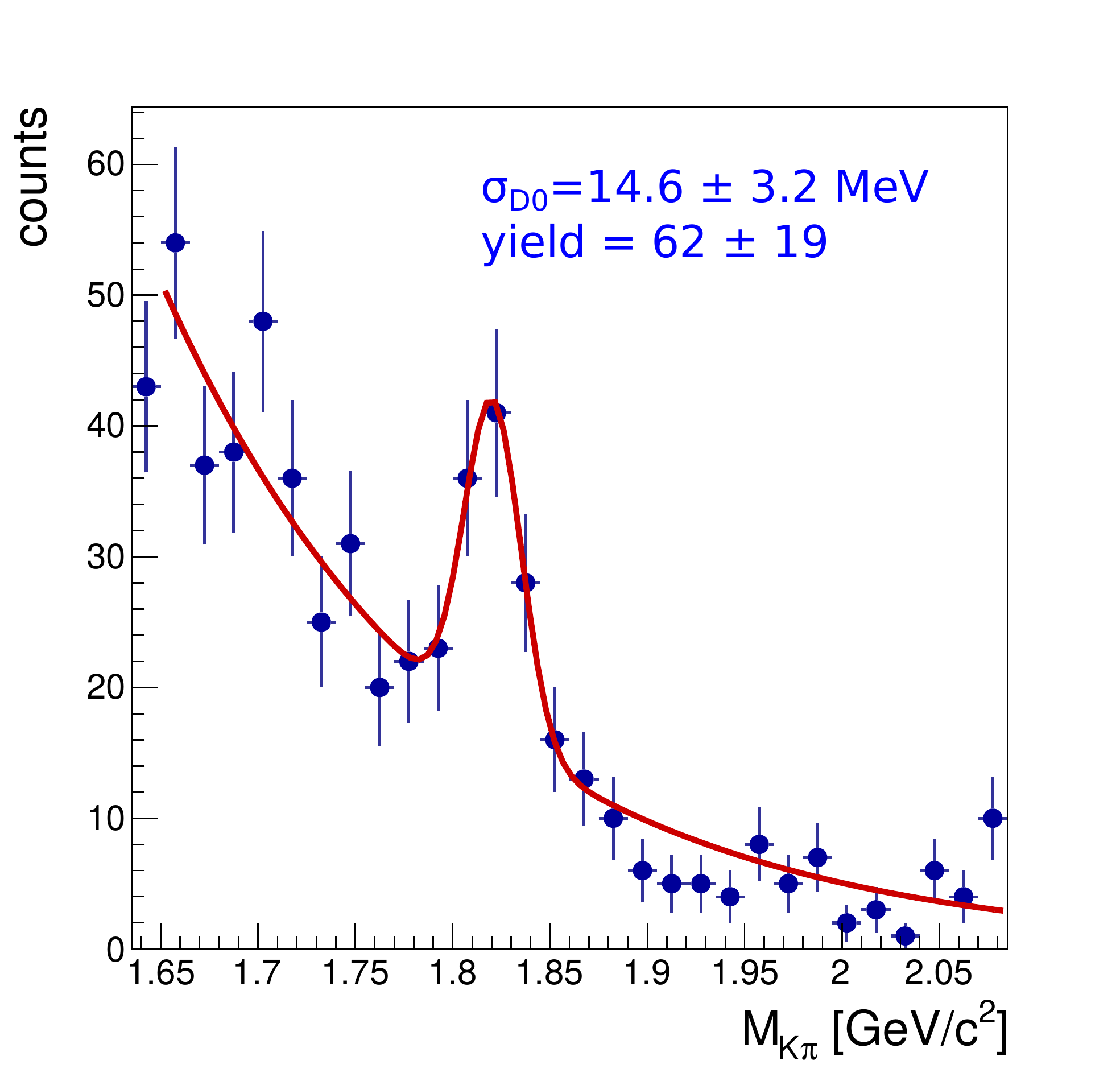}
\vspace{-0.2cm}
\caption[]{\footnotesize Performance result on the $D^0$ and $\overline{D^0}$ peak measured in central Pb+Pb collisions at 150$A$ GeV/c. The invariant mass distribution of reconstructed secondary tracks was obtained by assuming pion and kaon mass.}
\label{D0_signal}
\end{figure}

\begin{figure}
\centering
\includegraphics[width=0.7\textwidth]{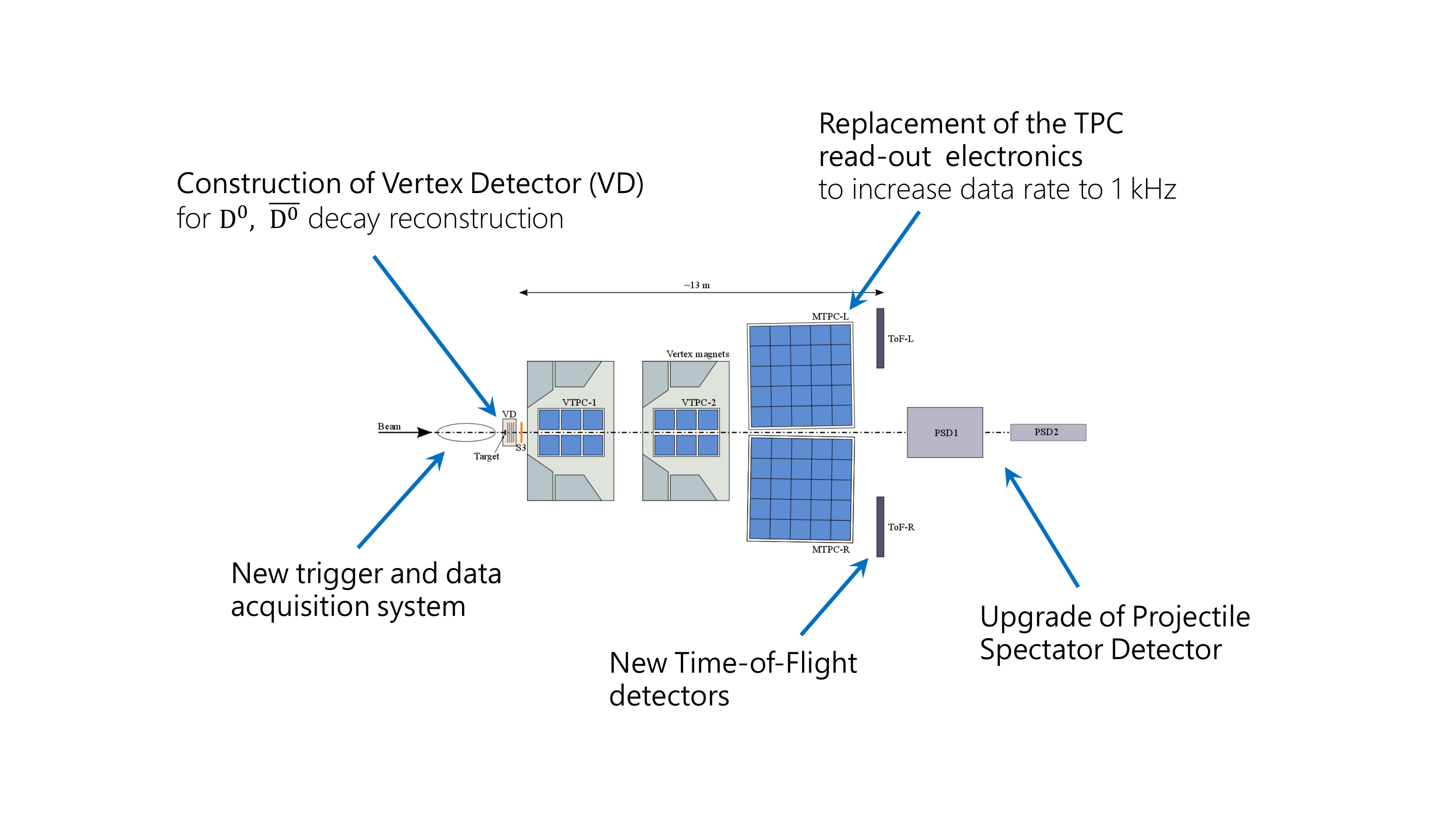}
\includegraphics[width=0.6\textwidth]{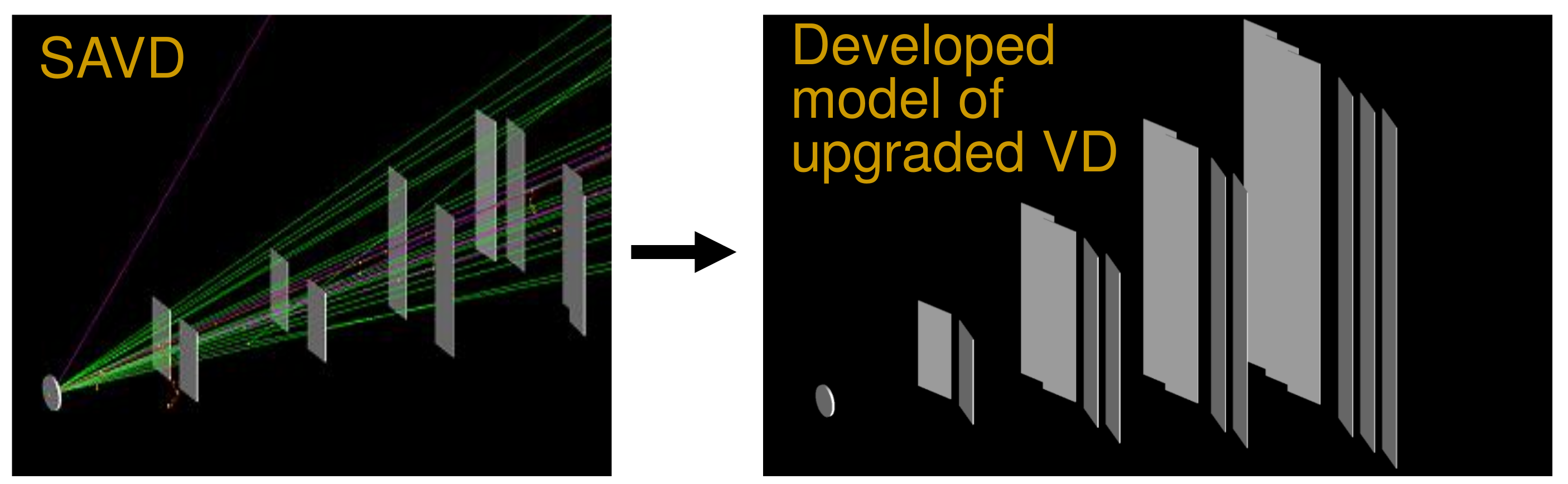}
\vspace{-0.1cm}
\caption[]{\footnotesize Upper diagram: upgrades of NA61/SHINE planned to be completed during LS2. Lower panel: 3D visualization of the SAVD and the upgraded Vertex Detector (VD).}
\label{na61upgrades_fig}
\end{figure}

The first indication of a $D^0$ and $\overline{D^0}$ peak, observed in central Pb+Pb collisions at 150$A$ GeV/c, is presented in Fig.~\ref{D0_signal}. The signal was obtained by matching Time Projection Chamber (TPC, see Fig.~\ref{na61upgrades_fig}, upper panel) tracks and SAVD tracks (interpolation procedures were used). The applied cuts included longitudinal distance of the pair vertex to the primary vertex. The particle identification of $V^0$ products was not used yet, but in future, it should reduce the background by a factor of~5.   

Motivated by the successful reconstruction of $D^0$ and $\overline{D^0}$ peak, the NA61/SHINE experiment plans to perform precise measurements of charm hadron production in 2021-2024. These will be possible with several detector upgrades (Fig.~\ref{na61upgrades_fig}) foreseen during Long Shutdown 2 (LS2) at CERN planned for 2019-2020. In principle, a new Vertex Detector (VD) will be constructed. The upgraded VD will be based on the same layout and mechanical support as the SAVD, but the 16 MIMOSA-26 sensors will be replaced by 46 ALPIDE sensors developed for the ALICE-Inner Tracking System (they produce much lower noise than MIMOSA-26 sensors). The total active surface, 32 cm$^\mathrm{2}$ of the SAVD, will be replaced by 180 cm$^\mathrm{2}$ for the VD, resulting in larger acceptance for each station. Finally, the time resolution of 115.2 $\mathrm{\mu s}$ (SAVD) will be improved to 10 $\mathrm{\mu s}$ (VD).    

\begin{figure}
\centering
\includegraphics[width=0.48\textwidth]{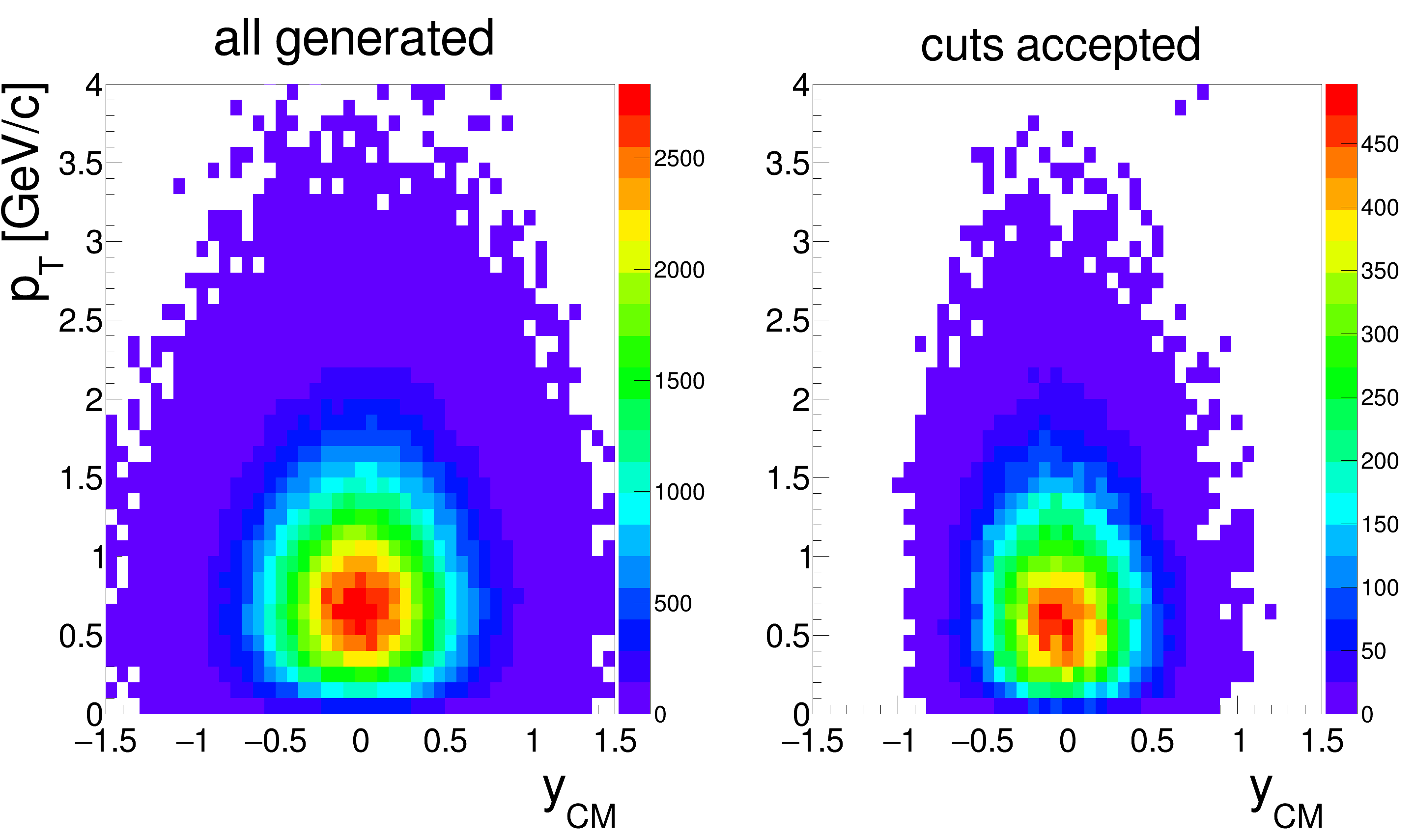}
\includegraphics[width=0.5\textwidth]{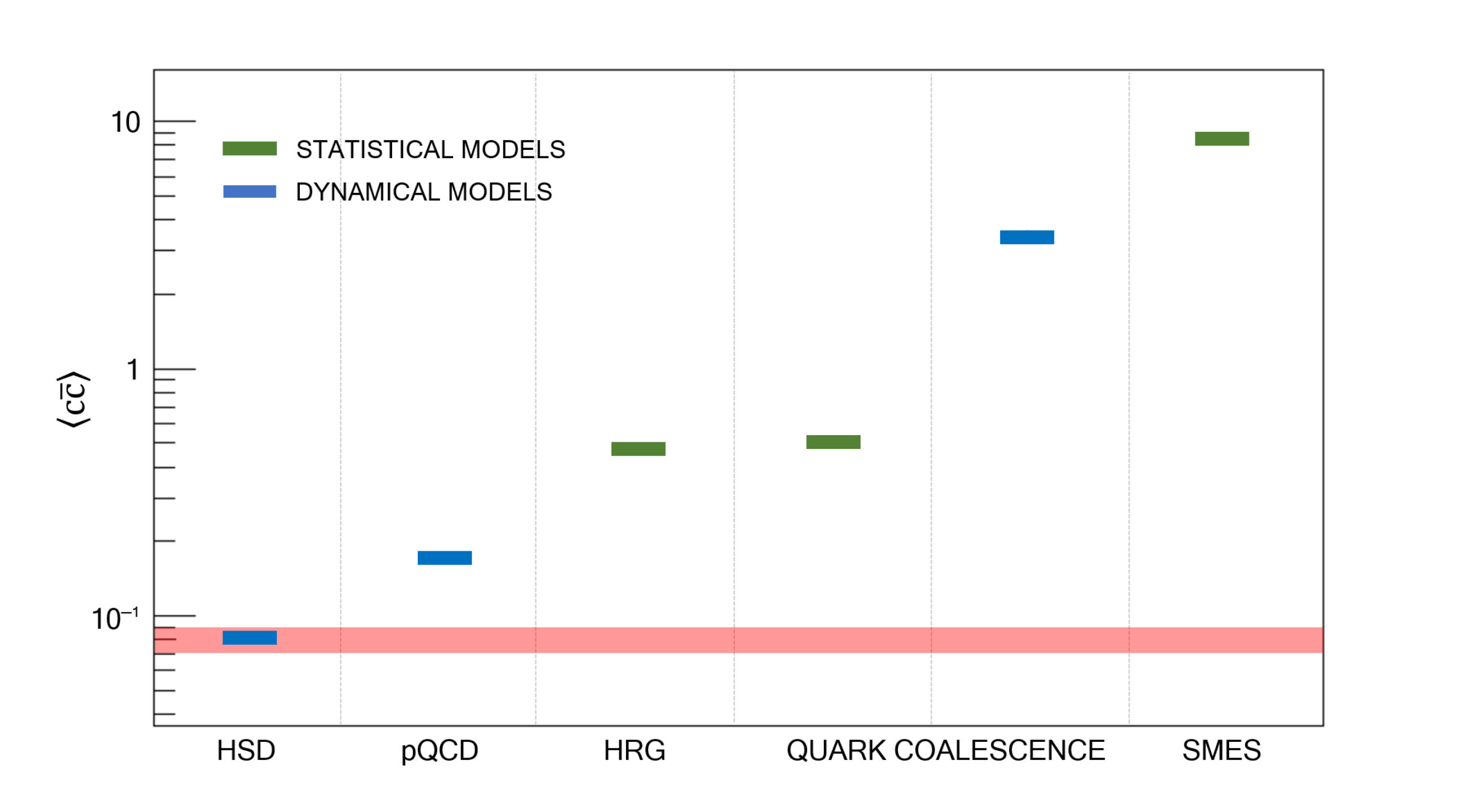}
\vspace{-0.2cm}
\caption[]{\footnotesize Left and middle: Monte Carlo (AMPT model~\cite{Lin:2004en}) simulation of transverse momentum versus center-of-mass rapidity distribution of $D^0$ and $\overline{D^0}$ mesons produced (left) in 500M Pb+Pb collisions at 150$A$ GeV/c and reconstructed (middle) by the VD. Right: same as Fig.~\ref{cc_models} but the width of the red band, at location assuming HSD prediction, shows expected accuracy of NA61/SHINE 2020+ result.}
\label{VD_acc_accuracy}
\end{figure}

Detection efficiency and expected accuracy for open charm measurements in the NA61/SHINE VD is presented in Fig.~\ref{VD_acc_accuracy}. The left and middle panels show that $D^0$ and $\overline{D^0}$ detection efficiency (including geometrical acceptance and analysis quality cuts) will be about 13\%
in the $D^0 \rightarrow \pi^{+}\, K^{-}$ channel (three times better than for the SAVD) and about 9\% in the $D^{+} \rightarrow \pi^{+}\, \pi^{+}\, K^{-}$ channel. The right panel of Fig.~\ref{VD_acc_accuracy} shows the expected accuracy of $\langle c \bar{c} \rangle$ determination in central Pb+Pb collisions at 150$A$ GeV/c. More details on the VD and open charm measurement in NA61/SHINE can be found in Ref.~\cite{Merzlaya_CPOD2018}.

\section{Can we measure high $p_T$ particles and multi-strange hadrons in NA61?}

Due to limited statistics the NA49 experiment measured transverse momentum spectra of identified hadrons only up to 4 GeV/c in Pb+Pb collisions and up to 2 GeV/c in p+p interactions~\cite{Alt:2007cd}. 
One of the goals of NA61/SHINE is to extend the measurements for p+p and obtain spectra in p+Pb collisions up to $\approx 4$ GeV/c, in order to determine the nuclear modification factors $R_{AA}$ and $R_{pA}$ at higher transverse momenta. These measurements will allow a more sensitive test for the presence of high $p_T$ particle suppression at SPS energy. 

\begin{figure}
\centering
\includegraphics[width=0.35\textwidth]{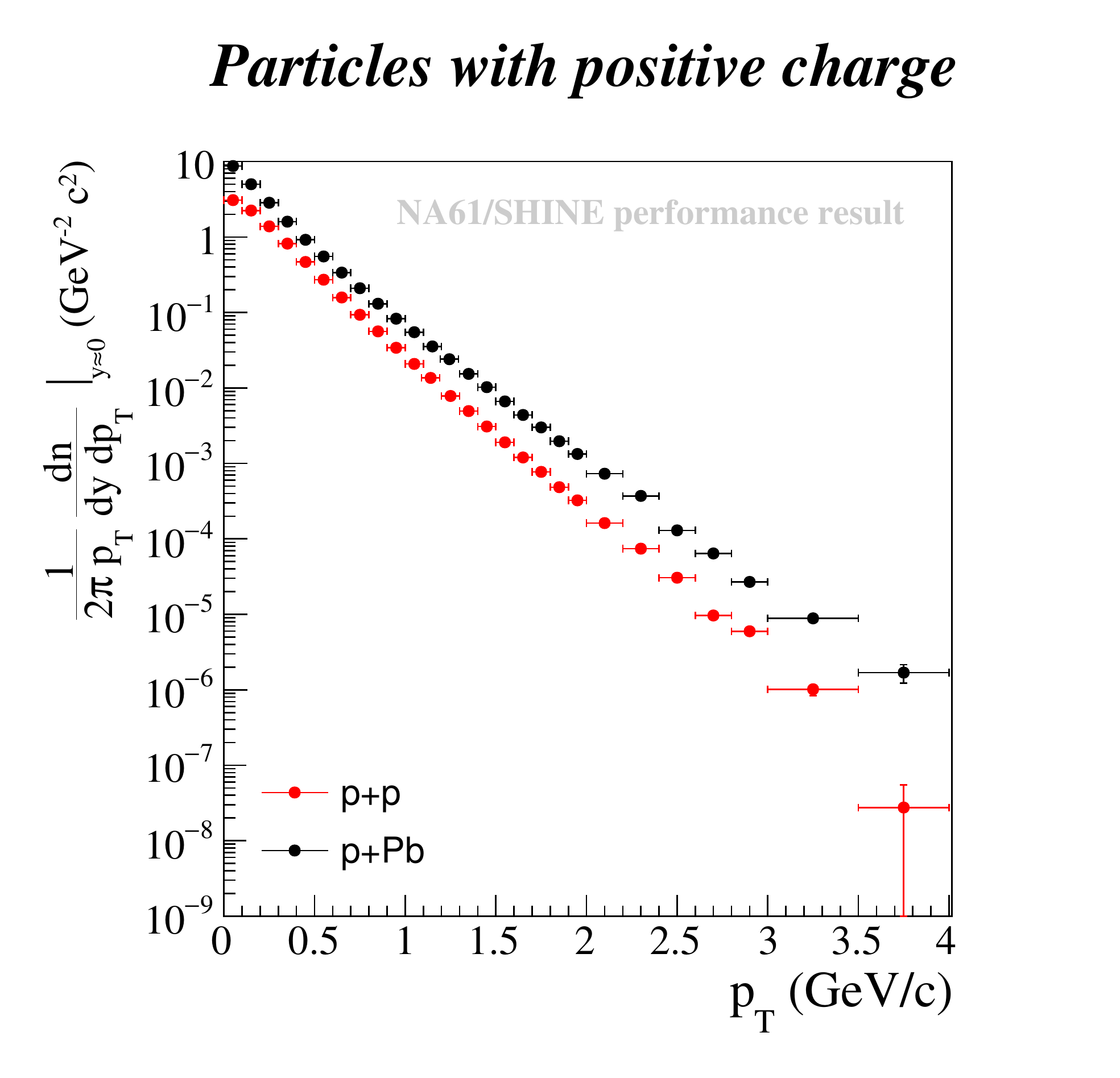}
\includegraphics[width=0.35\textwidth]{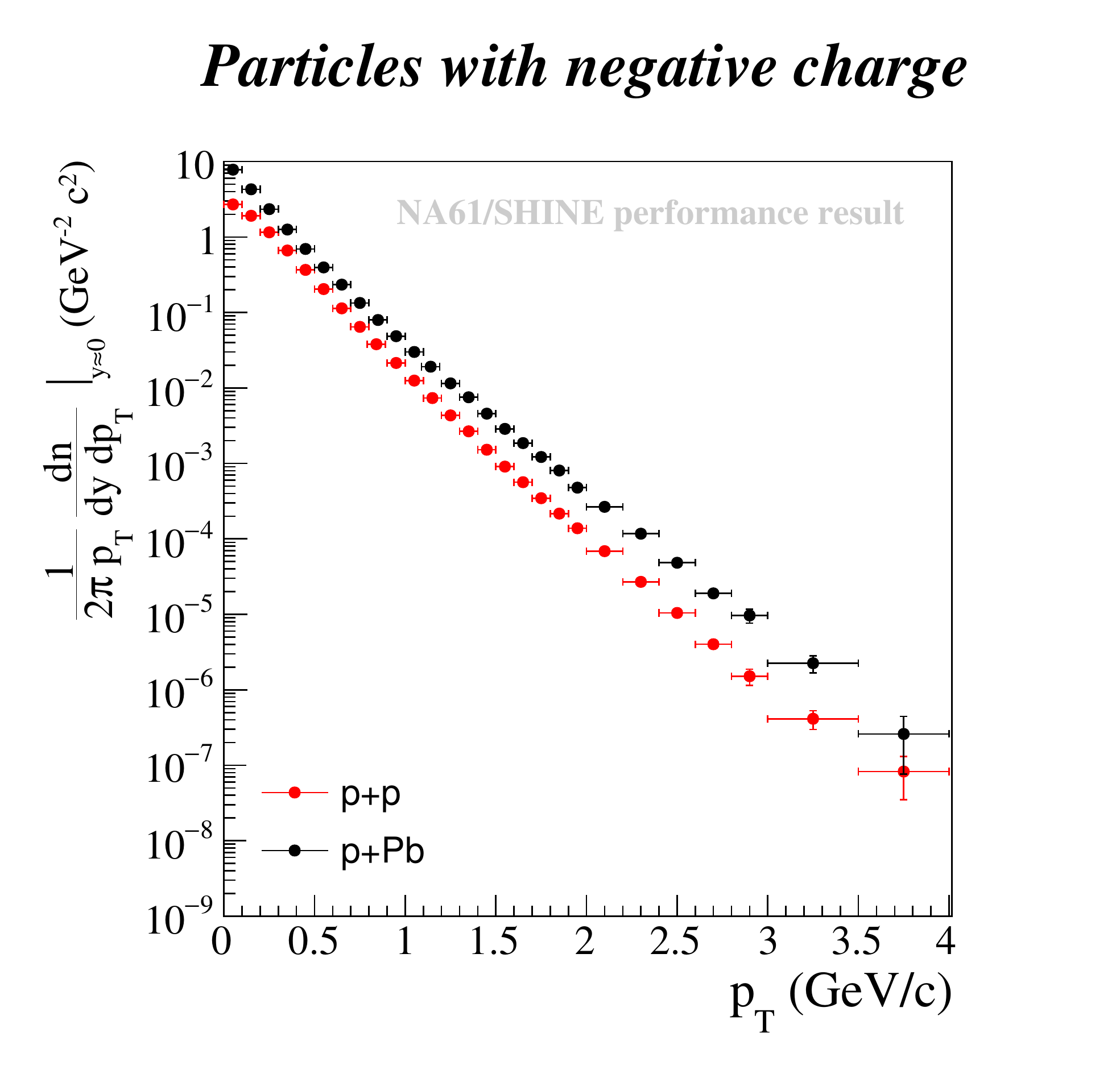}
\includegraphics[width=0.35\textwidth]{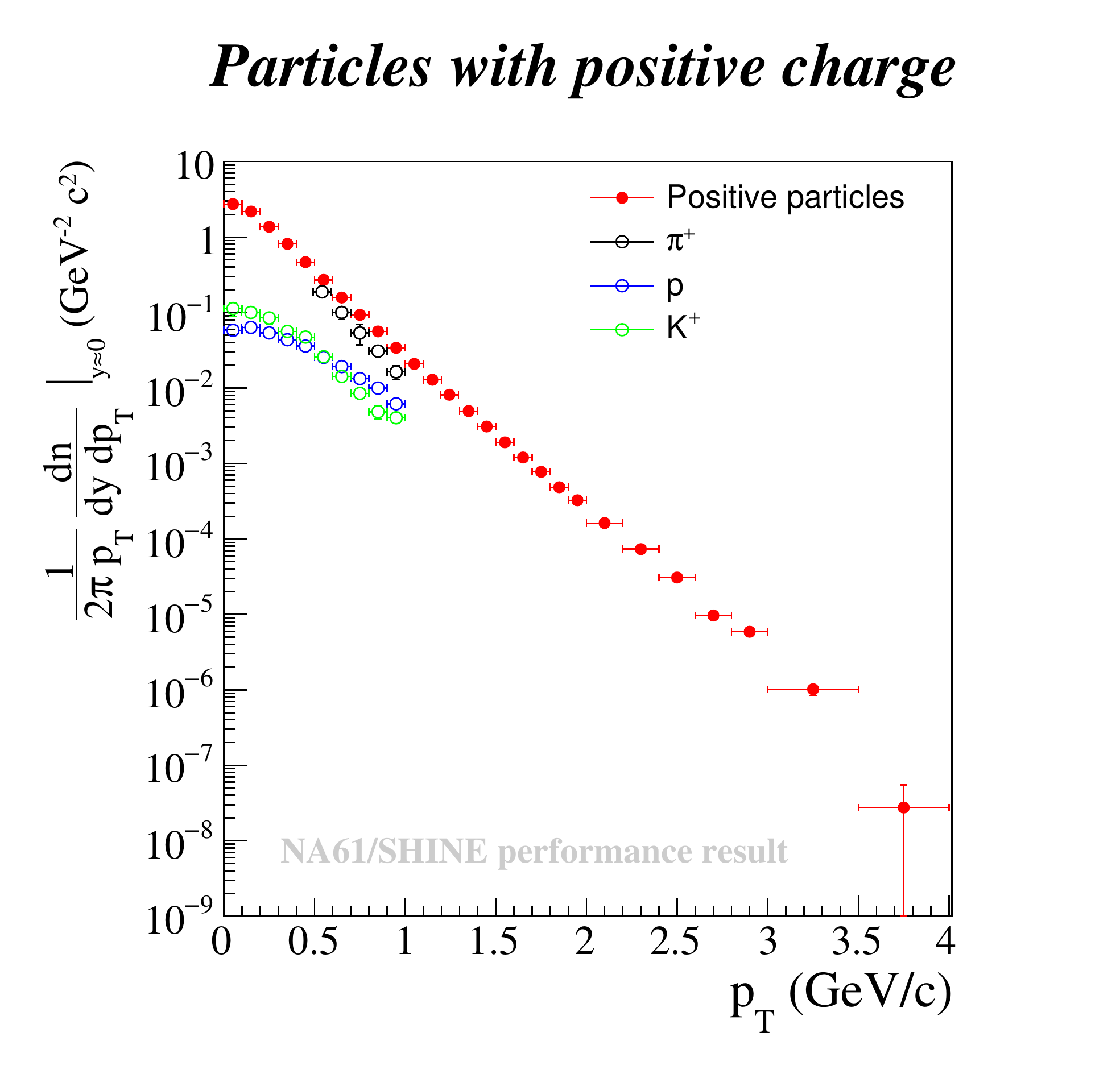}
\includegraphics[width=0.35\textwidth]{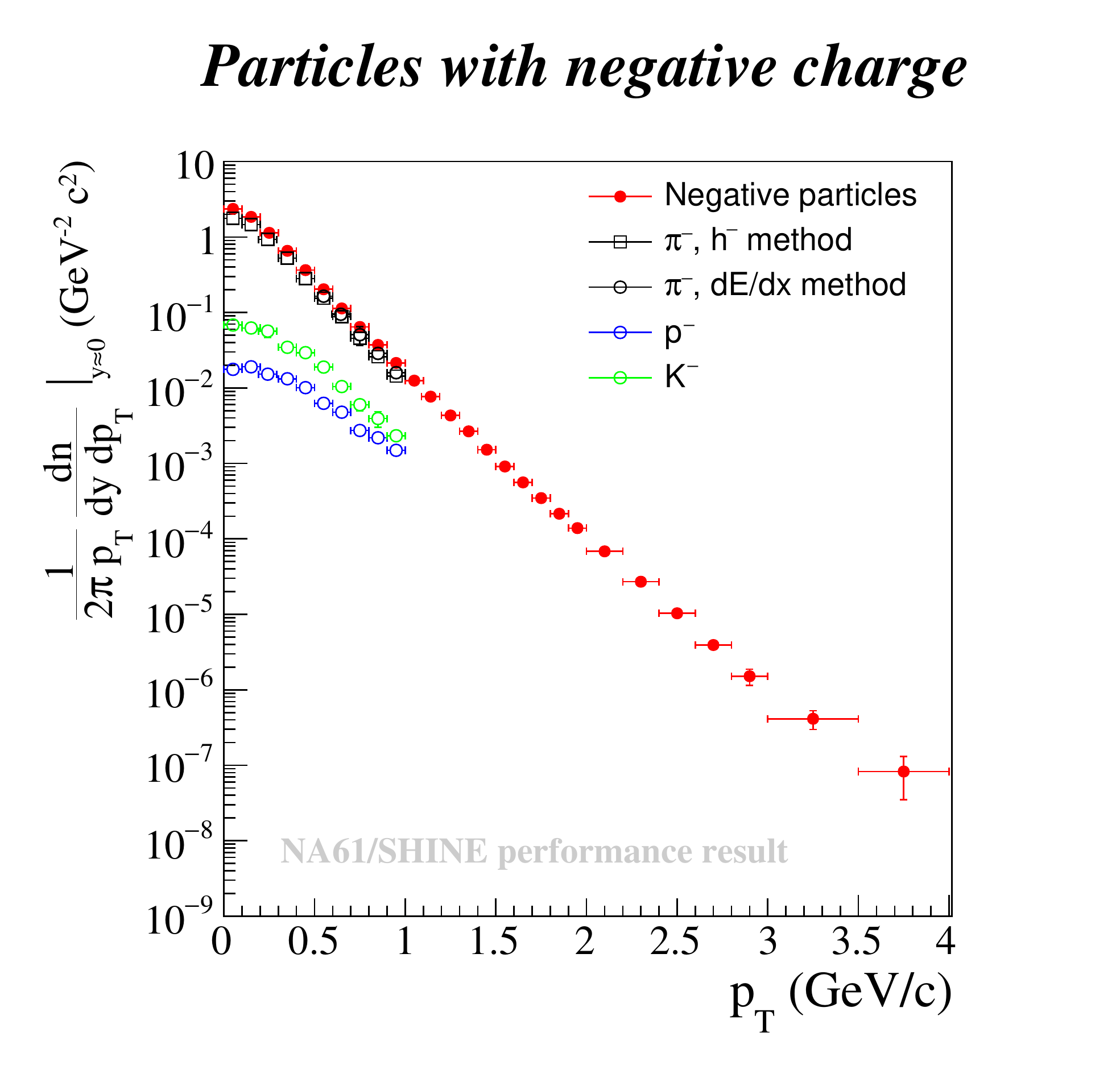}
\vspace{-0.3cm}
\caption[]{\footnotesize Upper panel: performance results on mid-rapidity spectra of positively and negatively charged particles in p+p and p+Pb collisions at 158 GeV/c. Lower panel: same as upper but compared to published~\cite{Aduszkiewicz:2017sei, Abgrall:2013qoa} NA61/SHINE results on identified particles.}
\label{highpt_krisztina}
\end{figure}

Figure~\ref{highpt_krisztina} shows the NA61/SHINE performance results on acceptance corrected spectra of positively and negatively charged particles at mid-rapidity in p+p and p+Pb interactions at 158 GeV/c beam momentum. A comparison of the new results with existing NA61/SHINE spectra~\cite{Aduszkiewicz:2017sei, Abgrall:2013qoa} of identified particles is shown in the lower panel of Fig.~\ref{highpt_krisztina}. The analysis of high $p_T$ identified particles is in progress.

\vspace{0.3cm}

\begin{SCfigure}
\includegraphics[width=0.5\textwidth]{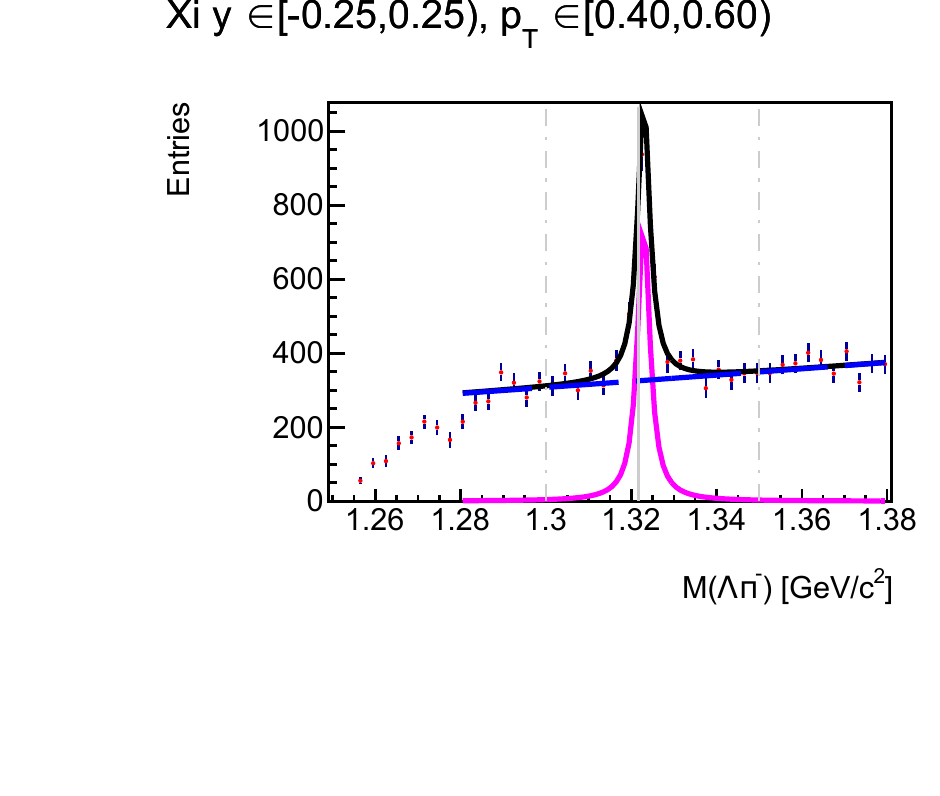}
\vspace{-2.1cm}
\caption{\footnotesize Example $\Lambda \pi^{-}$ invariant mass distribution in inelastic p+p interactions at 158 GeV/c, calculated assuming the $\Xi^{-}$ decay hypothesis.}
\label{xim_minv}
\end{SCfigure}

Similarly to the analysis of $K^{*}(892)^0$ resonance and high $p_T$ particle spectra, a very large statistics (approx. 60 milion of recorded events) of p+p collisions at 158 GeV/c were used to obtain the first NA61/SHINE results on $\Xi^{-}$ hyperon production. The analysis was performed in $\Lambda \pi^{-}$ decay channel. $\Lambda$ candidates were searched via their decays into $p\, \pi^{-}$ pairs. An example invariant mass distribution of $\Xi^{-}$ hyperon candidates, obtained by combining $\Lambda$ candidates with $\pi^{-}$ mesons, is presented in Fig.~\ref{xim_minv}. The black curve shows the fitted function consisting of a Lorentzian for the signal and a polynomial (up to 4th order) for the background. Additionally, the signal component of the fit is shown in magenta and the background as a blue line.       


\begin{figure}
\begin{tikzpicture}
	\begin{scope} [xshift=-2cm]
	\node {\includegraphics[width=0.4\textwidth]{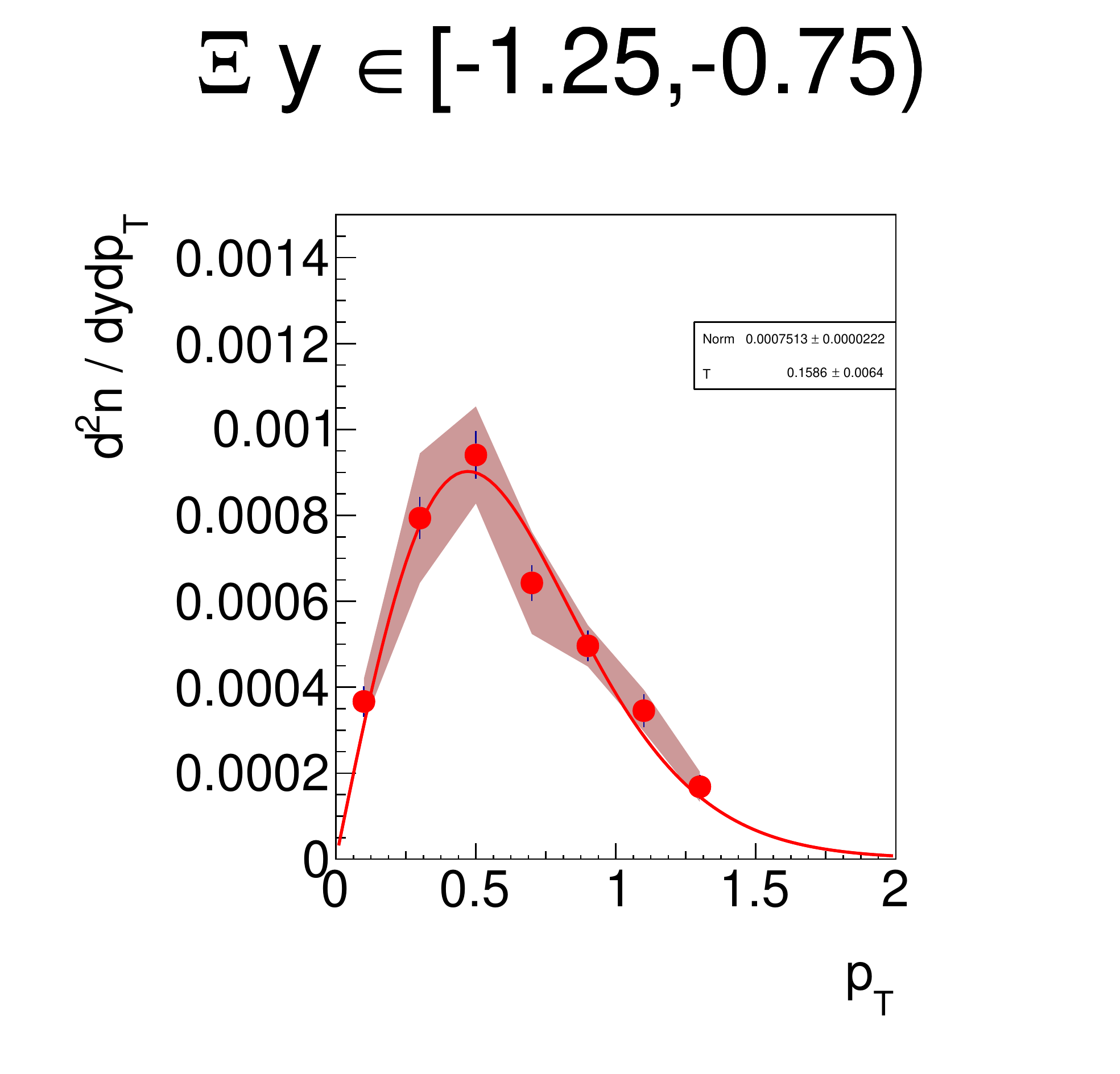}};
	\end{scope}
	\begin{scope} [xshift=3cm]
	\node {\includegraphics[width=0.4\textwidth]{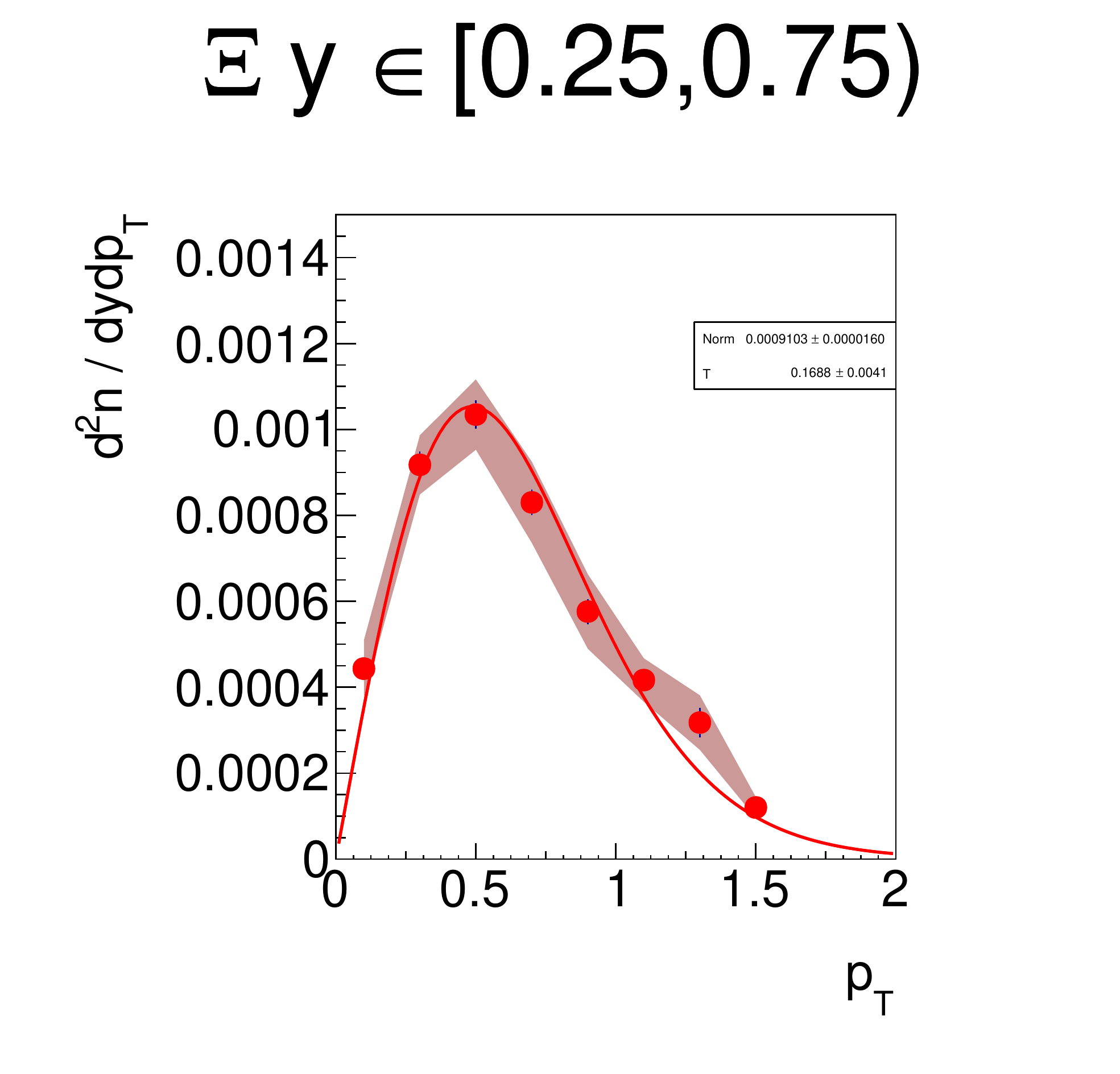}};
	\end{scope}
	\begin{scope} [xshift=8cm]
	\node {\includegraphics[width=0.4\textwidth]{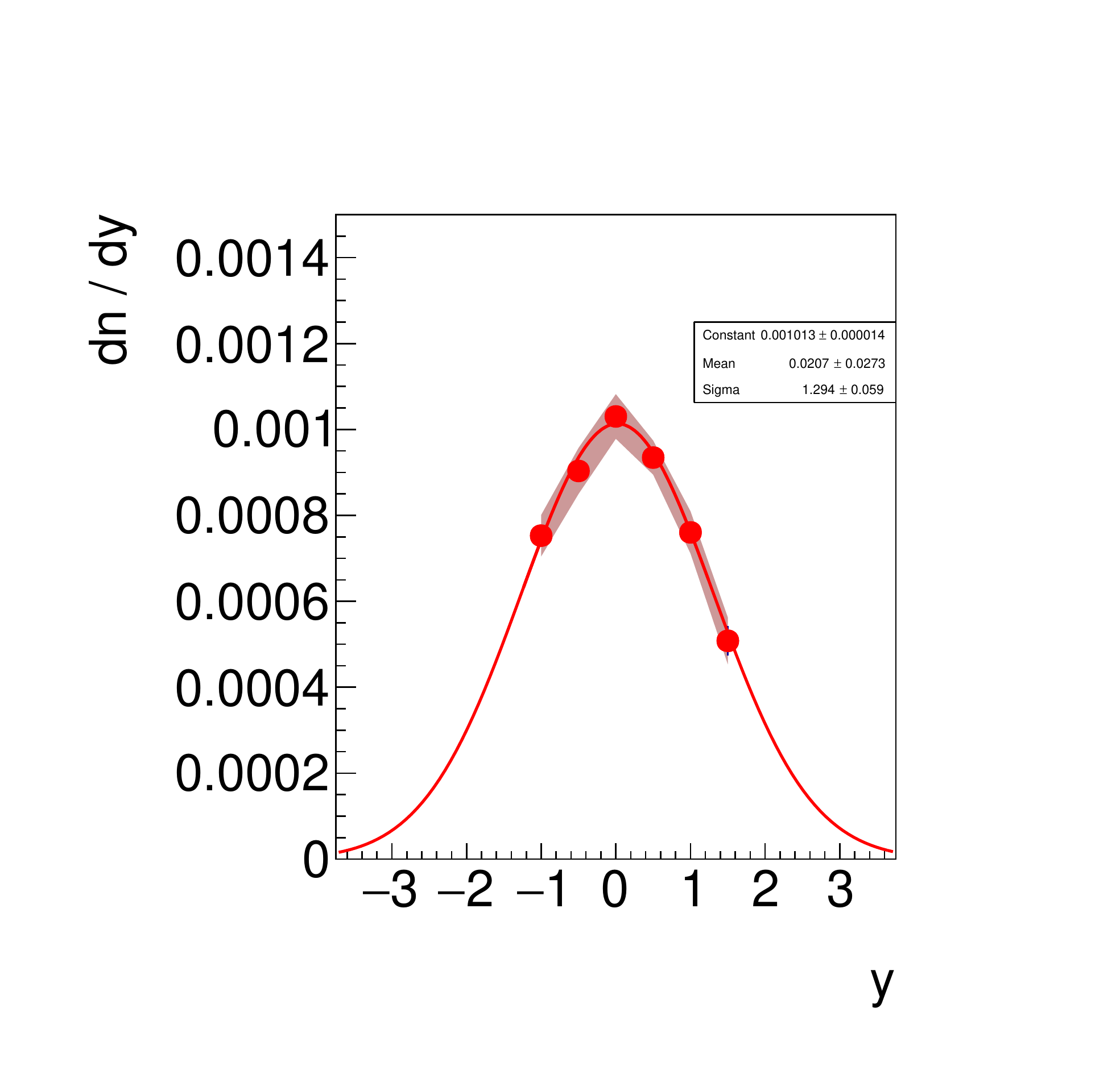}};
	\end{scope}
\end{tikzpicture}
\vspace{-1.2cm}
\caption[]{\footnotesize Left and middle: preliminary results on transverse momentum spectra (in GeV/c) of $\Xi^{-}$ hyperons produced in inelastic p+p interactions at 158 GeV/c. Two selected rapidity bins are shown. Right: $p_T$-extrapolated and $p_T$-integrated rapidity spectrum of $\Xi^{-}$ hyperons. In both panels statistical uncertainties are shown as vertical bars and systematic ones as shaded regions.}
\label{xim_pT_dndy}
\end{figure}

Preliminary results, derived from two dimensional ($y, p_T$) spectra, are presented as transverse momentum distributions in two example bins of rapidity in the left and middle panel of Fig.~\ref{xim_pT_dndy}. Statistical uncertainties are shown as vertical bars and preliminary estimates of systematic uncertainties are indicated by shaded bands.
The right panel of Fig.~\ref{xim_pT_dndy} presents the $p_T$-extrapolated and $p_T$-integrated rapidity spectrum of $\Xi^{-}$ hyperons, from which the $4\pi$ multiplicity can be obtained. The preliminary value of the mean multiplicity of $\Xi^{-}$ hyperons was found to be $0.0033 \pm 0.0001 \pm 0.0006$, where the first uncertainty is statistical and the second systematic. 


\vspace{0.15cm}

The analysis of $\overline{\Xi}^{+}$ production is ongoing and will be released in the near future. Moreover, we also plan to combine $\Xi$ hyperons with charged pions. This will allow to repeat and check the previous pentaquark search of NA49~\cite{Alt:2003vb}.

\section{Summary}

Several new NA61/SHINE results were presented during the CPOD 2018 conference. Among them charged pion, kaon and proton spectra in central Be+Be collisions. The energy dependencies of $K^{+}/\pi^{+}$ ratio and inverse slope parameter of $m_T$ spectra are similar to those in p+p collisions. 
The rapidity distributions of protons confirm the similarity between p+p and Be+Be collisions, suggesting that the \textit{onset of fireball} is occuring for systems larger than Be+Be.

The first NA61/SHINE results on anisotropic flow in Pb+Pb collisions at 30$A$ GeV/c were obtained. Proton $dv_{1}/dy$ changes sign for centrality of about 50\%. The centrality dependence of protons $dv_{1}/dy$ serves as an another tool to study the properties of the onset of deconfinement. 

We also studied the effects of electromagnetic interactions between produced pions and spectators. Such an effect is clearly seen for intermediate centrality Ar+Sc collisions at 150$A$ GeV/c and qualitatively comparable with the effect observed by NA49 for peripheral Pb+Pb collisions at 158$A$ GeV/c.   

For this conference the first, high precision results on $K^{*}(892)^0$ production in inelastic p+p collisions at 158 GeV/c were obtained. Combined with the data for heavier systems they can be used to estimate time interval between chemical and kinetic freeze-out.

The most intriguing result shown by NA61/SHINE at this conference was an indication of an intermittency effect in mid-central Ar+Sc collisions at 150$A$ GeV/c. This is the first possible evidence of a critical point signal in NA61/SHINE. 

We also presented the $\Delta \eta$ dependence of fluctuations. They are studied in order to extend the phase diagram scan. A non-trivial $\Delta[P_T, N]$ dependence on the width of the $\Delta \eta$ window is seen for p+p and Be+Be interactions at 158/150$A$ GeV/c.

Moreover, we presented the prospects of open charm and high $p_T$ particle measurements in NA61/SHINE. Finally, the analysis of multi-strange hadrons in p+p collisions is well advanced. The first results on $\Xi^{-}$ hyperon spectra are already released.

\vspace{1cm}

\footnotesize{
\textbf{Acknowledgements:} 
This work was partially supported by the National Science Centre, Poland (grants no. 2015/18/M/ST2/00125, 2017/25/N/ST2/02575, 2014/14/E/ST2/00018), the Ministry of Science and Higher Education, Poland (DIR/WK/2016/2017/10-1), the Ministry of Science and Higher Education of the Russian Federation (grant No. 3.3380.2017/4.6), and the National Research Nuclear University MEPhI in the framework of the Russian Academic Excellence Project (contract No. 02.a03.21.0005, 27.08.2013).
}



\begin{thebibliography}{99}

\bibitem{Gazdzicki:1998vd} 
  M.~Gazdzicki and M.~I.~Gorenstein,
  Acta Phys.\ Polon.\ B {\bf 30}, 2705 (1999)
  [hep-ph/9803462].


\bibitem{Gazdzicki:2014sva} 
  M.~Gazdzicki, M.~I.~Gorenstein and P.~Seyboth,
  Int.\ J.\ Mod.\ Phys.\ E {\bf 23}, 1430008 (2014)
  doi:10.1142/S0218301314300082
  [arXiv:1404.3567 [nucl-ex]].


\bibitem{Aduszkiewicz:2017sei} 
  A.~Aduszkiewicz {\it et al.} [NA61/SHINE Collaboration],
  Eur.\ Phys.\ J.\ C {\bf 77}, no. 10, 671 (2017)
  doi:10.1140/epjc/s10052-017-5260-4
  [arXiv:1705.02467 [nucl-ex]].


\bibitem{Pulawski:2015tka} 
  S.~Pulawski [NA61 Collaboration],
  PoS CPOD {\bf 2014}, 010 (2015)
  doi:10.22323/1.217.0010
  [arXiv:1502.07916 [nucl-ex]].


\bibitem{Alt:2007aa} 
  C.~Alt {\it et al.} [NA49 Collaboration],
  Phys.\ Rev.\ C {\bf 77}, 024903 (2008)
  doi:10.1103/PhysRevC.77.024903
  [arXiv:0710.0118 [nucl-ex]].


\bibitem{Anticic:2012ay} 
  T.~Anticic {\it et al.} [NA49 Collaboration],
  Phys.\ Rev.\ C {\bf 86}, 054903 (2012)
  doi:10.1103/PhysRevC.86.054903
  [arXiv:1207.0348 [nucl-ex]].


\bibitem{Motornenko:2018gdc} 
  A.~Motornenko, V.~V.~Begun, V.~Vovchenko, M.~I.~Gorenstein and H.~Stoecker,
  Phys.\ Rev.\ C {\bf 99}, no. 3, 034909 (2019)
  doi:10.1103/PhysRevC.99.034909
  [arXiv:1811.10645 [nucl-th]].


\bibitem{Lewicki_CPOD2018} M.~Lewicki, these proceedings.


\bibitem{Aduszkiewicz:2017mei} 
  A.~Aduszkiewicz [NA61/SHINE Collaboration],
  Nucl.\ Phys.\ A {\bf 967}, 35 (2017)
  doi:10.1016/j.nuclphysa.2017.04.046
  [arXiv:1704.08071 [hep-ex]].


\bibitem{Blume:2008zza} 
  C.~Blume,
  J.\ Phys.\ G {\bf 35}, 044004 (2008).
  doi:10.1088/0954-3899/35/4/044004



\bibitem{Kaptur:2015xzu} 
  E.~Kaptur [NA61/SHINE Collaboration],
  PoS CPOD {\bf 2014}, 053 (2015).
  doi:10.22323/1.217.0053


\bibitem{Csernai:1999nf} 
  L.~P.~Csernai and D.~Rohrich,
  Phys.\ Lett.\ B {\bf 458}, 454 (1999)
  doi:10.1016/S0370-2693(99)00615-2
  [nucl-th/9908034].


\bibitem{Stoecker:2004qu} 
  H.~Stoecker,
  Nucl.\ Phys.\ A {\bf 750}, 121 (2005)
  doi:10.1016/j.nuclphysa.2004.12.074
  [nucl-th/0406018].


\bibitem{Brachmann:1999xt} 
  J.~Brachmann, S.~Soff, A.~Dumitru, H.~Stoecker, J.~A.~Maruhn, W.~Greiner, L.~V.~Bravina and D.~H.~Rischke,
  Phys.\ Rev.\ C {\bf 61}, 024909 (2000)
  doi:10.1103/PhysRevC.61.024909
  [nucl-th/9908010].


\bibitem{Alt:2003ab} 
  C.~Alt {\it et al.} [NA49 Collaboration],
  Phys.\ Rev.\ C {\bf 68}, 034903 (2003)
  doi:10.1103/PhysRevC.68.034903
  [nucl-ex/0303001].


\bibitem{Adamczyk:2014ipa} 
  L.~Adamczyk {\it et al.} [STAR Collaboration],
  Phys.\ Rev.\ Lett.\  {\bf 112}, no. 16, 162301 (2014)
  doi:10.1103/PhysRevLett.112.162301(2014), 10.1103/PhysRevLett.112.162301
  [arXiv:1401.3043 [nucl-ex]].


\bibitem{Adamczyk:2017nxg} 
  L.~Adamczyk {\it et al.} [STAR Collaboration],
  Phys.\ Rev.\ Lett.\  {\bf 120}, no. 6, 062301 (2018)
  doi:10.1103/PhysRevLett.120.062301
  [arXiv:1708.07132 [hep-ex]].


\bibitem{Meehan:2017cum} 
  K.~Meehan [STAR Collaboration],
  Nucl.\ Phys.\ A {\bf 967}, 808 (2017)
  doi:10.1016/j.nuclphysa.2017.06.007
  [arXiv:1704.06342 [nucl-ex]].


\bibitem{Shanmuganathan:2015qxb} 
  P.~Shanmuganathan [STAR Collaboration],
  Nucl.\ Phys.\ A {\bf 956}, 260 (2016)
  doi:10.1016/j.nuclphysa.2016.04.006
  [arXiv:1512.09009 [nucl-ex]].


\bibitem{Klochkov:2018xvw} 
  V.~Klochkov {\it et al.} [NA61/SHINE Collaboration],
  Nucl.\ Phys.\ A {\bf 982}, 439 (2019)
  doi:10.1016/j.nuclphysa.2018.10.058
  [arXiv:1810.07579 [nucl-ex]].


\bibitem{Selyuzhenkov_CPOD2018} I.~Selyuzhenkov, these proceedings.


\bibitem{Rybicki:2006qm} 
  A.~Rybicki and A.~Szczurek,
  Phys.\ Rev.\ C {\bf 75}, 054903 (2007)
  doi:10.1103/PhysRevC.75.054903
  [nucl-th/0610036].


\bibitem{Rybicki:2013qla} 
  A.~Rybicki and A.~Szczurek,
  Phys.\ Rev.\ C {\bf 87}, no. 5, 054909 (2013)
  doi:10.1103/PhysRevC.87.054909
  [arXiv:1303.7354 [nucl-th]].


\bibitem{Rybicki:2009zz} 
  A.~Rybicki [NA49 Collaboration],
  PoS EPS {\bf -HEP2009}, 031 (2009).
  doi:10.22323/1.084.0031

\bibitem{Davis_CPOD2018} N.~Davis, these proceedings (on electromegnetic effects).


\bibitem{Markert:2002rw} 
  C.~Markert, G.~Torrieri and J.~Rafelski,
  AIP Conf.\ Proc.\  {\bf 631}, no. 1, 533 (2002)
  doi:10.1063/1.1513698
  [hep-ph/0206260].


\bibitem{Adams:2004ep} 
  J.~Adams {\it et al.} [STAR Collaboration],
  Phys.\ Rev.\ C {\bf 71}, 064902 (2005)
  doi:10.1103/PhysRevC.71.064902
  [nucl-ex/0412019].


\bibitem{Tanabashi:2018oca} 
  M.~Tanabashi {\it et al.} [Particle Data Group],
  Phys.\ Rev.\ D {\bf 98}, no. 3, 030001 (2018).
  doi:10.1103/PhysRevD.98.030001



\bibitem{Tefelska_CPOD2018} A.~Tefelska, these proceedings.


\bibitem{Anticic:2011zr} 
  T.~Anticic {\it et al.} [NA49 Collaboration],
  Phys.\ Rev.\ C {\bf 84}, 064909 (2011)
  doi:10.1103/PhysRevC.84.064909
  [arXiv:1105.3109 [nucl-ex]].



\bibitem{Anticic:2010yg} 
  T.~Anticic {\it et al.} [NA49 Collaboration],
  Eur.\ Phys.\ J.\ C {\bf 68}, 1 (2010)
  doi:10.1140/epjc/s10052-010-1328-0
  [arXiv:1004.1889 [hep-ex]].


\bibitem{Alt:2004wc} 
  C.~Alt {\it et al.} [NA49 Collaboration],
  Phys.\ Rev.\ Lett.\  {\bf 94}, 052301 (2005)
  doi:10.1103/PhysRevLett.94.052301
  [nucl-ex/0406031].



\bibitem{Afanasiev:2002mx} 
  S.~V.~Afanasiev {\it et al.} [NA49 Collaboration],
  Phys.\ Rev.\ C {\bf 66}, 054902 (2002)
  doi:10.1103/PhysRevC.66.054902
  [nucl-ex/0205002].



\bibitem{Becattini:2005xt} 
  F.~Becattini, J.~Manninen and M.~Gazdzicki,
  Phys.\ Rev.\ C {\bf 73}, 044905 (2006)
  doi:10.1103/PhysRevC.73.044905
  [hep-ph/0511092].


\bibitem{Begun:2018qkw} 
  V.~V.~Begun, V.~Vovchenko, M.~I.~Gorenstein and H.~Stoecker,
  Phys.\ Rev.\ C {\bf 98}, no. 5, 054909 (2018)
  doi:10.1103/PhysRevC.98.054909
  [arXiv:1805.01901 [nucl-th]].


\bibitem{Bialas:1985jb} 
  A.~Bialas and R.~B.~Peschanski,
  Nucl.\ Phys.\ B {\bf 273}, 703 (1986).
  doi:10.1016/0550-3213(86)90386-X


\bibitem{Turko:1989dc} 
  L.~Turko,
  Phys.\ Lett.\ B {\bf 227}, 149 (1989).
  doi:10.1016/0370-2693(89)91298-7


\bibitem{Diakonos:2006zz} 
  F.~K.~Diakonos, N.~G.~Antoniou and G.~Mavromanolakis,
  PoS CPOD {\bf 2006}, 010 (2006).
  doi:10.22323/1.029.0010


\bibitem{Antoniou:2006zb} 
  N.~G.~Antoniou, F.~K.~Diakonos, A.~S.~Kapoyannis and K.~S.~Kousouris,
  Phys.\ Rev.\ Lett.\  {\bf 97}, 032002 (2006)
  doi:10.1103/PhysRevLett.97.032002
  [hep-ph/0602051].


\bibitem{Davis_CPOD2018_inter} N.~Davis, these proceedings (on intermittency).


\bibitem{Anticic:2012xb} 
  T.~Anticic {\it et al.} [NA49 Collaboration],
  Eur.\ Phys.\ J.\ C {\bf 75}, no. 12, 587 (2015)
  doi:10.1140/epjc/s10052-015-3738-5
  [arXiv:1208.5292 [nucl-ex]].


\bibitem{Anticic:2009pe} 
  T.~Anticic {\it et al.} [NA49 Collaboration],
  Phys.\ Rev.\ C {\bf 81}, 064907 (2010)
  doi:10.1103/PhysRevC.81.064907
  [arXiv:0912.4198 [nucl-ex]].


\bibitem{Grebieszkow:2009jr} 
  K.~Grebieszkow [NA49 Collaboration],
  Nucl.\ Phys.\ A {\bf 830}, 547C (2009)
  doi:10.1016/j.nuclphysa.2009.09.044
  [arXiv:0907.4101 [nucl-ex]].


\bibitem{Anticic:2015fla} 
  T.~Anticic {\it et al.} [NA49 Collaboration],
  Phys.\ Rev.\ C {\bf 92}, no. 4, 044905 (2015)
  doi:10.1103/PhysRevC.92.044905
  [arXiv:1509.04633 [nucl-ex]].


\bibitem{Aduszkiewicz:2015jna} 
  A.~Aduszkiewicz {\it et al.} [NA61/SHINE Collaboration],
  Eur.\ Phys.\ J.\ C {\bf 76}, no. 11, 635 (2016)
  doi:10.1140/epjc/s10052-016-4450-9
  [arXiv:1510.00163 [hep-ex]].


\bibitem{Czopowicz:2015mfa} 
  T.~Czopowicz [NA61/SHINE Collaboration],
  PoS CPOD {\bf 2014}, 054 (2015)
  doi:10.22323/1.217.0054
  [arXiv:1503.01619 [nucl-ex]].



\bibitem{Gorenstein:2013nea} 
  M.~I.~Gorenstein and K.~Grebieszkow,
  Phys.\ Rev.\ C {\bf 89}, no. 3, 034903 (2014)
  doi:10.1103/PhysRevC.89.034903
  [arXiv:1309.7878 [nucl-th]].



\bibitem{evgeny_cpod16_slides} E.~Andronov, slides from Critical Point and Onset of Deconfinement 2016, 
https://indico.cern.ch/event/449173/contributions/2167165/attachments/1280888/1902801/EA\_cpod.pdf


\bibitem{Grebieszkow:2017gqx} 
  K.~Grebieszkow [NA61/SHINE Collaboration],
  PoS EPS {\bf -HEP2017}, 167 (2017)
  doi:10.22323/1.314.0167
  [arXiv:1709.10397 [nucl-ex]].


\bibitem{EPOS_CRMC_www} T.~Pierog, R.~Ulrich, EPOS 1.99 in CRMC (Cosmic Ray Monte Carlo package),
https://web.ikp.kit.edu/rulrich/crmc.html



\bibitem{Prokhorova:2018tcl} 
  D.~Prokhorova [NA61/SHINE Collaboration],
  KnE Energ.\ Phys.\  {\bf 3}, 217 (2018)
  doi:10.18502/ken.v3i1.1747
  [arXiv:1801.06690 [nucl-ex]].


\bibitem{Andronov:2018ukf} 
  E.~Andronov [NA61/SHINE Collaboration],
  EPJ Web Conf.\  {\bf 191}, 05002 (2018).
  doi:10.1051/epjconf/201819105002


\bibitem{Satz:2006kba} 
  H.~Satz,
  Nucl.\ Phys.\ A {\bf 783}, 249 (2007)
  doi:10.1016/j.nuclphysa.2006.11.026
  [hep-ph/0609197].


\bibitem{Kostyuk:2001zd} 
  A.~P.~Kostyuk, M.~I.~Gorenstein, H.~Stoecker and W.~Greiner,
  Phys.\ Lett.\ B {\bf 531}, 195 (2002)
  doi:10.1016/S0370-2693(02)01488-0
  [hep-ph/0110269].


\bibitem{Linnyk:2008hp} 
  O.~Linnyk, E.~L.~Bratkovskaya and W.~Cassing,
  Int.\ J.\ Mod.\ Phys.\ E {\bf 17}, 1367 (2008)
  doi:10.1142/S0218301308010507
  [arXiv:0808.1504 [nucl-th]].


\bibitem{Song_priv} T.~Song, private communication.


\bibitem{Gavai:1994gb} 
  R.~V.~Gavai, S.~Gupta, P.~L.~McGaughey, E.~Quack, P.~V.~Ruuskanen, R.~Vogt and X.~N.~Wang,
  Int.\ J.\ Mod.\ Phys.\ A {\bf 10}, 2999 (1995)
  doi:10.1142/S0217751X95001431
  [hep-ph/9411438].


\bibitem{BraunMunzinger:2000px} 
  P.~Braun-Munzinger and J.~Stachel,
  Phys.\ Lett.\ B {\bf 490}, 196 (2000)
  doi:10.1016/S0370-2693(00)00991-6
  [nucl-th/0007059].


\bibitem{Levai:2000ne} 
  P.~Levai, T.~S.~Biro, P.~Csizmadia, T.~Csorgo and J.~Zimanyi,
  J.\ Phys.\ G {\bf 27}, 703 (2001)
  doi:10.1088/0954-3899/27/3/357
  [nucl-th/0011023].


\bibitem{Lin:2004en} 
  Z.~W.~Lin, C.~M.~Ko, B.~A.~Li, B.~Zhang and S.~Pal,
  Phys.\ Rev.\ C {\bf 72}, 064901 (2005)
  doi:10.1103/PhysRevC.72.064901
  [nucl-th/0411110].


\bibitem{Merzlaya_CPOD2018} A.~Merzlaya, these proceedings.



\bibitem{Alt:2007cd} 
  C.~Alt {\it et al.} [NA49 Collaboration],
  Phys.\ Rev.\ C {\bf 77}, 034906 (2008)
  doi:10.1103/PhysRevC.77.034906
  [arXiv:0711.0547 [nucl-ex]].



\bibitem{Abgrall:2013qoa} 
  N.~Abgrall {\it et al.} [NA61/SHINE Collaboration],
  Eur.\ Phys.\ J.\ C {\bf 74}, no. 3, 2794 (2014)
  doi:10.1140/epjc/s10052-014-2794-6
  [arXiv:1310.2417 [hep-ex]].



\bibitem{Alt:2003vb} 
  C.~Alt {\it et al.} [NA49 Collaboration],
  Phys.\ Rev.\ Lett.\  {\bf 92}, 042003 (2004)
  doi:10.1103/PhysRevLett.92.042003
  [hep-ex/0310014].



\end{thebibliography}
\end{document}